\documentclass[reprint,aps,prd,onecolumn,nofootinbib,notitlepage]{revtex4-1}

\usepackage[utf8]{inputenc}
\usepackage[T1]{fontenc}
\usepackage{mathtools}

\usepackage{mathrsfs} 
\usepackage{amssymb} 
\usepackage{anyfontsize} 
\usepackage{natbib}

\usepackage{xcolor}
\usepackage[ocgcolorlinks, colorlinks = true, linkcolor = blue, citecolor = green, urlcolor = blue, filecolor = red, pdftoolbar = true, pdfstartview = FitH, pdfmenubar = true]{hyperref}

{

	\hypersetup{pdftitle=1805.12197 Family of scalar-nonmetricity theories of gravity,pdfauthor=Mihkel R\"unkla and Ott Vilson,pdfkeywords=nonmetricity; scalar-nonmetricity gravity}
}

\newlength{\mytextsize}


\newcommand*\tinierTextFont[1]{\fontsize{ 0.35\mytextsize }{ 0.42\mytextsize }\selectfont{}#1{}\normalsize\selectfont}

\newcommand*\tinyTextforTop[1]{\text{\centering\tinierTextFont{#1}}}

	\newcommand*{\GLC}{ \overset{ \tinyTextforTop{LC} }{\Gamma}{}}
	\newcommand*{\GSTP}{ \overset{ \tinyTextforTop{STP} }{\Gamma}{}}

	\newcommand*{\DLC}{ \overset{ \tinyTextforTop{LC} }{\nabla}{}}
	\newcommand*{\DSTP}{ \overset{ \tinyTextforTop{STP} }{\nabla}{}}

	\newcommand*{\RLC}{ \overset{ \tinyTextforTop{LC}}{R}{} }
	\newcommand*{\RSTP}{ \overset{ \tinyTextforTop{STP} }{R}{} }

	\newcommand*{\TSTP}{ \overset{ \tinyTextforTop{STP} }{T}{}}

	\newcommand*{\QSTP}{ \overset{ \tinyTextforTop{STP} }{Q}{}}
	\newcommand*{\QtSTP}{ \overset{ \tinyTextforTop{STP} }{\tilde{Q}}{}}

	\newcommand*{\LSTP}{ \overset{ \tinyTextforTop{STP} }{L}{}}
	\newcommand*{\PSTP}{ \overset{  \tinyTextforTop{STP} }{P}{}}
	\newcommand*{\PcalSTP}{ \overset{  \tinyTextforTop{STP} }{\mathcal{P}}{}}

\begin{document}

\title{Family of scalar-nonmetricity theories of gravity}

\author{Mihkel R\"unkla}
\email{mrynkla@ut.ee}

\author{Ott Vilson}
\email{ovilson@ut.ee}

\affiliation{Laboratory of Theoretical Physics, Institute of Physics, University of Tartu, W. Ostwaldi Str 1, 50411 Tartu, Estonia}



\begin{abstract}
	We extend the class of recently formulated scalar-nonmetricity theories by coupling a five-parameter nonmetricity  scalar to a scalar field and considering a mixed kinetic term between the metric and the scalar field. The symmetric teleparallel constraint is invoked by Lagrange multipliers or by inertial variation. The equivalents for the general relativity and ordinary (curvature-based) scalar-tensor theories are obtained as particular cases. We derive the field equations, discuss some technical details, e.g., debraiding, and formulate the Hamilton-like approach.
\end{abstract}


\maketitle

\twocolumngrid

\tableofcontents

\section{Introduction}

Both the success and failure of general relativity (GR) motivate community to conduct the study of gravity theories in two directions. The first direction focuses on finding alternative formulations of general relativity, and a well-known example of this kind is teleparallel gravity \cite{Aldrovandi:2013wha}. The latter imposes a zero curvature constraint which yields to an alternative interpretation of  gravity: it is torsion \cite{Aldrovandi:2013wha,Maluf:2013gaa} or nonmetricity \cite{Nester:1998mp,BeltranJimenez:2017tkd} rather than curvature that mediates gravitational interaction. Though a mere rephrasing should not extend the scope of the theory, it might give new insights and deeper understanding than the original formulation.
For example in classical mechanics the Noether theorem does not reveal anything that could not be deduced from the equations of motion. The theorem is nevertheless useful as it points out what to look for.

The second direction in the study of gravity theories involves extensions of general relativity. Perhaps the simplest extension is given by including a scalar field in the gravity sector yielding to scalar-tensor gravity \cite{PhysRev.124.925,Flanagan:2004bz}. The first generation of scalar-tensor theories without derivative couplings or higher derivative terms involves a non-minimal coupling between the scalar field and the curvature scalar and therefore these theories are dubbed also as scalar-curvature theories. Although one could consider multiple scalar fields \cite{Damour:1992we} and higher generations of scalar-tensor theories such as Horndeski \cite{Horndeski:1974wa} and beyond \cite{Gleyzes:2014dya}, the simplest scalar-curvature theories exhibit inflationary solutions \cite{Belinchon:2016lwr}, and are powerful enough to explain phenomenologically the early inflationary epoch \cite{Ade:2015lrj} or the current accelerated expansion of the universe.

In this paper our route encompasses both of the aforementioned directions: we reformulate general relativity using the symmetric teleparallel connection and extend the theory by allowing arbitrary coefficients in the quadratic nonmetricity scalar (referred to as the newer general relativity in \cite{BeltranJimenez:2017tkd}) which is nonminimally coupled to a scalar field. This generalizes the theories formulated in \cite{Jarv:2018bgs} where the quadratic nonmetricity scalar was simply the quadratic Einstein Lagrangian, which without nonminimal coupling would yield to the symmetric teleparallel equivalent of general relativity. 

Considering affine connection as an independent variable in addition to the metric is referred to as the so-called Palatini variation or working in the metric-affine framework. The research directions involving nonmetricity are not new and there are several studies in this field mainly in the context of metric-affine gravity and possible microstructure of spacetime \cite{Hehl:1994ue,Neeman:1996zcr,Puetzfeld:2007hr,Vitagliano,Latorre:2017uve,Ariki:2017qov}. General affine connection contains additional structures to the Levi-Civita connection such as torsion and nonmetricity. As the latter are tensorial, one can argue at a textbook level that including them yields to just a theory with some additional fields \cite{Carroll:2004st}. However, from the gauge theory perspective one may ascribe to torsion and nonmetricity a more fundamental meaning and thus provide a further motivation for their inclusion \cite{Blagojevic:2012bc}. A related issue is whether the connection is coupled to other matter fields and whether it is constrained. A well-known example with the gravitational Lagrangian given by the Ricci scalar is the case  where a symmetric connection is neither coupled to matter fields nor invoking any other constraints, then the Palatini variation yields to no modification of the Levi-Civita connection. One can motivate the introduction of constraints from similar considerations in mechanics where constraints play a very useful role (e.g., describing the motion of a simple pendulum). In the current work we thus impose the symmetric teleparallel constraint, for previous studies involving symmetric teleparallelism consider \cite{Nester:1998mp,Adak:2004uh,Adak:2005cd,Adak:2008gd,Mol2017,BeltranJimenez:2017tkd,Conroy:2017yln,Koivisto:2018aip,BeltranJimenez:2018vdo,Hohmann:2018xnb,Harko:2018gxr,Heisenberg:2018vsk,Hohmann:2018wxu,Adak,Soudi:2018dhv}.

The symmetric teleparallel connection relies only on nonmetricity and  does not possess neither curvature nor torsion which yields to some interesting corollaries. One can transform to a zero connection gauge and thereby covariantize the partial derivatives as well as the split of the Einstein-Hilbert action into the Einstein Lagrangian density and a boundary term \cite{Nester:1998mp,BeltranJimenez:2017tkd}. The symmetric teleparallel covariant derivatives commute, this property can be for example used in order to eliminate the Lagrange multipliers from the connection equation \cite{Jarv:2018bgs}. Instead of introducing the Lagrange multipliers, one could alternatively assume the symmetric inertial connection from the beginning and perform the so-called inertial variation, both methods yield the same equations for the connection (for similar calculations in the torsion-based teleparallel framework see \cite{Golovnev:2017dox,Hohmann:2018rwf}). 

As this paper accompanies the work of \cite{Jarv:2018bgs} we look in more detail some of the issues discussed there but also use a different perspective. Thus in addition to the nonminimally coupled quadratic nonmetricity scalar we add to the action a mixed kinetic term and discuss its role in relation to scalar-curvature theories. In fact the particular expression is motivated by the boundary term in general relativity, and hence we are actually including a disguised curvature-based scalar-tensor theory. It is worth to pay attention that in principle one could consider modified or exotic matter fields which are coupled to symmetric teleparallel connection and yield to nonvanishing hypermomentum. In the latter case we would not obtain a simple scalar-tensor (or general relativity) equivalent since the matter sector is deformed.

A new perspective is the classical mechanics viewpoint of the quadratic nonmetricity theory. One can interpret the metric $g$ as the \textquotedblleft{}generalized coordinates\textquotedblright{} and its covariant derivative $Q$, which by definition is the nonmetricity, as the \textquotedblleft{}generalized velocity\textquotedblright{}. In the simplest case, by \textquotedblleft{}lowering the index\textquotedblright{} with the geometric object $\mathcal{G}$, which is \textquotedblleft{}the metric\textquotedblright{} in the kinetic term, one obtains the conjugate momentum (or superpotential). One can further transform to the Hamilton-like formulation and define the field space metric $\mathscr{G}$. It is noteworthy that the objects $\mathcal{G}$ and $\mathscr{G}$ possess several interesting properties from which one could obtain some physical insights (e.g., the initial value formulation).       

We adopt the conventions
\begin{subequations}
	\begin{align}
	K_{[\mu\nu]} &\equiv \frac{1}{2} \left( K_{\mu\nu} - K_{\nu\mu} \right) \,, \\
	K_{[\mu|\lambda|\nu]} &\equiv \frac{1}{2} \left( K_{\mu\lambda\nu} - K_{\nu\lambda\mu} \right) \,, \\
	K_{(\mu\nu)} &\equiv \frac{1}{2} \left( K_{\mu\nu} + K_{\nu\mu} \right) \,, \\ 
	K_{(\mu|\lambda|\nu)} &\equiv \frac{1}{2} \left( K_{\mu\lambda\nu} + K_{\nu\lambda\mu} \right)
	\end{align}
\end{subequations}
for (anti)symmetrization. We use the mostly plus signature of the metric and set $c=1$.

The paper is organized as follows. In the Section \ref{sec:Foreknowledge} we revise the concepts of nonmetricity and symmetric teleparallel connection (in that section stressed by $\mathrm{STP}$ on top of quantities, e.g., $\DSTP$), write down the quadratic kinetic term for the metric, and recall the contracted second Bianchi identity. Section \ref{sec:Action_and_field_equations} is devoted to postulating the action and deriving the field equations for the metric tensor $g^{\mu\nu}$, the scalar field $\Phi$, and for the connection $\Gamma^{\lambda}{}_{\mu\nu}$. In the Section \ref{sec:Hamilton_like_approach} we make use of $\DSTP_{\lambda} g_{\mu\nu} \neq 0$ in order to formulate a manifestly covariant Hamilton-like approach. Section \ref{sec:Summary} concludes the paper. The main body of the paper is followed by Appendixes \ref{sec:AppendixA}-\ref{app:metric_eom}, which contain further mathematical details.


\section{Foreknowledge\label{sec:Foreknowledge}}


\subsection{\texorpdfstring{Nonmetricity $Q_{\omega\mu\nu}$}{Nonmetricity}}

The nonmetricity
\begin{subequations}
\label{nonmetricity}
\begin{equation}
\tag{\theequation}
Q_{\omega\mu\nu} \equiv \nabla_{\omega} g_{\mu\nu} = Q_{\omega(\mu\nu)} \,, \quad
Q_{\omega}{}^{\sigma\rho} = - \nabla_{\omega} g^{\sigma\rho} \,,
\end{equation}
\end{subequations}
enters the coefficients of the affine connection as
\begin{align}
\Gamma^{\lambda}{}_{\mu\nu} = \GLC^{\lambda}{}_{\mu\nu} + L^{\lambda}{}_{\mu\nu} + K^{\lambda}{}_{\mu\nu} \,,
\end{align}
where
\begin{equation}
\tag{\theequation a}
\GLC^{\lambda}{}_{\mu\nu} \equiv \frac{1}{2} g^{\lambda\omega} \left( 2 \partial_{(\mu} g_{|\omega|\nu)} - \partial_{\omega} g_{\mu\nu} \right) = \GLC^{\lambda}{}_{(\mu\nu)}
\end{equation}
is the Levi-Civita part of the connection,
\begin{equation}
\label{definition_L}
\tag{\theequation b}
L^{\lambda}{}_{\mu\nu} \equiv -\frac{1}{2} g^{\lambda\omega} \left( 2 Q_{(\mu| \omega| \nu)} - Q_{\omega \mu \nu} \right) = L^{\lambda}{}_{(\mu\nu)} \,,
\end{equation}
and
\begin{equation}
\tag{\theequation c}
K^{\lambda}{}_{\mu\nu} \equiv \frac{1}{2} g^{\lambda\omega} \left( 2T_{(\mu|\omega|\nu)} + T_{\omega\mu\nu} \right) = g^{\lambda\omega} K_{[\omega|\mu|\nu]} \,.
\end{equation}
Here $T^{\lambda}{}_{\mu\nu} = T^{\lambda}{}_{[\mu\nu]}$ is the torsion. (Note that the torsion has been included for completeness. Actually, in the following sections we assume it to vanish.)

The nonmetricity tensor (\ref{nonmetricity}) possesses two independent contractions
\begin{subequations}
\label{contractions}
\begin{equation}
\tag{\theequation}
Q_{\omega} \equiv Q_{\omega\mu\nu} g^{\mu\nu} \,, \quad
\tilde{Q}_{\mu} \equiv Q_{\omega\mu\nu} g^{\omega\nu} \,.
\end{equation}
\end{subequations}
The first of them is related to the invariant volume form as
\begin{equation}
\nabla_{\omega} \sqrt{-g} = \frac{1}{2} \sqrt{-g} g^{\mu\nu} \nabla_{\omega} g_{\mu\nu} = \frac{1}{2} \sqrt{-g} Q_{\omega} \,.
\end{equation}
A straightforward calculation leads us further to
\begin{subequations}
\begin{align}
\sqrt{-g} R^{\sigma}{}_{\sigma\mu\nu} &= - 2\nabla_{[\mu} \nabla_{\nu]} \sqrt{-g} - T^{\lambda}{}_{\mu\nu} \nabla_{\lambda} \sqrt{-g} \\
\tag{\theequation$'$}
&= \sqrt{-g} \nabla_{[\nu} Q_{\mu]} - \frac{1}{2} \sqrt{-g} T^{\lambda}{}_{\mu\nu} Q_{\lambda} \\
\label{DQ_symmetry}
\tag{\theequation$''$}
&= \sqrt{-g} \DLC_{[\nu} Q_{\mu]} = \sqrt{-g} \partial_{[\nu} Q_{\mu]} \,,
\end{align}
\end{subequations}
which is the  homothetic or segmental curvature [cf.\ Eq.\ ({\hypersetup{urlcolor=red}\href{https://arxiv.org/pdf/0911.0334.pdf#page=22}{1.3.34}}) in Ref.\ \cite{Poplawski:2009fb}].


\subsection{\texorpdfstring{Symmetric teleparallel connection $\protect\GSTP^{\lambda}{}_{\mu\nu}$}{Symmetric teleparallel connection}}

In the current paper we shall utilize the symmetric teleparallel (STP) connection $\GSTP^{\lambda}{}_{\mu\nu}$ by imposing, in addition to symmetricity
\begin{subequations}
\label{RT0}
\begin{align}
\label{vanishing_T}
\GSTP^{\lambda}{}_{\mu\nu} &= \GSTP^{\lambda}{}_{(\mu\nu)} \quad
\Leftrightarrow \quad \TSTP^{\sigma}{}_{\mu\nu} \equiv 2 \GSTP^{\sigma}{}_{[\mu\nu]}  \overset{!}{=} 0 \,, 
\phantomsection
\hypertarget{vanishing_R_place}{}\\
\shortintertext{also flatness}
\label{vanishing_R}
\RSTP^{\sigma}{}_{\rho\mu\nu} &\equiv 2 \partial_{[\mu} \GSTP^{\sigma}{}_{\nu]\rho} + 2 \GSTP^{\sigma}{}_{[\mu|\lambda|}\GSTP^{\lambda}{}_{\nu]\rho} \overset{!}{=} 0 \,.
\end{align}
\end{subequations}
In that case, based on the Proposition 10.4.1.\ in Ref.\ \cite{Bao:2000}, there exists a coordinate system $\lbrace \xi^{\sigma} \rbrace$ where the connection coefficients $\GSTP^{\lambda}{}_{\mu\nu}$ vanish, i.e.,
\begin{subequations}
\label{zero_Gamma}
\begin{equation}
\tag{\theequation}
\exists \, \lbrace \xi^{\sigma} \rbrace\,: \quad \GSTP^{\lambda}{}_{\mu\nu}(\xi^{\sigma}) = 0 \quad \Rightarrow \quad \left. \DSTP_{\mu}()\right|_{\lbrace \xi^{\sigma} \rbrace} = \partial_{\mu}() \,,
\end{equation}
\end{subequations}
provided that the considered covariant derivative is partial derivative plus additive terms multiplied by the coefficients $\GSTP^{\lambda}{}_{\mu\nu}$. The result (\ref{zero_Gamma}) leads us to interesting corollaries. In particular, firstly, the covariant derivatives commute \cite{Nester:1998mp} [cf.\ Eqs.\ (1.28) and (1.29) in Ref.\ \cite{Ortin:2015hya}]
\begin{subequations}
\label{symmetry_of_covariant_derivatives}
\begin{equation}
\tag{\theequation}
\left. \DSTP_{\mu}\DSTP_{\nu} \mathbb{T} \right|_{ \lbrace \xi^{\sigma} \rbrace } = \partial_{\mu} \partial_{\nu} \mathbb{T} = \partial_{\nu} \partial_{\mu} \mathbb{T} = \left. \DSTP_{\nu} \DSTP_{\mu} \mathbb{T} \right|_{ \lbrace \xi^{\sigma} \rbrace } \,,
\end{equation}
\end{subequations}
where $\mathbb{T}$ is a tensor (density) of arbitrary rank (and weight). Secondly, in an arbitrary coordinate system $\lbrace x^{\mu} \rbrace$, the connection coefficients read \cite{BeltranJimenez:2017tkd}
\begin{subequations}
\label{STP_connection_via_xi}
\begin{equation}
\tag{\theequation}
\GSTP^{\lambda}{}_{\mu\nu} = \frac{\partial x^{\lambda}}{\partial \xi^{\sigma}} \frac{\partial}{\partial x^{\mu}} \left( \frac{\partial \xi^{\sigma}}{\partial x^{\nu}} \right) \,,
\end{equation}
\end{subequations}
where $\lbrace \xi^{\sigma} \rbrace$ are the coordinates for which (\ref{zero_Gamma}) holds.

Thirdly, one can covariantize the split \cite{Nester:1998mp}
\begin{subequations}
\label{split}
\begin{equation}
\tag{\theequation}
\sqrt{-g} \RLC = \sqrt{-g} \mathcal{L}_{\mathrm{E}} - \partial_{\sigma} \left( \sqrt{-g} \mathscr{B}^{\sigma} \right) 
\end{equation}
\end{subequations}

\phantomsection
\label{Einstein_L}

\vspace{-\parskip}

\noindent{}where [see Eq.\ ({\hypersetup{urlcolor=red}\href{https://arxiv.org/pdf/gr-qc/9809049.pdf#page=4}{8}}) in Ref.\ \cite{Nester:1998mp}, and also, e.g., Eq.\ ({\hypersetup{urlcolor=red}\href{https://arxiv.org/pdf/1710.03116v2.pdf#page=4}{28}}) in Ref.\ \cite{BeltranJimenez:2017tkd}]
\begin{align}
\tag{\theequation a}
\label{Einstein_Lagrangian}
\mathcal{L}_{\mathrm{E}} ={}& \GLC^{\rho}{}_{\lambda\sigma} g^{\lambda\nu} \GLC^{\sigma}{}_{\nu\rho} - \GLC^{\lambda}{}_{\sigma\lambda} \GLC^{\sigma}{}_{\nu\rho} g^{\nu\rho} 
\phantomsection
\hypertarget{Einstein_L_prime}{}\\
\nonumber
={}& \partial_{\lambda} g^{\mu\nu} \Bigg( -\frac{1}{4} g^{\lambda\omega} g_{\mu \sigma } g_{\nu \rho} + \frac{1}{2} \delta^{\omega}_{\nu} g_{\mu \sigma } \delta^{\lambda}_{\rho}  \\
\tag{\theequation a$'$}
\label{Einstein_Lagrangian_prime}
&{}+ \frac{1}{4} g_{\mu\nu} g^{\lambda \omega } g_{\sigma\rho} -\frac{1}{2} g_{\mu\nu} \delta^{\lambda}_{\rho } \delta^{\omega}_{\sigma} \Bigg)  \partial_{\omega} g^{\sigma\rho} 
\end{align}

\phantomsection
\label{boundary_term_place}
\vspace{-\parskip}

\noindent{}is the quadratic Einstein Lagrangian, and
\begin{align}
\label{boundary_term}
\tag{\theequation b}
\mathscr{B}^{\sigma} &= g^{\sigma\rho} \GLC^{\nu}{}_{\nu\rho} - \GLC^{\sigma}{}_{\nu\rho} g^{\nu\rho} 
\phantomsection
\hypertarget{boundary_term_prime_place}{}
\\
\tag{\theequation b$'$}
\label{boundary_term_prime}
&= g^{\sigma\rho} \left(\partial_{\rho} g_{\mu\nu} \right) g^{\mu\nu} - g^{\sigma\rho} \left( \partial_{\mu} g_{\rho\nu} \right) g^{\mu\nu}
\end{align}
is the boundary term, hosting the second derivatives of the metric that reside in $\RLC$. From the viewpoint of the Levi-Civita connection, neither (\hyperref[Einstein_L]{\ref*{Einstein_Lagrangian}}) nor (\hyperref[boundary_term_place]{\ref*{boundary_term}}) is a tensor. However, both terms can be covariantized by considering the symmetric teleparallel connection and promoting the partial derivatives in (\hyperlink{Einstein_L_prime}{\ref*{Einstein_Lagrangian_prime}}) and (\hyperlink{boundary_term_prime_place}{\ref*{boundary_term_prime}}) to covariant ones, thus reversing the line of thought that underlies (\ref{zero_Gamma}). The Einstein quadratic Lagrangian (\hyperlink{Einstein_L_prime}{\ref*{Einstein_Lagrangian_prime}}) yields [see, e.g., Eq.\ ({\hypersetup{urlcolor=red}\href{https://arxiv.org/pdf/1710.03116.pdf#page=2}{17}}) in Ref.\ \cite{BeltranJimenez:2017tkd}, as well as Eq.\ ({\hypersetup{urlcolor=red}\href{https://arxiv.org/pdf/1802.00492.pdf#page=3}{18}}) in Ref.\ \cite{Jarv:2018bgs}]
\begin{subequations}
\label{definition_of_GRQ}
\begin{align}
\nonumber
\mathcal{L}_{\mathrm{E},\mathrm{cov}} = \QSTP \equiv{}&{} -\frac{1}{4}\QSTP_{\lambda\mu\nu} \QSTP^{\lambda\mu\nu} + \frac{1}{2} \QSTP_{\lambda\mu\nu} \QSTP^{\nu\mu\lambda} \\
\tag{\theequation}
&{} + \frac{1}{4} \QSTP_{\mu} \QSTP^{\mu} - \frac{1}{2} \QSTP_{\mu} \QtSTP^{\mu} \,,
\end{align}
\end{subequations} 
while [cf.\ Eq.\ ({\hypersetup{urlcolor=red}\href{https://arxiv.org/pdf/1802.00492.pdf#page=3}{17}}) in Ref.\ \cite{Jarv:2018bgs}]
\begin{subequations}
\label{covariant_boundary}
\begin{equation}
\tag{\theequation}
\mathscr{B}^{\sigma}_{\mathrm{cov}} = \QSTP^{\sigma} - \QtSTP^{\sigma}
\end{equation}
\end{subequations}
is the covariantized version of the boundary term (\hyperlink{boundary_term_prime_place}{\ref*{boundary_term_prime}}), as
\begin{equation}
\left. \mathcal{L}_{\mathrm{E},\mathrm{cov}} \right|_{ \lbrace \xi^{\tau} \rbrace } \overset{(\ref{zero_Gamma})}{=} \mathcal{L}_{\mathrm{E}} \,, \quad 
\left. \mathscr{B}^{\sigma}_{\mathrm{cov}} \right|_{ \lbrace \xi^{\tau} \rbrace } \overset{(\ref{zero_Gamma})}{=} \mathscr{B}^{\sigma} \,.
\end{equation}


\subsection{\texorpdfstring{Kinetic term for the metric $g^{\mu\nu}$}{Kinetic term for the metric}}

The nonvanishing covariant derivative of the metric $g^{\mu\nu}$ allows us to consider the kinetic term for the metric indeed analogously to the kinetic energy in classical mechanics. Let us define\footnote{Note that in this section we actually do not need to assume the symmetric teleparallel connection, we just need the nonmetricity. Thus, the quantities $Q_{\lambda\mu\nu}$, etc., will not be equipped with \textquoteleft{}$\mathrm{STP}$\textquoteright{} on top. Concerning notation, see also Subsubsec.\ \ref{subsubsec:notation} and footnote \ref{footnote:notation}.}
\begin{subequations}
\label{definition_of_Q}
\begin{equation}
\tag{\theequation}
\mathcal{Q} \equiv Q_{\lambda}{}^{\mu\nu} \mathcal{G}^{\lambda}{}_{\mu\nu}{}^{\omega}{}_{\sigma\rho} Q_{\omega}{}^{\sigma\rho} \,,
\end{equation}
\end{subequations}
where\footnote{The form $\delta^{\hphantom{(}\alpha}_{(\mu} g_{\nu)\beta} g^{\lambda\omega} \delta^{\hphantom{(}\beta}_{(\sigma} g_{\rho)\alpha}$ (multiplied by $c_1$) in the first line of Eq.\ (\ref{definition_of_G}) emphasizes the symmetry (\ref{symmetry}) but for practical calculations $g_{\mu(\rho} g_{\sigma)\nu} g^{\lambda\omega} = \delta^{\hphantom{(}\alpha}_{(\mu} g_{\nu)\beta} \delta^{\hphantom{(}\beta}_{(\sigma} g_{\rho)\alpha} g^{\lambda\omega} = g_{\rho(\mu} g_{\nu)\sigma}g^{\lambda\omega}$ is more suitable.}
\begin{subequations}
\label{definition_of_G}
\begin{align}
\nonumber
\mathcal{G}^{\lambda}{}_{ \mu\nu }{}^{\omega}{}_{ \sigma\rho } \equiv{}& c_1 \delta^{\hphantom{(}\alpha}_{(\mu} g_{\nu)\beta} g^{\lambda\omega} \delta^{\hphantom{(}\beta}_{(\sigma} g_{\rho)\alpha} + c_2 \delta^{\hphantom{(}\omega}_{(\nu} g_{\mu)(\sigma} \delta^{\lambda}_{\rho)} \\
\nonumber
&{} + c_3 g_{\mu\nu} g^{\lambda\omega} g_{\sigma\rho} + c_4 \delta^{\hphantom{(}\lambda}_{(\nu} g_{\mu)(\sigma} \delta^{\omega}_{\rho)} \\
\tag{\theequation}
&{} + \frac{c_5}{2} g_{\mu\nu} \delta^{\hphantom{(}\lambda}_{(\sigma} \delta^{\omega}_{\rho)} + \frac{c_5}{2} g_{\sigma\rho} \delta^{\hphantom{(}\omega}_{(\mu} \delta^{\lambda}_{\nu)} \,,
\end{align}
\end{subequations}
with constants $c_1$, $\ldots$, $c_5$, and definitions (\ref{nonmetricity}), (\ref{contractions}), contracts in Eq.\ (\ref{definition_of_Q}) to give \cite{BeltranJimenez:2017tkd}
\begin{subequations}
\label{definition_of_Q_explicit}
\begin{align}
\nonumber
\mathcal{Q} ={}& c_1 Q_{\lambda\mu\nu} Q^{\lambda\mu\nu} + c_2 Q_{\lambda\mu\nu} Q^{\nu\mu\lambda} + c_3 Q_{\lambda} Q^{\lambda} \\
\tag{\theequation}
&{} + c_4 \tilde{Q}_{\mu} \tilde{Q}^{\mu} + c_5 Q_{\mu} \tilde{Q}^{\mu} \,.
\end{align}
\end{subequations}
Let us point out that in addition to the symmetries
\begin{subequations}
\label{G_symmetries}
\begin{align}
	\mathcal{G}^{\lambda}{}_{ \mu\nu }{}^{\omega}{}_{ \sigma\rho } &= \mathcal{G}^{\lambda}{}_{ \nu\mu }{}^{\omega}{}_{ \sigma\rho } = \mathcal{G}^{\lambda}{}_{ (\mu\nu) }{}^{\omega}{}_{ \sigma\rho } \\
	&= \mathcal{G}^{\lambda}{}_{ \mu\nu }{}^{\omega}{}_{ \rho\sigma } = \mathcal{G}^{\lambda}{}_{ \mu\nu }{}^{\omega}{}_{ (\sigma\rho) } \\
	\shortintertext{the tensor $\mathcal{G}^{\lambda}{}_{ \mu\nu }{}^{\omega}{}_{ \sigma\rho }$ is symmetric}
	\label{symmetry}
	\mathcal{G}^{\lambda}{}_{ \mu\nu }{}^{\omega}{}_{ \sigma\rho } &= \mathcal{G}^{\omega}{}_{ \sigma\rho }{}^{\lambda}{}_{ \mu\nu }
\end{align}
\end{subequations}
in the sense of the Definition {\hypersetup{urlcolor=red}\href{https://arxiv.org/pdf/1109.3830.pdf#page=6}{3.9}} in Ref.\ \cite{Brazell}. Precisely the quality (\ref{symmetry}) furnishes the result [see definitions ({\hypersetup{urlcolor=red}\href{https://arxiv.org/pdf/1802.00650.pdf#page=5}{12}}) in Ref.\ \cite{Koivisto:2018aip} and ({\hypersetup{urlcolor=red}\href{https://arxiv.org/pdf/1710.03116.pdf#page=2}{18}}) in Ref.\ \cite{BeltranJimenez:2017tkd}]
\begin{subequations}
	\label{definition_of_P_all}
	\begin{align}
	\label{definition_of_P}
	\mathcal{P}^{\lambda}{}_{\mu\nu} \equiv{}& \frac{1}{2} \frac{\partial \mathcal{Q}}{\partial Q_{\lambda}{}^{\mu\nu}} = \mathcal{G}^{\lambda}{}_{ \mu\nu }{}^{\omega}{}_{ \sigma\rho } Q_{\omega}{}^{\sigma\rho} 
	\phantomsection
	\hypertarget{def_P_ex_place}{}
	\\
	\nonumber
	={}& c_1 Q^{\lambda}_{\hphantom{\lambda}\mu\nu} + c_2 Q^{\hphantom{(\mu}\lambda}_{(\mu \hphantom{\lambda}\nu)} + c_3 Q^{\lambda} g_{\mu\nu} \\
	\label{definition_of_P_explicit}
	&{} + c_4 \delta^{\hphantom{(}\lambda}_{( \mu} \tilde{Q}_{\nu)} + \frac{c_5}{2} \left( \tilde{Q}^{\lambda} g_{\mu\nu} + \delta^{\hphantom{(}\lambda}_{(\mu} Q_{\nu)} \right) \,.
	\end{align}
\end{subequations}
From (\ref{definition_of_P}) one can clearly see a similarity to classical mechanics. In terms of an analogy, for the simplest case, the free particle, the \textquotedblleft{}generalized momentum\textquotedblright{} $\mathcal{P}^{\lambda}{}_{\mu\nu}$ is obtained by taking the derivative of the \textquotedblleft{}kinetic energy\textquotedblright{} $\frac{1}{2} \mathcal{Q}$ with respect to the \textquotedblleft{}generalized velocity\textquotedblright{} $Q_{\lambda}{}^{\mu\nu}$. \textquotedblleft{}Lowering the index\textquotedblright{} of the \textquotedblleft{}generalized velocity\textquotedblright{} with the \textquotedblleft{}metric\textquotedblright{} $\mathcal{G}^{\lambda}{}_{\mu\nu}{}^{\omega}{}_{\sigma\rho}$ yields the \textquotedblleft{}generalized momentum\textquotedblright{}.


\subsubsection{\texorpdfstring{Varying $\mathcal{G}^{\lambda}{}_{\mu\nu}{}^{\omega}{}_{\sigma\rho}$}{Varying G}}

\phantomsection
\label{varying_G_place}
A straightforward calculation shows that the variation of (\ref{definition_of_G}) yields
\begin{align}
\label{varying_G}
\delta \mathcal{G}^{\lambda \hphantom{\mu\nu} \omega }_{\hphantom{\lambda}\mu\nu \hphantom{\omega} \sigma\rho } \equiv  \left( \Delta \mathcal{G}^{\lambda \hphantom{\mu\nu} \omega }_{\hphantom{\lambda}\mu\nu \hphantom{\omega} \sigma\rho } \right)_{\beta\alpha} \delta g^{\alpha\beta}\,,
\end{align}
where
\begin{align}
\nonumber
&\left( \Delta \mathcal{G}^{\lambda}{}_{\mu\nu}{}^{ \omega }{}_{\sigma\rho } \right)_{\beta\alpha} = \frac{1}{2} \Big\lbrace \delta^{\lambda}_{\beta} g_{\alpha\tau} \mathcal{G}^{\tau}{}_{\mu\nu}{}^{ \omega }{}_{\sigma\rho } + \delta^{\omega}_{\beta} g_{\alpha\tau} \mathcal{G}^{\tau}{}_{\sigma\rho}{}^{ \lambda }{}_{\mu\nu } \\
\tag{\theequation a}
&\quad{} - 2 g_{\alpha(\mu} \mathcal{G}^{\lambda}{}_{\nu)\beta}{}^{ \omega }{}_{\sigma\rho } - 2 g_{\alpha(\sigma} \mathcal{G}^{\omega}{}_{\rho)\beta}{}^{ \lambda }{}_{\mu\nu } \Big\rbrace \,.
\end{align}
The positioning of the indices emphasizes that the variation respects the symmetries (\ref{G_symmetries}) of $\mathcal{G}^{\lambda}{}_{\mu\nu}{}^{\omega}{}_{\sigma\rho}$, i.e.,
\begin{align}
\tag{\theequation b}
\left( \Delta \mathcal{G}^{\lambda}{}_{\mu\nu}{}^{\omega}{}_{\sigma\rho} \right)_{\beta\alpha} = \left( \Delta \mathcal{G}^{\lambda}{}_{(\mu\nu)}{}^{\omega}{}_{(\sigma\rho)} \right)_{\beta\alpha} = \left( \Delta \mathcal{G}^{\omega}{}_{\sigma\rho}{}^{\lambda}{}_{\mu\nu} \right)_{\beta\alpha} \,.
\end{align}
While it is clear that varying with respect to a symmetric object $g^{\alpha\beta}$ must yield a symmetric result, a straightforward calculation verifies
\vspace{-\baselineskip}
\phantomsection
\label{sym_G_var_place}
\vspace{\baselineskip}
\begin{equation}
\tag{\theequation c}
\label{symmetry_of_G_variation}
\left( \Delta \mathcal{G}^{\lambda}{}_{\mu\nu}{}^{\omega}{}_{\sigma\rho} \right)_{\beta\alpha} = \left( \Delta \mathcal{G}^{\lambda}{}_{\mu\nu}{}^{\omega}{}_{\sigma\rho} \right)_{(\beta\alpha)} \,,
\end{equation}
and therefore there is no need to invoke the symmetrizing brackets. Analogously
\begin{equation}
\tag{\theequation d}
\nabla_{\xi} \mathcal{G}^{\lambda \hphantom{\mu\nu} \omega }_{\hphantom{\lambda}\mu\nu \hphantom{\omega} \sigma\rho } = - \left( \Delta \mathcal{G}^{\lambda \hphantom{\mu\nu} \omega }_{\hphantom{\lambda}\mu\nu \hphantom{\omega} \sigma\rho } \right)_{\beta\alpha} Q_{\xi}{}^{\alpha\beta} \,,
\end{equation}
where the minus sign appears due to the convention (\ref{nonmetricity}).


\subsubsection{Equivalent of general relativity}

By comparing Eqs.\ (\ref{definition_of_GRQ}) and (\ref{definition_of_Q_explicit}), we conclude that the symmetric teleparallel equivalent of general relativity is covered by the coefficients
\begin{subequations}
\label{GR_coefficients}
	\begin{alignat}{3}
	c_1 &= -\frac{1}{4} \,, \quad &
	c_2 &=  \hphantom{-}\frac{1}{2} \,, \quad
	c_3 = \frac{1}{4} \,, \\
	c_4 &= \hphantom{-}0 \,, \quad &
	c_5 &= -\frac{1}{2} \,.
	\end{alignat}
\end{subequations}
Expression (\ref{definition_of_G}) reduces to
\begin{subequations}
\label{definition_of_GGR}
\begin{align}
\nonumber
G^{\lambda}{}_{ \mu\nu }{}^{\omega}{}_{ \sigma\rho } \equiv{}&{} -\frac{1}{4} \delta^{\hphantom{(}\alpha}_{(\mu} g_{\nu)\beta} g^{\lambda\omega} \delta^{\hphantom{(}\beta}_{(\sigma} g_{\rho)\alpha} + \frac{1}{2} \delta^{\hphantom{(}\omega}_{(\nu} g_{\mu)(\sigma} \delta^{\lambda}_{\rho)} \\
\nonumber
&{} + \frac{1}{4} g_{\mu\nu} g^{\lambda\omega} g_{\sigma\rho} - \frac{1}{4} g_{\mu\nu} \delta^{\hphantom{(}\lambda}_{(\sigma} \delta^{\omega}_{\rho)} \\ 
\tag{\theequation}
&{} - \frac{1}{4} g_{\sigma\rho} \delta^{\hphantom{(}\omega}_{(\mu} \delta^{\lambda}_{\nu)} \,,
\end{align}
\end{subequations}
which is the contracting object in (\hyperlink{Einstein_L_prime}{\ref*{Einstein_Lagrangian_prime}}), symmetrized with respect to (\ref{G_symmetries}). In particular the splitting of the last term appears due to (\ref{symmetry}). Let us point out that the variation (\hyperref[varying_G_place]{\ref*{varying_G}}), applied to (\ref{definition_of_GGR}), is useful also in the context of general relativity, if one plugs the Einstein Lagrangian (\hyperlink{Einstein_L_prime}{\ref*{Einstein_Lagrangian_prime}}) into the Euler-Lagrange equations. Definition (\hyperlink{def_P_ex_place}{\ref*{definition_of_P_explicit}}) yields
\begin{subequations}
\label{definition_of_GRP}
\begin{align}
\nonumber
P^{\lambda}{}_{\mu\nu} \equiv{}&{} - \frac{1}{4} Q^{\lambda}_{\hphantom{\lambda}\mu\nu} + \frac{1}{2} Q^{\hphantom{(\mu}\lambda}_{(\mu \hphantom{\lambda}\nu)} \\
\tag{\theequation}
&{} + \frac{1}{4} \left( Q^{\lambda} - \tilde{Q}^{\lambda} \right) g_{\mu\nu} - \frac{1}{4} \delta^{\hphantom{(}\lambda}_{(\mu} Q_{\nu)} 
\end{align}
\end{subequations}
[cf.\ definition ({\hypersetup{urlcolor=red}\href{https://arxiv.org/pdf/1802.00492.pdf#page=3}{24}}) in Ref.\ \cite{Jarv:2018bgs}].


\subsection{Bianchi identity\label{Sec:Bianchi_identity}}

If we impose (\ref{RT0}), then
\begin{subequations}
	\label{Riemann_and_Ricci}
	\begin{align}
	\nonumber
	\RLC^{\omega}{}_{\rho\mu\nu} ={}&{} - \DSTP_{\mu} \LSTP^{\omega}{}_{ \nu \rho} + \DSTP_{\nu} \LSTP^{\omega}{}_{\mu \rho} \\
	\label{Riemann}
	&{} - \LSTP^{\lambda}{}_{ \mu \rho} \LSTP^{\omega}{}_{ \nu \lambda} + \LSTP^{\lambda}{}_{\nu \rho} \LSTP^{\omega}{}_{ \mu \lambda} \,, 
	\phantomsection
	\hypertarget{Ricci_tensor_place}{}\\
	\nonumber
	\RLC^{\sigma}{}_{\nu} ={}& \frac{2}{\sqrt{-g}} \DSTP_{\lambda} \left( \sqrt{-g} \PSTP^{\lambda\sigma}{}_{\nu} \right) + \PSTP^{\sigma}{}_{\omega\lambda} \QSTP_{\nu}{}^{\omega\lambda} \\
	\label{Ricci_tensor}
	&{} - \frac{1}{2} \DLC_{\omega} \left( \QSTP^{\omega} - \QtSTP^{\omega} \right) \delta^{\sigma}_{\nu} \,, \\
	\RLC ={}& \QSTP - \DLC_{\omega} \left( \QSTP^{\omega} - \QtSTP^{\omega} \right) \,.
	\end{align}
\end{subequations}
\phantomsection
\label{Einstein_tensor_place}
Therefore, by making use of the definitions (\ref{definition_L}), (\ref{definition_of_GRQ}), (\ref{definition_of_GRP}), and the result (\ref{Riemann_and_Ricci}),
\begin{align}
\nonumber
&E^{\sigma}{}_{\nu} \equiv \RLC^{\sigma}{}_{\nu} - \frac{1}{2} \delta^{\sigma}_{\nu} \RLC \\
\label{Einstein_tensor}
&\quad = \frac{2}{\sqrt{-g}} \DSTP_{\lambda} \left( \sqrt{-g} \PSTP^{\lambda\sigma}{}_{\nu} \right) + \PSTP^{\sigma}{}_{\omega\lambda} \QSTP_{\nu}{}^{\omega\lambda} - \frac{1}{2} \delta^{\sigma}_{\nu} \QSTP  
\end{align}
is the Einstein tensor.

One can show that for a symmetric tensor $E_{\mu\nu} = E_{(\mu\nu)}$
\begin{subequations}
\label{derivative_LC_to_STP}
\begin{align}
\nonumber
&\DLC_{\sigma} \left( \sqrt{-g} E^{\sigma}{}_{\nu} \right) = \DSTP_{\sigma} \left( \sqrt{-g} E^{\sigma}{}_{\nu} \right) + \sqrt{-g} \LSTP^{\lambda\sigma}{}_{\nu} E_{\sigma\lambda} \\
\tag{\theequation}
&\qquad = \DSTP_{\sigma} \left( \sqrt{-g} E^{\sigma}{}_{\nu} \right) - \frac{1}{2} \sqrt{-g} \QSTP_{\nu}{}^{\lambda\sigma} E_{\sigma\lambda} \,.
\end{align}
\end{subequations}
By a straightforward calculation
\begin{align}
\nonumber
\DSTP_{\sigma} &\left( \sqrt{-g} E^{\sigma}{}_{\nu} \right) \\
&= 2 \DSTP_{\sigma} \DSTP_{\lambda} \left( \sqrt{-g} \PSTP^{\lambda\sigma}{}_{\nu} \right) + \frac{1}{2} \sqrt{-g} \QSTP_{\nu}{}^{\lambda\sigma} E_{\lambda\sigma} \,,
\end{align}
where in addition to (\hyperref[varying_G_place]{\ref*{varying_G}}) we made use of
\begin{subequations}
	\begin{align}
	\nonumber
	&\DSTP_{\sigma} \left( \sqrt{-g} \PSTP^{\sigma}{}_{\omega\lambda} \right) g^{\omega\mu} = \\
	&\quad= \DSTP_{\sigma} \left( \sqrt{-g} \PSTP^{\sigma\mu}{}_{\lambda} \right) + \sqrt{-g} \PSTP^{\sigma}{}_{\omega\lambda} \QSTP_{\sigma}{}^{\omega\mu} \,, \\
	&\DSTP_{\mu} \QSTP_{\nu\sigma\rho} = \DSTP_{\nu} \QSTP_{\mu\sigma\rho} \,, \qquad
	\DSTP_{\mu} \QSTP_{\nu}^{\hphantom{\nu} \sigma\rho} = \DSTP_{\nu} \QSTP_{\mu}^{\hphantom{\mu} \sigma\rho } \,.
	\end{align}
\end{subequations}
Hence
\begin{subequations}
\label{BianchiId}
\begin{equation}
\tag{\theequation}
\DLC_{\sigma} \left( \sqrt{-g} E^{\sigma}{}_{\nu} \right) = 2 \DSTP_{\sigma} \DSTP_{\lambda} \left( \sqrt{-g} \PSTP^{\lambda\sigma}{}_{\nu} \right) = 0  \,.
\end{equation}
\end{subequations}
The obtained result also follows from the symmetries of the index structure of the included objects. In particular, based on (\ref{definition_of_GRP}),
\begin{subequations}
\label{GRP_symmetric}
\begin{align}
\tag{\theequation}
&2\sqrt{-g} \PSTP^{(\lambda\sigma)}{}_{\nu} =  \DSTP_{\omega} \left( \sqrt{-g} \delta^{(\lambda}_{\hphantom{(}\rho} g^{\sigma)[\omega} \delta^{\rho]}_{\nu} \right) \\
\nonumber
&\quad =  \frac{\sqrt{-g}}{2} \left( \QSTP_{\nu}{}^{\lambda\sigma} + \frac{1}{2} \QSTP^{(\lambda} \delta^{\sigma)}_{\nu} - \QtSTP^{(\lambda} \delta^{\sigma)}_{\nu} - \frac{1}{2} \QSTP_{\nu} g^{\lambda\sigma} \right) \,.
\end{align}
\end{subequations}
Hence, acting on (\ref{GRP_symmetric}) with $\DSTP_{\sigma} \DSTP_{\lambda}$,
\begin{subequations}
\label{Bianchi_identity_form}
\begin{equation}
\tag{\theequation}
 2 \DSTP_{\sigma} \DSTP_{\lambda} \left( \sqrt{-g} \PSTP^{(\lambda\sigma)}{}_{\nu} \right) = \DSTP_{\sigma} \DSTP_{\lambda} \DSTP_{\omega} \left( \sqrt{-g} \delta^{(\lambda}_{\hphantom{(}\rho} g^{\sigma)[\omega} \delta^{\rho]}_{\nu} \right),
\end{equation}
\end{subequations}
and taking into account that the covariant derivatives commute (\ref{symmetry_of_covariant_derivatives}) yields to the zero result (\ref{BianchiId}).


\subsubsection{Bianchi identity backwards}

Yet another possibility for obtaining the general relativity motivated coefficients (\ref{GR_coefficients}) is the following. Let us consider generic coefficients $c_1$, $\ldots$, $c_5$ and the definition (\hyperlink{def_P_ex_place}{\ref*{definition_of_P_explicit}}). By imposing
\begin{subequations}
\label{Deformed_Bianchi_identity}
\begin{equation}
\tag{\theequation}
\DSTP_{\sigma} \DSTP_{\lambda} \left( \sqrt{-g} \PcalSTP^{\lambda\sigma}{}_{ \nu} \right) \overset{!}{=} 0
\end{equation}
\end{subequations}
we obtain $62$ different terms, which vanish identically, if
\begin{subequations}
	\label{system}
	\begin{alignat}{3}
	2 c_1 + c_2 &= 0 \,, \qquad &
	2 c_3 + c_5 &= 0 \,, \\
	c_2 + c_5  &= 0 \,, \qquad &
	c_4 &= 0 \,.
	\end{alignat}
\end{subequations}
Hence, up to an overall multiplier, we obtain the general relativity motivated coefficients (\ref{GR_coefficients}).

One can loosen the conditions by demanding only the second derivatives of $\QSTP_{\lambda\mu\nu}$ to vanish. The explicit terms in (\ref{Deformed_Bianchi_identity}) are
\begin{subequations}
\begin{align}
\frac{1}{2} \left(2 c_1 + c_2 + c_4 \right) \sqrt{-g} g^{\mu\lambda} g^{\sigma\rho} \DSTP_{\mu} \DSTP_{\sigma} \QSTP_{\rho \lambda \nu} &= 0 \,, \\
\frac{1}{2} \left( c_2 + c_4 + c_5 \right) \sqrt{-g} g^{\mu\lambda} g^{\sigma\rho} \DSTP_{\mu} \DSTP_{\sigma} \QSTP_{\nu \rho \lambda} &= 0 \,, \\
\frac{1}{2} \left( 2 c_3 + c_5 \right) \sqrt{-g} g^{\mu\lambda} \DSTP_{\mu} \DSTP_{\lambda} \QSTP_{\nu \sigma\rho } g^{\sigma\rho} &= 0 \,,
\end{align}
\end{subequations}
which are the three independent possibilities for placing indices. Hence, we slightly deform the system (\ref{system}) to yield
\begin{subequations}
	\begin{align}
	2 c_1 + \tilde{c}_2 &= 0 \,, \qquad
	2 c_3 + c_5 = 0 \,, \\
	\tilde{c}_2 + c_5 &= 0 \,,
	\end{align}
\end{subequations}
where
\begin{subequations}
\label{c_tilde}
\begin{equation}
\tag{\theequation}
\tilde{c}_2 = c_2 + c_4 \,.
\end{equation}
\end{subequations}
It is interesting to note that the sum (\ref{c_tilde}) is mentioned in \cite{BeltranJimenez:2017tkd} after Eq.\ ({\hypersetup{urlcolor=red}\href{https://arxiv.org/pdf/1710.03116.pdf#page=3}{23}}). Whatever deviation from the coefficients (\ref{GR_coefficients}), however, instantly introduces dozens of terms into (\ref{Deformed_Bianchi_identity}).


\subsection{Remark\label{ss:remark}}

Let us point out that many of the presented results are actually valid in the usual curvature-based general relativity as well. Namely, Eqs.\ (\ref{Riemann_and_Ricci}) are rather the usual definitions in the symmetric teleparallel disguise, than links between different geometries. Intuitively, if we consider the coincident gauge (\ref{zero_Gamma}) then $\DSTP = \partial$ and
\begin{equation}
\GSTP^{\lambda}{}_{\mu\nu} = \GLC^{\lambda}{}_{\mu\nu} + \LSTP^{\lambda}{}_{\mu\nu} = 0 \quad \Rightarrow \quad \LSTP^{\lambda}{}_{\mu\nu} = -\GLC^{\lambda}{}_{\mu\nu} \,.
\end{equation}
Therefore, one recognizes that in the coincident gauge (\ref{zero_Gamma}) the result (\hyperref[Riemann_and_Ricci]{\ref*{Riemann}}) is just the usual definition of the Riemann curvature tensor in terms of the Levi-Civita connection. A straightforward calculation verifies that the same holds in an arbitrary coordinate system -- the connection coefficients $\GSTP^{\lambda}{}_{\mu\nu}$ for the symmetric teleparallel connection simply drop out. No connection is introduced while contracting, and hence none of Eqs.\ (\ref{Riemann_and_Ricci}) actually contain the symmetric teleparallel connection. The symmetric teleparallel version of the Einstein tensor (\hyperref[Einstein_tensor_place]{\ref*{Einstein_tensor}}) is also just a disguise.

The same holds for the Bianchi identity. In the case of a coordinate transformation 
\begin{equation}
x^{\lambda} = x^{\lambda} \left( x^{\lambda^\prime} \right) \,, \quad g_{\mu\nu} = \frac{\partial x^{\mu^\prime}}{\partial x^{\mu}} \bar{g}_{\mu^\prime \nu^\prime} \frac{\partial x^{\nu^\prime}}{\partial x^\nu} \,,
\end{equation}
for (\ref{Bianchi_identity_form}) one can show
\begin{subequations}
\label{transformation}
\begin{align}
\nonumber
&\partial_{\sigma} \partial_{\lambda} \partial_{\omega} \left( \sqrt{-g} \delta^{(\lambda}_{\hphantom{(}\rho} g^{\sigma)[\omega} \delta^{\rho]}_{\nu} \right) \\
\tag{\theequation}
& = \det \left| \frac{\partial x^\prime}{\partial x} \right| \frac{\partial x^{\nu^\prime}}{\partial x^{\nu}} \partial_{\sigma^\prime} \partial_{\lambda^\prime} \partial_{\omega^\prime} \left( \sqrt{-\bar{g}} \delta^{(\lambda^\prime}_{\hphantom{(}\rho^\prime} \bar{g}^{\sigma^\prime)[\omega^\prime} \delta^{\rho^\prime]}_{\nu^\prime} \right) 
\end{align}
\end{subequations}
which verifies that the Bianchi identity has nothing to do with the symmetric teleparallel connection. In the coincident gauge $\DSTP = \partial$ and due to (\ref{transformation}) a change of coordinates actually does not introduce symmetric teleparallel connection coefficients into (\ref{Bianchi_identity_form}). Partial derivatives as well as symmetric teleparallel covariant derivatives commute. Hence, the part with partial derivatives vanishes separately, and thus, the other half with connection coefficients $\GSTP^{\lambda}{}_{\mu\nu}$ must vanish separately as well.

The field equations for the symmetric teleparallel equivalent of general relativity are given by the Einstein tensor (\hyperref[Einstein_tensor_place]{\ref*{Einstein_tensor}}) which is sourced by the usual energy-momentum tensor, and the Bianchi identity (\ref{Bianchi_identity_form}). Hence, in that theory and on that level the basic geometrical object, the nonmetricity tensor $Q_{\lambda\mu\nu}$ is left undetermined, as we have the freedom to declare whatever coordinate system to be the coincident gauge (\ref{zero_Gamma}). Note that a similar result was obtained for a slightly more general case in Ref.\ \cite{Burton:1997sj}. We conclude that on the level of the field equations the symmetric teleparallel equivalent of general relativity is rather just the general relativity, based on the curvature of the Levi-Civita connection, but disguised as a symmetric teleparallel theory. The situation, however, changes drastically, once we extend the theory.


\section{Action and field equations\label{sec:Action_and_field_equations}}


\subsection{Action}

Let us postulate an action for the metric $g^{\mu\nu}$, scalar field $\Phi$, connection $\Gamma^{\omega}{}_{\sigma\rho}$, and matter fields, collectively denoted by $\chi$, as
\begin{subequations}
\label{action}
\begin{equation}
\tag{\theequation}
S = \int_{M_4} \mathrm{d}^4x \sqrt{-g} \left\lbrace \mathcal{L}_{\mathrm{g}} + \mathcal{L}_{\Phi} + \mathcal{L}_{\mathrm{b}} + \mathcal{L}_{\mathrm{L}} + \mathcal{L}_{\mathrm{m}} \right\rbrace \,, 
\end{equation}
\end{subequations}
composed of the following components. 

The kinetic term for the metric $g^{\mu\nu}$,
{
	%
	%
	%
	%
	%
	%
	%
	%
	%
}
\begin{subequations}
	\label{Lagrangians}
	\begin{flalign}
	\label{L_g}
	\qquad \mathcal{L}_{\mathrm{g}} &\equiv \mathcal{L}_{\mathrm{g}} \left[ g_{\mu\nu}, \Gamma^{\lambda}{}_{\sigma\rho}, \Phi \right] \equiv \frac{1}{2\kappa^2} \mathcal{A}(\Phi) \mathcal{Q} \,, &
	\end{flalign}
\end{subequations}
contains in addition to the nonmetricity scalar $\mathcal{Q}$, defined by (\ref{definition_of_Q}), also the dimensionless nonminimal coupling function $\mathcal{A}(\Phi)$. Roughly speaking, as in scalar-curvature theories \cite{PhysRev.124.925}, the latter introduces a scalar field dependent gravitational \textquotedblleft{}constant\textquotedblright{} $\propto \kappa^2 / \mathcal{A}(\Phi)$. Here the constant $\kappa^2$ wields the dimension, and its numerical value must be determined from the Newtonian limit.

The kinetic term with noncanonical kinetic coupling function $\mathcal{B}(\Phi)$, and self-interaction potential $\mathcal{V}(\Phi)$ for the scalar field $\Phi$ are described by
\vspace{-\baselineskip}
\phantomsection
\label{L_Phi_place}
\vspace{\baselineskip}
\begin{flalign}
\nonumber
\qquad \mathcal{L}_{\Phi} &\equiv \mathcal{L}_{\Phi}\left[g_{\mu\nu},\Phi \right]  \\
\label{L_Phi}
\tag{\theequation b}
&\equiv \mathrlap{{}-\frac{1}{2\kappa^2} \left( \mathcal{B}(\Phi) g^{\mu\nu} \partial_{\mu} \Phi \partial_{\nu} \Phi + 2 \ell^{-2} \mathcal{V}(\Phi) \right) . } & 
\end{flalign}
%
%
The scalar field $\Phi$, as well as the functions $\mathcal{B}(\Phi)$ and $\mathcal{V}(\Phi)$ are considered to be dimensionless. Note that we have introduced yet another dimensionful constant $\left[ \ell^{-2} \right] = \mathrm{length}^{-2} = \left[ \partial^2 \right]$.

\phantomsection
\label{L_b_place}
In addition to pure kinetic terms, one can include mixed term for the metric $g^{\mu\nu}$ and scalar field $\Phi$ as
\begin{flalign}
\nonumber
\qquad \mathcal{L}_{\mathrm{b}} &\equiv \mathcal{L}_{\mathrm{b}} \left[ g_{\mu\nu}, \Gamma^{\lambda}{}_{\sigma\rho}, \Phi \right] & \\
\label{L_b}
\tag{\theequation c}
&\equiv \frac{\epsilon}{2\kappa^2} \partial_{\mu} \mathcal{A}(\Phi) \left( Q^{\mu} - \tilde{Q}^{\mu} \right) \,.
\end{flalign}
In principle, by making use of (\ref{covariant_boundary}), we have just integrated the boundary term in (\ref{split}) by parts. Let us point out that the latter is indeed only a motivation, because we do not have to consider any boundary terms explicitly when postulating the action (\ref{action}). The term (\hyperref[L_b_place]{\ref*{L_b}}) has been introduced with a constant parameter $\epsilon$. 

If the matter Lagrangian $\mathcal{L}_{\mathrm{m}}$ is directly imported from general relativity, i.e., without any alterations\footnote{Note that invoking  the usual minimal coupling principle in general relativity would yield to an  additional nonminimal coupling in the teleparallel framework \cite{So:2006pm}.}, then there are two particularly interesting subcases.
\begin{itemize}
	\item[ i)] If $\epsilon = 0$, and the coefficients $c_1$, $\ldots$, $c_5$ are given by (\ref{GR_coefficients}), then the action (\ref{action}) is equivalent to the action ({\hypersetup{urlcolor=red}\href{https://arxiv.org/pdf/1802.00492.pdf#page=3}{20}}) in Ref.\ \cite{Jarv:2018bgs}.

	\item[ii)] If $\epsilon=1$, and the coefficients are again those originating from general relativity (\ref{GR_coefficients}), then the action (\ref{action}) is equivalent to the action in scalar-curvature theories, see, e.g., action ({\hypersetup{urlcolor=red}\href{https://arxiv.org/pdf/gr-qc/0403063.pdf#page=2}{2.2}}) in Ref.\ \cite{Flanagan:2004bz}, but without the boundary term.
\end{itemize}

\phantomsection
\label{L_L_place}
The symmetric teleparallel conditions (\ref{RT0}) are enforced by making use of the Lagrange multipliers
\begin{flalign}
\nonumber
\qquad \mathcal{L}_{\mathrm{L}} &\equiv \mathcal{L}_{\mathrm{L}} \left[ \Gamma^{\lambda}{}_{\sigma\rho} , \lambda_{\lambda}{}^{\nu\rho\mu}, \lambda_{\lambda}{}^{\mu\nu} \right] & \\
\label{L_L}
\tag{\theequation d}
&\equiv \kappa^{-2}\left( \lambda_{\lambda}{}^{\nu\rho\mu} R^{\lambda}{}_{\nu\rho\mu} +  \lambda_{\lambda}{}^{\mu\nu} T^{\lambda}{}_{\mu\nu} \right) \,,
\end{flalign}

\vspace{-\parskip}
\phantomsection
\label{antisym_L_mult_place}

\noindent{}where by assumption
\begin{flalign}
\label{antisymmetry_of_Lagrange_multipliers}
\tag{\theequation d$'$}
\quad \lambda_{\lambda}{}^{\nu\rho\mu} &= \lambda_{\lambda}{}^{\nu[\rho\mu]} \,, \quad 
\lambda_{\lambda}{}^{\mu\nu} = \lambda_{\lambda}{}^{[\mu\nu]} \,. &
\end{flalign}

\phantomsection
\label{L_m_place}
Finally,
\begin{flalign}
\label{L_m}
\tag{\theequation e}
\qquad \mathcal{L}_{\mathrm{m}} &\equiv \mathcal{L}_{\mathrm{m}} \left[ g_{\mu\nu}, \Gamma^{\lambda}{}_{\sigma\rho}, \chi \right] \,, & \\
\tag{\theequation e$'$}
\label{matter_action}
S_{\mathrm{m}} &= \int_{M_4} \mathrm{d}^4x \sqrt{-g} \mathcal{L}_{\mathrm{m}} \,,&
\end{flalign}
describes the matter fields $\chi$. Note that $\mathcal{L}_{\mathrm{m}}$ may depend on the connection coefficients $\Gamma^{\lambda}{}_{\sigma\rho}$.


\subsubsection{Concerning notation}
\label{subsubsec:notation}

First, we vary the action (\ref{action}) with respect to the Lagrange multipliers and in what follows, we already assume the symmetric teleparallel connection (\ref{RT0}), unless stated otherwise. Therefore, due to narrower scope, we will omit some of the notational specifications used in \cite{Jarv:2018bgs} and also in the previous parts of the current paper. In particular, we omit the $\mathrm{STP}$ on top of quantities, and keep the notation somewhat simpler. Nevertheless, occasionally it is neater to use the Levi-Civita connection, which in that case would be denoted by $\mathrm{LC}$ on top of the quantities.

Second, we drop the arguments of the functions $\mathcal{A}$, $\mathcal{B}$ and $\mathcal{V}$. In addition to taking spacetime derivatives of these functions, we introduce the derivative with respect to the scalar field $\Phi$ as
\begin{equation}
\mathcal{A}^{\prime} \equiv \frac{\mathrm{d} \mathcal{A}}{\mathrm{d} \Phi} \,, \quad
\mathcal{B}^{\prime} \equiv \frac{\mathrm{d} \mathcal{B}}{\mathrm{d} \Phi}\,, \quad
\mathcal{V}^{\prime} \equiv \frac{\mathrm{d} \mathcal{V}}{\mathrm{d} \Phi}  \,.
\end{equation}


\subsection{\texorpdfstring{Field equation for the metric $g^{\mu \nu}$}{Field equation for the metric}}

Varying the action (\ref{action}) with respect to the metric $g^{\mu\nu}$ leads us to the expression
\begin{equation}
\delta_{g} S = \frac{1}{2\kappa^2} \int_{M_4} \mathrm{d}^4x \left\lbrace \sqrt{-g} E^{(g)}_{\mu\nu} \delta g^{\mu\nu} + \partial_{\sigma} \left( \sqrt{ -g } \mathscr{B}_{(g)}^{\sigma} \right) \right\rbrace \,.
\end{equation}
Therefore, the equation of motion for the metric $g^{\mu\nu}$ is
\begin{subequations}
\label{eom_for_metric}
\begin{align}
\nonumber
E^{(g)}_{\mu\nu} ={}& \frac{2}{\sqrt{-g}} \nabla_{\lambda} \left( \sqrt{-g}\mathcal{A} \mathcal{P}^{\lambda}_{\hphantom{\lambda}\mu\nu} \right) - \frac{1}{2} g_{\mu\nu} \mathcal A \mathcal{Q} \\
\nonumber
&{} + \mathcal{A} \left(  \mathcal{P}_{\mu \sigma\rho} Q_{\nu}^{\hphantom{\nu}\sigma\rho} - 2 Q_{\rho\mu}{}^{\sigma} \mathcal{P}^{\rho}{}_{\nu\sigma} \right) \\
\nonumber
&{} + \epsilon \left( g_{\mu\nu} \DLC^{\sigma} \DLC_{\sigma} \mathcal{A} - \DLC_{\mu} \DLC_{\nu} \mathcal{A} - 2 P^{\lambda}{}_{\mu\nu} \partial_{\lambda}\mathcal{A} \right)
\\
\nonumber
&{} + \frac{1}{2} g_{\mu\nu} \mathcal{B} g^{\sigma\rho} \partial_{\sigma} \Phi \partial_{\rho} \Phi - \mathcal{B} \partial_{\mu} \Phi \partial_{\nu} \Phi \\
\tag{\theequation}
&{} + \ell^{-2} g_{\mu\nu} {\mathcal V} - \kappa^2 \mathcal{T}_{\mu\nu} = 0 \,,
\end{align}
\end{subequations}
where the energy-momentum tensor $\mathcal{T}_{\mu\nu}$ is defined as
\begin{equation}
\tag{\theequation a}
\mathcal{T}_{\mu \nu} \equiv -\frac{2}{\sqrt{-g}} \frac{\delta S_\mathrm{m}}{\delta g^{\mu \nu}} \,.
\end{equation}
Due to (\hyperref[sym_G_var_place]{\ref*{symmetry_of_G_variation}})
\begin{subequations}
\label{result_of_variation_of_G_place}
\begin{align}
\nonumber
\mathcal{P}_{\mu \sigma\rho} Q_{\nu}{}^{\sigma\rho} &{}- 2 Q_{\rho\mu}{}^{\sigma} \mathcal{P}^{\rho}{}_{\nu\sigma} \\
\label{result_of_variation_of_G}
={}& \mathcal{P}_{(\mu |\sigma\rho|} Q_{\nu)}{}^{\sigma\rho} - 2 Q_{\rho(\mu}{}^{\sigma} \mathcal{P}^{\rho}{}_{\nu)\sigma} \\
\nonumber
={}& c_1 \left( Q_{\mu\sigma\rho} Q_{\nu}{}^{\sigma\rho} - 2 Q_{\rho\mu}{}^{\sigma} Q^{\rho}{}_{\nu\sigma} \right) \\
\nonumber
&{} - c_2 Q_{\rho\mu}{}^{\sigma} Q_{\sigma\nu}{}^{\rho} + c_3 \left( Q_{\mu} Q_{\nu} - 2 Q^{\sigma} Q_{\sigma\mu\nu}  \right) \\
&{} - c_4 \tilde{Q}_{\mu} \tilde{Q}_{\nu} - c_5 \tilde{Q}^{\sigma} Q_{\sigma\mu\nu} = -q_{\mu\nu} \,, 
\end{align}
\end{subequations}
where the tensor $q_{\mu\nu}$ is defined by Eqs.\ ({\hypersetup{urlcolor=red}\href{https://arxiv.org/pdf/1710.03116.pdf#page=3}{21}}), ({\hypersetup{urlcolor=red}\href{https://arxiv.org/pdf/1803.10185.pdf#page=11}{98}}) and also ({\hypersetup{urlcolor=red}\href{https://arxiv.org/pdf/1808.02894v1.pdf#page=3}{13}}) in Refs.\ \cite{BeltranJimenez:2017tkd}, \cite{BeltranJimenez:2018vdo} and in the first version of \cite{Hohmann:2018wxu}, respectively. Let us point out that on the third line of (\ref{eom_for_metric}), $P^{\lambda}{}_{\mu\nu}$ is indeed the quantity (\ref{definition_of_GRP}), corresponding to general relativity, and not the generic $\mathcal{P}^{\lambda}{}_{\mu\nu}$, defined by (\ref{definition_of_P_all}). This, and also the appearance of the Levi-Civita covariant derivatives on the same line, is due to the fact that Eq.\ (\hyperref[L_b_place]{\ref*{L_b}}), the Lagrangian $\mathcal{L}_{\mathrm{b}}$ is related to general relativity. One can write down different versions of the same equation and some of those can be found in Appendix \ref{app:metric_eom}. For completeness, we include the boundary term
\begin{subequations}
\label{boundary_for_metric}
\begin{align}
\nonumber
\mathscr{B}_{(g)}^{\sigma} \equiv - \Big[& \epsilon \left(g_{\mu\nu} g^{\sigma\lambda} \partial_{\lambda} \mathcal{A} - \delta^{\sigma}_{\mu} \partial_{\nu} \mathcal{A} \right) \\
\tag{\theequation}
&{} + 2 \mathcal{A} \mathcal{P}^{\sigma}{}_{\mu\nu} \Big] \delta g^{\mu\nu} + \mathscr{B}^{\sigma}_{(\mathrm{m},g)} \,,
\end{align}
\end{subequations}
where $\mathscr{B}^{\sigma}_{(\mathrm{m},g)} $ is the part that in principle may arise from the unspecified matter action $S_{\mathrm{m}}$. The boundary term (\ref{boundary_for_metric}) does not contribute to the field equations, and contains only the variation $\delta g^{\mu\nu}$ of the metric, and not its derivative [cf.\ Eq.\ ({\hypersetup{urlcolor=red}\href{https://arxiv.org/pdf/1504.02686.pdf#page=4}{6}}) in Ref.\ \cite{Jarv:2015kga}].


\subsubsection{\texorpdfstring{Further comments on equation for $g^{\mu\nu}$}{Further comments}}

From (\ref{eom_for_metric}), the field equation for the metric tensor $g^{\mu\nu}$, one obtains that the second order derivatives of the metric are contracted by $\mathcal{G}^{\lambda}{}_{\mu\nu}{}^{\omega}{}_{\sigma\rho}$ as
\begin{equation}
E^{(g)}_{\mu\nu} = -2\mathcal{A} \mathcal{G}^{\lambda}{}_{\mu\nu}{}^{\omega}{}_{\sigma\rho} \nabla_{\lambda} \nabla_{\omega} g^{\sigma\rho} + \ldots \,.
\end{equation}
It remains for further study, how this observation is related to the initial value problem. See Theorem on page {\hypersetup{urlcolor=red}\href{https://arxiv.org/pdf/1111.4824.pdf#page=13}{13}} in Ref.\ \cite{Schuller:2011nh}.

Contracting (\ref{eom_for_metric}) yields
\begin{subequations}
\label{contraction_of_metric_eom}
\begin{align}
\nonumber
g^{\mu\nu} E^{(g)}_{\mu\nu} ={}& \partial_{\lambda} \mathcal{A} \left[ \left( 2 C_1 - \epsilon \right) Q^{\lambda} + \left( 2 C_2 + \epsilon \right) \tilde{Q}^{\lambda} \right] \\
\nonumber
&{} + 2 \mathcal{A} \DLC_{\lambda} \left( C_1 Q^{\lambda} + C_2 \tilde{Q}^{\lambda} \right) - \mathcal{A} \mathcal{Q} \\
\nonumber
&{}  + \epsilon 3 \DLC^{\lambda} \DLC_{\lambda} \mathcal{A} + \mathcal{B} g^{\sigma\rho} \partial_{\sigma} \Phi \partial_{\rho} \Phi \\
\tag{\theequation}
&{} + 4 \ell^{-2} \mathcal{V} - \kappa^2 \mathcal{T} \,,
\end{align}
\end{subequations}
where $\mathcal{T} \equiv g^{\mu\nu} \mathcal{T}_{\mu\nu}$, and the constants $C_1$ and $C_2$ are defined by (\hyperref[def_C_1_C_2_place]{\ref*{definition_C_1_C_2}}).


\subsection{\texorpdfstring{Field equation for the scalar field $\Phi$}{Field equation for the scalar field}}

Varying action (\ref{action}) with respect to the scalar field $\Phi$ reads
\begin{equation}
\delta_{\Phi} S = \frac{1}{2\kappa^2} \int_{M_4} \mathrm{d}^4x \left\lbrace \sqrt{-g} E^{(\Phi)} \delta \Phi + \partial_{\sigma} \left( \sqrt{-g} \mathscr{B}^{\sigma}_{(\Phi)} \right)\right\rbrace \,.
\end{equation}
Hence, the dynamics for the scalar field is governed by 
\begin{subequations}
\label{scalar_field_eom}
\begin{align}
\nonumber
E^{(\Phi)} \equiv{}& 2\mathcal B \DLC_{\sigma} \DLC^{\sigma} \Phi 
+\mathcal{B}^{\prime} g^{\mu \nu}\partial_{\mu}\Phi\partial_{\nu}\Phi - 2\ell^{-2} \mathcal{V}^{\prime} \\
\tag{\theequation}
&{} + \mathcal{A}^{\prime} \mathcal{Q} - \epsilon \mathcal{A}^{\prime} \DLC_{\sigma} \left( Q^{\sigma} - \tilde{Q}^{\sigma} \right) = 0 \,,
\end{align}
\end{subequations}
while
\begin{equation}
\mathscr{B}^{\sigma}_{(\Phi)} \equiv \left[ \epsilon \mathcal{A}^{\prime} \left( Q^{\sigma} - \tilde{Q}^{\sigma} \right) - 2 \mathcal{B} g^{\sigma\nu} \partial_{\nu}\Phi \right] \delta \Phi
\end{equation}
[cf.\ Eq.\ ({\hypersetup{urlcolor=red}\href{https://arxiv.org/pdf/1504.02686.pdf#page=4}{7}}) in Ref.\ \cite{Jarv:2015kga}].

Adding (\ref{contraction_of_metric_eom}) to (\ref{scalar_field_eom}) yields
\begin{subequations}
\label{scalar_field_eom_with_trace}
\begin{align}
\nonumber
&\quad \mathcal{A} E^{(\Phi)} + \mathcal{A}^{\prime} g^{\mu\nu} E^{(g)}_{\mu\nu} = \\
\nonumber
={}& 4\mathcal{A}^2 \mathcal{F}(\epsilon) \DLC^{\sigma} \DLC_{\sigma} \Phi + \left( 2\mathcal{A}^2 \mathcal{F}(\epsilon) \right)^{\prime} g^{\mu\nu} \partial_{\mu} \Phi \partial_{\nu} \Phi \\
\nonumber
&{} - 2\ell^{-2} \left( \mathcal{V}^{\prime} \mathcal{A} - 2 \mathcal{A}^{\prime} \mathcal{V} \right) - \kappa^2 \mathcal{A}^{\prime} \mathcal{T} \\
\nonumber
&{} + \left( \mathcal{A}^{\prime} \right)^2 \partial_{\lambda}\Phi \left[ \left( 2C_1 - \epsilon \right) Q^{\lambda} + \left( 2C_2 + \epsilon \right) \tilde{Q}^{\lambda} \right] \\
\tag{\theequation}
&{} + \mathcal{A} \mathcal{A}^{\prime} \DLC_{\lambda} \left[ \left( 2C_1 - \epsilon \right) Q^{\lambda} + \left( 2C_2 + \epsilon \right) \tilde{Q}^{\lambda} \right] \,,
\end{align}
\end{subequations}

\vspace{-\parskip}

\phantomsection
\label{def_F_place}
\noindent{}where
\begin{equation}
\tag{\theequation a}
\label{definition_of_F}
4\mathcal{A}^2 \mathcal{F}(\epsilon) \equiv 2 \mathcal{A} \mathcal{B} + \epsilon 3 \left( \mathcal{A}^{\prime} \right)^2 \,.
\end{equation}


\subsection{\texorpdfstring{Debraiding the equations (\ref{eom_for_metric}) and (\ref{scalar_field_eom})}{Debraiding the equations}\label{debraiding_the_equations}}

For solving the field equations (\ref{eom_for_metric}) and (\ref{scalar_field_eom}) or equivalently (\ref{scalar_field_eom_with_trace}), it would be good to have them debraided \cite{Bettoni:2015wta}. Let us consider two distinct cases.
\begin{itemize}
	\item[ i)] If
	\begin{equation}
	\epsilon = 0 \,,
	\end{equation}
	then, with respect to spacetime coordinates, (\ref{eom_for_metric}) contains second order derivatives of only the metric, and (\ref{scalar_field_eom}) contains second derivatives of only the scalar field. Hence the equations (\ref{eom_for_metric}) and (\ref{scalar_field_eom}) are in that case naturally debraided. Let us recall that this means dropping the boundary-term-motivated Lagrangian $\mathcal{L}_{\mathrm{b}}$, defined by (\hyperref[L_b_place]{\ref*{L_b}}). This observation holds for each choice of the coefficients $c_1$, $\ldots$, $c_5$. In the scalar-tensor extension of general relativity [corresponding to the coefficients (\ref{GR_coefficients}), and $\epsilon=1$], one would have to transform to the Einstein frame, in order to obtain the situation, where the equations are debraided \cite{Jarv:2015kga}. Thus, one could argue, that if $\epsilon = 0$, then the theory under consideration is postulated in the Einstein frame. On the other hand, the matter fields couple to the metric residing in geometry Lagrangian, and hence, it is the Jordan frame. Therefore, contrary to the scalar-curvature case, one could say that for the theory with $\epsilon=0$ (see, e.g., \cite{Jarv:2018bgs}), the Einstein and Jordan frames coincide, exactly as in general relativity. In other words, the matter fields couple to the propagating tensorial degree of freedom. However, to be more conservative, we follow Ref.\ \cite{Hohmann:2018ijr} and refer to the frame as the debraiding frame (see Section {\hypersetup{urlcolor=red}\href{https://arxiv.org/pdf/1801.06531.pdf#page=9}{VI.C}} in Ref.\ \cite{Hohmann:2018ijr}).

	Let us point out that in this case, adding (\ref{contraction_of_metric_eom}) and (\ref{scalar_field_eom}) to yield (\ref{scalar_field_eom_with_trace}) actually introduces second derivatives of the metric to the equation for the scalar field.

	\item[ii)] If
	\begin{equation}
	\epsilon \neq 0 \,,
	\end{equation}
	then the equation (\ref{eom_for_metric}) for the metric $g^{\mu\nu}$ inevitably contains the second derivatives of the scalar field $\Phi$ (note the $\DLC_{\mu} \DLC_{\nu} \mathcal{A}$ term, which is not a scalar). One may, however, ease finding solutions by trying to debraid the equation for the scalar field $\Phi$. From (\ref{scalar_field_eom_with_trace}) it follows that sufficient conditions are
	\begin{subequations}
	\label{conditions_for_scalar_field_eq_with_trace}
	\begin{equation}
	\tag{\theequation}
	2 C_1 - \epsilon \overset{!}{=} 0 \,, \qquad 2 C_2 + \epsilon \overset{!}{=} 0 \,.
	\end{equation}
	\end{subequations}
\end{itemize}


\subsection{\texorpdfstring{Field equation for the connection $\Gamma^{\lambda}{}_{\mu\nu}$}{Field equation for the connection}}

Varying the action (\ref{action}) with respect to the connection $\Gamma^{\lambda}{}_{\mu\nu}$ reveals
\begin{align}
\nonumber
\delta_{\Gamma}S = \frac{1}{2\kappa^2} \int_{M_4} \mathrm{d}^{4}x \Bigg\lbrace &\sqrt{-g} \left( E^{(\Gamma)} \right){}_{\lambda}{}^{\mu\nu} \delta \Gamma^{\lambda}{}_{\mu\nu} \\
&{} + \partial_{\sigma} \left( \sqrt{-g} \mathscr{B}^{\sigma}_{(\Gamma)} \right) \Bigg\rbrace \,.
\end{align}
Thus,
\begin{subequations}
\label{ConnectionEq}
\begin{align}
\nonumber
&\qquad\frac{\sqrt{-g}}{4}\left( E^{(\Gamma)} \right){}_{\lambda}{}^{\mu\nu} \equiv \\
\nonumber
\equiv{}& \nabla_{\rho}\left( \sqrt{-g} \lambda_{\lambda}{}^{\nu \mu \rho}\right) + \sqrt{-g} \lambda_{\lambda}{}^{\mu \nu} - \sqrt{-g} \mathcal{A} \mathcal{P}^{\mu \nu}{}_{\lambda} - \kappa^2 \mathcal{H}_{\lambda}{}^{\mu \nu} \\
\tag{\theequation}
&{} - \sqrt{-g} \epsilon \partial_{\omega} \mathcal{A} \delta^{(\omega}_{\hphantom{(}\sigma} g^{\mu)[\sigma}  \delta^{\nu]}_{\lambda} = 0  \,,
\end{align}
\end{subequations}
and
\begin{equation}
\mathscr{B}^{\sigma}_{(\Gamma)} \equiv -4 \lambda_{\lambda}{}^{\nu\mu\sigma} \delta\Gamma^{\lambda}{}_{\mu\nu} + \mathscr{B}^{\sigma}_{(\mathrm{m},\Gamma)} \,,
\end{equation}
where, as in the variation with respect to the metric, $\mathscr{B}^{\sigma}_{(\mathrm{m},\Gamma)}$ is the part which in principle may arise from the unspecified matter Lagrangian (\hyperref[L_m_place]{\ref*{L_m}}). The hypermomentum density is defined as 
\begin{equation}
\label{Hypermomentum}
\mathcal{H}_{\lambda}{}^{\mu \nu} \equiv -\frac{1}{2} \frac{\delta S_\mathrm{m}}{\delta \Gamma^{\lambda}{}_{\mu \nu}} \,,
\end{equation}
and at this point it may have antisymmetric part, but this will not contribute into what follows. Due to (\ref{symmetry_of_covariant_derivatives}) and (\hyperref[antisym_L_mult_place]{\ref*{antisymmetry_of_Lagrange_multipliers}})
\begin{subequations}
\label{connection_equation_with_derivatives}
\begin{align}
\nonumber
&\qquad -\frac{1}{4} \nabla_{\nu} \nabla_{\mu} \left[ \sqrt{-g}\left( E^{(\Gamma)} \right){}_{\lambda}{}^{\mu\nu}\right] = \\
\nonumber
&\quad = \nabla_{\nu} \nabla_{\mu} \left[ \sqrt{-g} \mathcal{A} \left( \mathcal{P}^{(\mu\nu)}{}_{\lambda} - \epsilon P^{(\mu\nu)}{}_{\lambda} \right) +  \kappa^2 \mathcal{H}_{\lambda}{}^{(\mu\nu)} \right] \\
\tag{\theequation}
&\quad = 0\,,
\end{align}
\end{subequations}
which can be easily proven, if one opens the symmetrizing parenthesis in (\ref{ConnectionEq}), and takes into account [cf.\ Eq.\ ({\hypersetup{urlcolor=red}\href{https://arxiv.org/pdf/1802.00492.pdf#page=4}{30}}) in Ref.\ \cite{Jarv:2018bgs}]
\begin{subequations}
\label{NMSSTEGR_seostuse_vorrand}
\begin{align}
\nonumber
&\left( \nabla_{\nu} \nabla_{\mu} \mathcal{A} \right) \sqrt{-g} P^{(\mu\nu)}{}_{\lambda} + 2 \left(\nabla_{\nu} \mathcal{A} \right) \nabla_{\mu} \left( \sqrt{-g} P^{(\mu\nu)}{}_{\lambda} \right) \\ 
\tag{\theequation}
& \quad = -\frac{1}{2} \nabla_{\mu} \left[ \left(\partial_{\nu} \mathcal{A} \right) \nabla_{\omega} \left( \sqrt{-g} g^{\mu[\nu} \delta^{\omega]}_{\lambda} \right) \right] \,,
\end{align}
\end{subequations}
and the Bianchi identity
\begin{subequations}
\label{Bianchi_identity}
\begin{equation}
\tag{\theequation}
\nabla_{\nu} \nabla_{\mu} \left( \sqrt{-g} P^{(\mu\nu)}{}_{\lambda} \right) = 0
\end{equation}
\end{subequations}
(see Subsec.\ \ref{Sec:Bianchi_identity}). The result (\ref{NMSSTEGR_seostuse_vorrand}) is easily derived from (\ref{GRP_symmetric}) and
\begin{equation}
\nabla_{\mu} \nabla_{\omega} \left( \sqrt{-g} g^{\nu[\mu} \delta^{\omega]}_{\lambda} \right) = 0 \,.
\end{equation}


\subsubsection{\texorpdfstring{Varying with respect to $\xi^{\sigma}$}{Varying with respect to xi}\label{Varying_wrt_xi}}

Instead of varying the action (\ref{action}) with respect to the generic connection $\Gamma^{\lambda}{}_{\mu\nu}$, and imposing flatness and torsionless conditions via the Lagrange multipliers (\hyperref[L_L_place]{\ref*{L_L}}), one may assume the form (\ref{STP_connection_via_xi}) and vary with respect to the coordinates $\xi^{\sigma}$ [see also discussion following Eq.\ ({\hypersetup{urlcolor=red}\href{https://arxiv.org/pdf/1802.00650.pdf#page=5}{13}}) in Ref.\ \cite{Koivisto:2018aip}]. Note that if this approach has been chosen, then the Lagrangian (\hyperref[L_L_place]{\ref*{L_L}}) vanishes and therefore no derivatives of the connection appear in the action (up to the possibility for introducing exotic matter). Let us note that\footnote{\label{footnote:notation}As previously, we will not use the $\mathrm{STP}$ notation, but we only consider the symmetric teleparallel connection.}
\begin{subequations}
	\begin{align}
	\delta_{\xi} \left( \frac{\partial x^{\lambda}}{\partial \xi^{\sigma} } \right) &= - \frac{\partial x^{\lambda}}{\partial \xi^{\omega}} \frac{\partial x^{\rho} }{\partial \xi^{\sigma}} \frac{\partial \delta \xi^{\omega}}{\partial x^{\rho}} \,, \\
	\delta_{\xi} \Gamma^{\lambda}{}_{\mu\nu} &= - \frac{\partial x^{\lambda}}{\partial \xi^{\sigma}} \Gamma^{\rho}{}_{\mu\nu} \frac{\partial \delta \xi^{\sigma}}{\partial x^{\rho}} + \frac{\partial x^{\lambda}}{\partial \xi^{\sigma}} \frac{\partial^2 \delta \xi^{\sigma}}{\partial x^{\mu} \partial x^{\nu}} \,.
	\end{align}
\end{subequations}
Therefore
\begin{subequations}
\label{varying_wrt_xi}
\begin{align}
\nonumber
\delta_{\xi} S = \frac{1}{2\kappa^2}\int_{M_4} \mathrm{d}^4x \Bigg\lbrace& \sqrt{-g} \left( E^{(\Gamma)} \right){}_{\lambda}{}^{\mu\nu} \delta_{\xi} \Gamma^{\lambda}{}_{\mu\nu} + \mathrm{b.t.} \Bigg\rbrace = \\
\nonumber
= \frac{1}{2\kappa^2} \int_{M_4} \mathrm{d}^4x \Bigg\lbrace& \nabla_{\nu} \nabla_{\mu} \left[ \sqrt{-g} \left( E^{(\Gamma)} \right){}_{\lambda}{}^{\mu\nu} \right] \frac{ \partial x^{\lambda} }{\partial \xi^{\sigma}} \delta \xi^{\sigma} \\
\tag{\theequation}
&{} + \partial_{\sigma} \left( \sqrt{-g} \mathscr{B}^{\sigma}_{(\xi)} \right) \Bigg\rbrace \,,
\end{align}
\end{subequations}
where
\begin{subequations}
\label{boundary_for_xi}
\begin{align}
\nonumber
&\sqrt{-g} \mathscr{B}^{\sigma}_{(\xi)} \equiv \sqrt{-g} \left( E^{(\Gamma)} \right){}_{\lambda}{}^{\sigma\nu} \frac{\partial x^{\lambda}}{\partial \xi^{\rho}} \frac{\partial \delta \xi^{\rho}}{\partial x^{\nu}} \\
\tag{\theequation}
&\quad - \nabla_{\mu} \left( \sqrt{-g} \left( E^{(\Gamma)} \right){}_{\lambda}{}^{\mu\sigma} \right) \frac{\partial x^{\lambda}}{\partial \xi^{\rho}} \delta \xi^{\rho} + \mathscr{B}^{\sigma}_{(\mathrm{m},\Gamma)} \,.
\end{align}
\end{subequations}
First, varying with respect to $\xi^{\sigma}$ indeed gave us Eq.\ (\ref{connection_equation_with_derivatives}). Second, from (\ref{varying_wrt_xi}) $(\partial x^{\lambda}/\partial \xi^{\sigma}) \delta \xi^{\sigma} = \delta x^{\lambda}$, which means that varying with respect to $\xi^{\sigma}$ is varying with respect to the coordinates $x^{\lambda}$. Third, the boundary term (\ref{boundary_for_xi}) contains $\partial_{\nu} \delta \xi^{\rho}$. Let us point out that the procedure was based on varying the connection coefficients $\Gamma^{\lambda}{}_{\mu\nu}$ with respect to $\xi^{\sigma}$, and hence the idea holds for arbitrary $\left( E^{(\Gamma)} \right){}_{\lambda}{}^{\mu\nu}$.


\subsubsection{Equation with GR motivated coefficients}

Let us consider the coefficients (\ref{GR_coefficients}), originating from general relativity, and matter action which does not contain generic connection. Then $\mathcal{P}^{(\mu\nu)}{}_{\lambda} = P^{(\mu\nu)}{}_{\lambda}$, and the equation for connection simplifies to
\begin{subequations}
\label{connection_equation_with_GR_coefficients}
\begin{align}
\tag{\theequation}
(1 - \epsilon) \nabla_{\mu} \left[ \left(\partial_{\nu} \mathcal{A} \right) \nabla_{\omega} \left( \sqrt{-g} g^{\mu[\nu} \delta^{\omega]}_{\lambda} \right) \right]  = 0 \,.
\end{align}
\end{subequations}
Hence, for the action where $\epsilon=0$, i.e., without the boundary-term-motivated Lagrangian (\hyperref[L_b_place]{\ref*{L_b}}), we obtain the equation ({\hypersetup{urlcolor=red}\href{https://arxiv.org/pdf/1802.00492.pdf#page=4}{30}}) in Ref.\ \cite{Jarv:2018bgs}. However, if $\epsilon=1$ and we are thus considering an action that is equivalent to the action in scalar-curvature tensor theories [see action ({\hypersetup{urlcolor=red}\href{https://arxiv.org/pdf/gr-qc/0403063.pdf#page=2}{2.2}}) in Ref.\ \cite{Flanagan:2004bz}], then the symmetric teleparallel connection is not constrained by this equation. It turns out that in that case, on the level of the field equations we are once more considering a curvature-based theory in symmetric teleparallel disguise -- the coefficients of the symmetric teleparallel connection do not appear in the equations. See also Subsec.\ \ref{ss:remark}.

The connection equation (\ref{connection_equation_with_GR_coefficients}) can be expressed as
\begin{subequations}
\label{connection_equation_transformation}
\begin{align}
\nonumber
&\left( 1 - \epsilon \right) \partial_{\mu^{\prime}} \left[ \left(\partial_{[\nu^{\prime}} \mathcal{A} \right) \partial_{\lambda^{\prime}]} \left( \sqrt{-\bar{g}} \bar{g}^{ \mu^{\prime} \nu^{\prime} } \right) \right] \\
\tag{\theequation}
&= \left(1 - \epsilon\right)\det\left| \frac{\partial x}{\partial \xi} \right| \frac{\partial x^{\lambda}}{\partial \xi^{\lambda^{\prime}}} \nabla_{\mu} \left[ \left(\partial_{[\nu} \mathcal{A} \right) \nabla_{\lambda]} \left( \sqrt{-g} g^{\mu\nu} \right) \right] \,,
\end{align}
\end{subequations}
where the left hand side is evaluated in $\xi^{\sigma^{\prime}}$ coordinates, stressed (only in this subsection) by adding a bar on top of $\bar{g}$, and a prime along the indices. The result (\ref{connection_equation_transformation}) just transforms the right hand side under a change of coordinates, convincing us that $\xi^{\sigma^{\prime}}$ are the coordinates in which the connection coefficients vanish.

In such theory, for particular \textit{ans\"atze} of the metric $g_{\mu\nu}$ and the scalar field $\Phi$, Eq.\ (\ref{connection_equation_transformation}) provides us a differential equation for determining the Jacobian matrix $\partial \xi^{\mu^{\prime}} \!/\partial x^{\mu} $ as
\begin{subequations}
\label{diff_equation_for_Jacobi_matrix}
\begin{align}
\nonumber
\frac{\partial x^{\mu}}{\partial \xi^{\mu^\prime}} \partial_{\mu} \Bigg[&  \partial_{\nu} \mathcal{A} \frac{\partial x^{\nu}}{\partial \xi^{[\nu^{\prime}}} \frac{\partial x^{\lambda}}{\partial \xi^{\lambda^{\prime}]}} \\
\tag{\theequation}
&{} \times \partial_{\lambda} \left( \det\left| \frac{\partial x}{\partial \xi} \right| \sqrt{-g} \frac{\partial \xi^{\mu^{\prime}} }{\partial x^{\sigma}} g^{\sigma\rho} \frac{\partial \xi^{\nu^{\prime}} }{\partial x^{\rho}}  \right) \Bigg] = 0 \,.
\end{align}
\end{subequations}

\vspace{-0.5cm}


\subsubsection{\texorpdfstring{Simple example of $\Gamma^{\lambda}{}_{\mu\nu} \neq 0$}{Simple example of nonvanishing connection coefficients}\label{s:Connection_example}}

Although the choice $\Gamma^{\lambda}{}_{\mu\nu} = 0$ is always consistent with the symmetric teleparallel conditions (\ref{RT0}), it might nevertheless lead to contradictions if a theory is presented in a particular coordinate system.

Let us consider the GR motivated coefficients (\ref{GR_coefficients}). The equation for the connection is then (\ref{connection_equation_with_GR_coefficients}) or analogously (\ref{connection_equation_transformation}). In Ref.\ \cite{Jarv:2018bgs} we studied spatially (Levi-Civita) flat Friedmann cosmology as an example (see Section {\hypersetup{urlcolor=red}\href{https://arxiv.org/pdf/1802.00492.pdf#page=5}{V}} in Ref.\ \cite{Jarv:2018bgs}). It turned out that vanishing connection coefficients $\bar{\Gamma}^{\lambda^\prime}{}_{\mu^\prime \nu^\prime} = 0$ lead to consistent results, if firstly the (Levi-Civita) flat Friedmann-Lema\^{\i}tre-Robertson-Walker (FLRW) line element is expressed in Cartesian coordinates $\xi^{0^{\prime}} \equiv t$, $\xi^{1^\prime} \equiv x$, $\xi^{2^\prime} \equiv y$, $\xi^{3^\prime} \equiv z$, i.e.,
\begin{subequations}
\label{FLRW_Cartesian_place}
\begin{align}
\label{FLRW_Cartesian}
\mathrm{d}s^2 & =-\left( \mathrm{d} \xi^{0^{\prime}} \right)^2+ \left( a(\xi^{0^{\prime}}) \right)^2 \bar{\delta}_{i^{\prime} j^{\prime}} \mathrm{d} \xi^{i^{\prime}} \mathrm{d} \xi^{j^{\prime}} \,, \\
\shortintertext{and secondly the scalar field is assumed to depend only on cosmological time, i.e.,}
\Phi &\equiv \Phi \left( \xi^{0^{\prime}} \right) \quad \Rightarrow \quad \mathcal{A} \equiv \mathcal{A}\left( \xi^{0^\prime} \right) \,.
\end{align}
\end{subequations}
Equation (\ref{connection_equation_transformation}) verifies that result immediately. Namely, both the metric $\bar{g}_{\mu\nu}$ and the scalar field $\Phi$ only depend on the cosmological time $t$ and hence the antisymmetrization on the first line yields zero. Reducing covariant derivatives to partial ones is in this case a consistent procedure. The nonvanishing components of the nonmetricity are
\begin{subequations}
\label{FLRW_nonmetricity}
\begin{equation}
\tag{\theequation}
\nabla_{0^{\prime}} \bar{g}_{i^{\prime}j^{\prime}} = \partial_{0^{\prime}} \bar{g}_{i^{\prime}j^{\prime}} = 2 H \bar{g}_{i^{\prime}j^{\prime}} \,,
\end{equation}
\end{subequations}
where $H \equiv \dot{a}/a$, and $\dot{a} \equiv \mathrm{d} a / \mathrm{d}t$.

Perhaps the simplest example of nonvanishing symmetric teleparallel connection coefficients arises, if one evaluates (\hyperref[FLRW_Cartesian_place]{\ref*{FLRW_Cartesian}}) in spherical coordinates $x^0 = t$, $x^1 = r$, $x^2 = \vartheta$, $x^3 = \varphi$
\begin{subequations}
	\begin{alignat}{3}
	\xi^0 &= x^0 \,, &\qquad
	\xi^{1} &= x^1 \sin x^2 \cos x^3 \,, \\
	\xi^{2} &= x^1 \sin x^2 \sin x^3 \,, &\qquad 
	\xi^{3} &= x^1 \cos x^2 \,,
	\end{alignat} 
\end{subequations}
resulting in
\begin{align}
\mathrm{d}s^2 &= - \left( \mathrm{d} x^{0} \right)^{2} + g_{ij} \mathrm{d}x^{i} \mathrm{d} x^{j} \,, 
\phantomsection
\hypertarget{spherical_metric_place}{}\\
\label{spherical_metric}
\tag{\theequation a}
\left( a(x^{0}) \right)^{-2} \left( g_{ij} \right)&= \begin{pmatrix}
1 & 0 & 0 \\ 0 & r^2 & 0 \\ 0 & 0 & r^2 \sin^2 \vartheta
\end{pmatrix} \,.
\end{align}
The corresponding Jacobian matrix
\begin{subequations}
\label{Jacobi_matrix}
\begin{equation}
\tag{\theequation}
\left( \frac{\partial \xi^{j^{\prime}}}{\partial x^{k}} \right) = \begin{pmatrix}
1 & 0 & 0 & 0 \\
0 & \sin\vartheta \cos\varphi & r\cos\vartheta\cos\varphi & - r\sin\vartheta\sin\varphi \\
0 & \sin\vartheta \sin\varphi & r\cos\vartheta \sin\varphi & r\sin\vartheta\cos\varphi \\
0 & \cos\vartheta & - r\sin\vartheta & 0
\end{pmatrix}
\end{equation}
\end{subequations}
and its inverse
\begin{equation}
\label{Jacobi_matrix_inv}
\left( \frac{\partial x^{i} }{\partial \xi^{j^\prime}} \right) = \begin{pmatrix}
1 & 0 & 0 & 0 \\
0 & \sin \vartheta \cos \varphi & \sin\vartheta \sin\varphi & \cos\vartheta \\
0 & \displaystyle\frac{\cos\vartheta \cos \varphi}{r} & \displaystyle\frac{\cos\vartheta \sin\varphi}{r} & - \displaystyle\frac{\sin\vartheta }{r} \\
0 & - \displaystyle\frac{\sin \varphi}{r \sin\vartheta} & \displaystyle\frac{\cos\varphi}{r\sin\vartheta} & 0
\end{pmatrix} \,,
\end{equation}
obviously satisfy (\ref{diff_equation_for_Jacobi_matrix}). Calculating the connection coefficients via (\ref{STP_connection_via_xi}) leads to
\begin{subequations}
	\label{coefficients}
	\begin{align}
	\Gamma^{1}{}_{22} &= -r \,, \quad \Gamma^{1}{}_{33} = -r \sin^2\vartheta \,, \quad \Gamma^{2}{}_{12} = \frac{1}{r} \,, \\
	\Gamma^{2}{}_{33} &= -\sin\vartheta \cos\vartheta \,, \quad \Gamma^{3}{}_{13} = \frac{1}{r} \,, \quad \Gamma^{3}{}_{32} = \cot \vartheta \,. 
	\end{align}
\end{subequations}
Expressions (\ref{coefficients}) are nothing else than the nonvanishing Christoffel symbols for (\hyperlink{spherical_metric_place}{\ref*{spherical_metric}}) [and thus possess metric compatibility with respect to (\hyperlink{spherical_metric_place}{\ref*{spherical_metric}})]. Applying the prescription (\ref{STP_connection_via_xi}) on the Jacobian matrix (\ref{Jacobi_matrix}) does not generate temporal components of the connection coefficients, such as $\Gamma^{1}{}_{01}$ [cf.\ Christoffel symbols for whole FLRW metric given for example by Eqs.\ (8.44) in Ref.\ \cite{Carroll:2004st}]. The covariant derivative with respect to the time direction thus reveals nonmetricity as
\begin{equation}
\nabla_{0} g_{ij} = \partial_{0} g_{ij} = 2 H g_{ij} \,,
\end{equation}
which corresponds to (\ref{FLRW_nonmetricity}). All other components of the covariant derivative yield zero also in the spherical coordinates.

\vspace{-0.4cm}


\subsection{Continuity equation\label{subs:continuity_equation}}

Let us consider the diffeomorphism invariance of the action (\ref{action})
\begin{align}
\nonumber
\delta_{\zeta} S ={}& \frac{1}{2\kappa^2} \int_{M_4} \mathrm{d}^4x \Bigg\lbrace \sqrt{-g} E^{(g)}_{\mu\nu} \mathscr{L}_{\zeta} g^{\mu\nu} + \sqrt{-g} E^{(\Phi)} \mathscr{L}_{\zeta} \Phi \\
&{} + \sqrt{-g} \left( E^{(\Gamma)} \right){}_{\lambda}{}^{\mu\nu} \mathscr{L}_{\zeta} \Gamma^{\lambda}{}_{\mu\nu} + \frac{\delta S_{\mathrm{m}}}{\delta \chi} \mathscr{L}_{\zeta} \chi \Bigg\rbrace = 0 \,,
\end{align}
where we have used (\ref{eom_for_metric}), (\ref{scalar_field_eom}), and (\ref{ConnectionEq}), respectively. By calculating the Lie derivatives, i.e., $\mathscr{L}_{\zeta} g^{\mu\nu}$, $\mathscr{L}_{\zeta}\Phi$ and $\mathscr{L}_{\zeta}\Gamma^{\lambda}{}_{\mu\nu}$ [see Ref.\ \cite{Hohmann:2015pva}, in particular Eq.\ ({\hypersetup{urlcolor=red}\href{https://arxiv.org/pdf/1505.07809.pdf#page=5}{10}}) for the Lie derivative of the connection], integrating by parts, neglecting matter equations and boundary terms, we obtain
\begin{subequations}
\label{continuity_action_level}
\begin{align}
\nonumber
\delta_{\zeta} S ={}& \frac{1}{2\kappa^2} \int_{M_4} \mathrm{d}^4x \Bigg\lbrace \sqrt{-g} \left[ 2 \DLC_{\omega} \left( g^{\omega\mu} E^{(g)}_{\mu\nu} \right) + E^{(\Phi)} \partial_{\nu} \Phi\right] \\
\tag{\theequation}
&{} + \nabla_{\omega} \nabla_{\lambda} \left[ \sqrt{-g} \left( E^{(\Gamma)} \right){}_{\nu}{}^{\lambda\omega}\right] \Bigg\rbrace \zeta^{\nu} = 0 \,.
\end{align}
\end{subequations}
In order to calculate the first line
\begin{subequations}
\label{continuity_metric_plus_scalar}
\begin{align}
\nonumber
&2 \sqrt{-g} \DLC_{\omega} \left( g^{\omega\mu} E^{(g)}_{\mu\nu} \right) + \sqrt{-g} E^{(\Phi)} \partial_{\nu} \Phi \\
\tag{\theequation}
& = 4 \nabla_{\omega} \nabla_{\lambda} \left[ \sqrt{-g} \mathcal{A} \left( \mathcal{P}^{\lambda\omega}{}_{\nu} - \epsilon P^{\lambda\omega}{}_{\nu} \right) \right] - \sqrt{-g} \kappa^2 2 \DLC_{\omega} \mathcal{T}^{\omega}{}_{\nu} \,,
\end{align}
\end{subequations}
we made use of (\ref{eom_for_metric_index_raised}), (\ref{derivative_LC_to_STP}), (\hyperlink{Ricci_tensor_place}{\ref*{Ricci_tensor}}), and (\ref{Bianchi_identity}). If the coefficients $c_1$, $\ldots$, $c_5$ are GR-motivated (\ref{GR_coefficients}), then for two particular cases the usual continuity equation $\DLC_{\omega} \mathcal{T}^{\omega}{}_{\nu} = 0$ is manifestly fulfilled. First, if $\mathcal{A} = 1$, i.e., we consider the symmetric teleparallel equivalent of general relativity (with minimally coupled scalar), second, if $\epsilon=1$, i.e., the equivalent to scalar-curvature theories (see, e.g., Ref.\ \cite{Flanagan:2004bz}). If this is not the case, then let us also include the third additive expression from (\ref{continuity_action_level}). Combining (\ref{continuity_metric_plus_scalar}) and (\ref{connection_equation_with_derivatives}) yields
\begin{equation}
-2 \kappa^2 \left( \sqrt{-g} \DLC_{\omega} \mathcal{T}^{\omega}{}_{\nu} + 2 \nabla_{\omega} \nabla_{\lambda} \mathcal{H}_{\nu}{}^{\lambda\omega} \right) = 0 \,,
\end{equation}
which also follows from 
\begin{equation}
2\kappa^2 \delta_{\zeta} S_{\mathrm{m}} = 0 \,,
\end{equation}
i.e., from the diffeomorphism invariance of the matter action (\ref{matter_action}).  


\section{Hamilton-like approach\label{sec:Hamilton_like_approach}}


\subsection{\texorpdfstring{Field space metric $\left( \mathscr{G}^{\lambda\omega} \right)$}{Field space metric}}

Let us define
\begin{subequations}
\label{Field_space_metric}
\begin{equation}
\tag{\theequation}
\left( \mathscr{G}^{\lambda\omega} \right) \equiv \begin{pmatrix}
\mathcal{A} \mathcal{G}^{\Lambda \Omega} & \epsilon \mathcal{A}^{\prime} \mathfrak{G}^{\Lambda\omega } \\
\epsilon \mathcal{A}^{\prime} \mathfrak{G}^{\lambda\Omega} & - \mathcal{B} g^{\lambda\omega} \end{pmatrix} \,,
\end{equation}
\end{subequations}
where in order to suppress some indices, we have used a convention where, e.g.,
\begin{equation}
\tag{\theequation a}
\mathcal{G}^{\Lambda \Omega} \equiv  \mathcal{G}^{\lambda}{}_{\mu\nu}{}^{\omega}{}_{\sigma\rho} \,, \qquad \mathfrak{G}^{\omega \Lambda} \equiv \mathfrak{G}^{\omega \lambda}{}_{\mu\nu} \,. 
\end{equation}
The capital Greek letter indicates the first small Greek letter. Here
\begin{subequations}
\label{off_diagonal_metric}
\begin{align}
\tag{\theequation}
\mathfrak{G}^{\xi \Lambda} &= \mathfrak{G}^{\xi \lambda}{}_{\mu\nu} \equiv -\frac{1}{2} \left( g^{\xi\lambda} g_{\mu\nu} - \delta^{\hphantom{(}\xi}_{(\mu} \delta^{\lambda}_{\nu)} \right) \\
\nonumber
&\equiv \mathfrak{G}^{\Lambda \xi} = \mathfrak{G}^{\lambda}{}_{\mu\nu}{}^{\xi}\,,
\end{align}
\end{subequations}
and thus the field space metric (\ref{Field_space_metric}) only depends on the usual metric $g_{\mu\nu}$ and on the scalar field $\Phi$ but not on their derivatives. By introducing
\begin{equation}
\Psi \equiv \begin{pmatrix}
g^{\mu\nu} \\ \Phi
\end{pmatrix} \,,
\end{equation}
we may write the kinetic terms in the action (\ref{action}) as
\begin{align}
\nonumber
&\mathcal{A} \mathcal{Q} - \mathcal{B}(\Phi) g^{\mu\nu} \partial_{\mu} \Phi \partial_{\nu} \Phi + \epsilon \partial_{\mu} \mathcal{A}(\Phi) \left( Q^{\mu} - \tilde{Q}^{\mu} \right) \\
&= \begin{pmatrix}
\nabla_{\lambda} g^{\mu\nu} & \nabla_{\lambda} \Phi
\end{pmatrix}
\begin{pmatrix}
\mathcal{A} \mathcal{G}^{\lambda}{}_{\mu\nu}{}^{\omega}{}_{\sigma\rho} & \epsilon \mathcal{A}^{\prime} \mathfrak{G}^{\lambda}{}_{\mu\nu}{}^{\omega} \\
\nonumber
\epsilon \mathcal{A}^{\prime} \mathfrak{G}^{\lambda\omega}{}_{\sigma\rho} & - \mathcal{B} g^{\lambda\omega} \end{pmatrix}
\begin{pmatrix}
\nabla_{\omega} g^{\sigma\rho} \\ \nabla_{\omega} \Phi
\end{pmatrix} \\
&= \nabla_{\lambda} \Psi \left( \mathscr{G}^{\lambda\omega} \right) \nabla_{\omega} \Psi \,.
\end{align}

\phantomsection
\label{generalized_velocities_place}

\vspace{-\parskip}

\noindent{}Here, in order to simplify the notation, we adopt
\begin{equation}
\tag{\theequation a}
\label{generalized_velocities}
\nabla_{\omega} \Psi = \begin{pmatrix}
\nabla_{\Omega} g \\ \nabla_{\omega} \Phi 
\end{pmatrix} = \begin{pmatrix}
\nabla_{\omega} g^{\sigma\rho} \\ \partial_{\omega} \Phi
\end{pmatrix} \,.
\end{equation}
One can thus write the whole Lagrangian (density) (\ref{Lagrangians}), a function of the metric $g^{\mu\nu}$, its \textquotedblleft{}generalized velocity\textquotedblright{}\footnote{
	Note that by convention we vary with respect to $g^{\mu\nu}$ and thus, due to (\ref{nonmetricity}), the \textquotedblleft{}generalized velocity\textquotedblright{} and also \textquotedblleft{}generalized momentum\textquotedblright{} gain a minus sign. One could also vary with respect to $g_{\mu\nu}$ and then the \textquotedblleft{}generalized velocity\textquotedblright{} would be $\nabla_{\lambda} g_{\mu\nu} \equiv + Q_{\lambda\mu\nu} $.} 
$\nabla_{\lambda} g^{\mu\nu} \equiv - Q_{\lambda}^{\hphantom{\lambda}\mu\nu}$, the scalar field $\Phi$, $\partial_{\lambda}\Phi$, and matter Lagrangian $\mathcal{L}_{\mathrm{m}}$ as
\begin{align}
\nonumber
\sqrt{-g} \mathcal{L} ={}& \frac{1}{2\kappa^2} \sqrt{-g} \nabla_{\lambda} \Psi \left( \mathscr{G}^{\lambda\omega} \right) \nabla_{\omega} \Psi \\
&{} - \kappa^{-2} \ell^{-2} \sqrt{-g} \mathcal{V} + \sqrt{-g} \mathcal{L}_{\mathrm{m}} \,.
\end{align}
Note that we have not included the Lagrangian (\hyperref[L_L_place]{\ref*{L_L}}) for the Lagrange multipliers. We assume the connection to have the symmetric teleparallel form (\ref{STP_connection_via_xi}), and in that case $\xi$ resides entirely in the \textquotedblleft{}generalized velocity\textquotedblright{} $\nabla_{\lambda} g^{\mu\nu}$. Hence, the whole Lagrangian is indeed only a function of the scalar field and the metric along with their \textquotedblleft{}generalized velocities\textquotedblright{}, and matter Lagrangian $\mathcal{L}_{\mathrm{m}}$.


\subsection{\textquotedblleft{}Generalized momenta\textquotedblright{}}

Based on analogy, let us define \textquotedblleft{}generalized momenta\textquotedblright{} as
\begin{subequations}
\label{gereralized_momenta}
\begin{align}
\nonumber
\Pi^{\Lambda}_{(g)} &\equiv \frac{ \partial \sqrt{-g} \mathcal{L}}{\partial \nabla_{\Lambda} g} \\
\nonumber
&=\sqrt{-g} \kappa^{-2} \left( \mathcal{A} \mathcal{G}^{\Lambda\Omega} \nabla_{\Omega} g + \epsilon \mathcal{A}^{\prime} \mathfrak{G}^{\Lambda\omega} \partial_{\omega} \Phi \right) \\
&= \sqrt{-g} \kappa^{-2} \left({} - \mathcal{A} \mathcal{P}^{\Lambda} + \epsilon \mathcal{A}^{\prime} \mathfrak{G}^{\Lambda\omega} \partial_{\omega} \Phi \right)  \,, \\
\nonumber
\Pi^{\lambda}_{(\Phi)} &\equiv \frac{ \partial \sqrt{-g} \mathcal{L}}{\partial \partial_{\lambda} \Phi} \\
&= \sqrt{-g} \kappa^{-2} \left( \epsilon \mathcal{A}^{\prime} \mathfrak{G}^{\lambda\Omega} \nabla_{\Omega} g - \mathcal{B} g^{\lambda\omega} \partial_{\omega}\Phi \right) \,.
\end{align}
\end{subequations}
In this section, for simplicity, we assume that the matter Lagrangian $\mathcal{L}_{\mathrm{m}}$ depends on the metric only algebraically. In principle one could also consider more generic cases, where these momenta also include, e.g., the Levi-Civita connection contribution to the matter Lagrangian $\mathcal{L}_{\mathrm{m}}$. The details of such calculations are beyond the scope of the current paper, but there does not seem to be any obvious reason, why the following results should not hold for the generic cases as well.

In order to construct a \textquotedblleft{}Hamiltonian\textquotedblright{}, one should invert $\left( \mathscr{G}^{\lambda\omega} \right)$. This fails in only two distinct cases. First, if the condition (\ref{determinant}) does not hold, and hence $\mathcal{G}^{\Lambda\Omega}$ is not invertible (at least not via such an \textit{ansatz}). Second, if the multiplier (\ref{multiplier}) vanishes. Of course we also assume that $\mathcal{A} \neq 0$. For all other cases $\left( \mathscr{G}^{\lambda\omega} \right)$ is invertible. See Appendix \ref{app:Inverting_field_G}.


\subsubsection{\textquotedblleft{}Generalized momenta\textquotedblright{} in distinct cases}

First, let us consider the case $\epsilon = 0$, then 
\begin{equation}
\Pi^{\lambda} = \sqrt{-g} \kappa^{-2} \begin{pmatrix}
 \mathcal{A} \mathcal{G}^{\Lambda\omega}{}_{\sigma\rho} \nabla_{\omega} g^{\sigma\rho}  \\
- \mathcal{B} g^{\lambda\omega} \partial_{\omega}\Phi
\end{pmatrix} \,,
\end{equation}
and we see that the fields are debraided as suggested in Subsection \ref{debraiding_the_equations}.

Second, in the case of the coefficients (\ref{GR_coefficients}) and $\epsilon = 1$, corresponding to the scalar-curvature \cite{Flanagan:2004bz} equivalent,
\begin{subequations}
\label{generalized_momenta_for_GR_coefficients}
\begin{equation}
\tag{\theequation}
\Pi^{\lambda} = \sqrt{-g} \kappa^{-2} \mathcal{A} \begin{pmatrix}
\mathcal{A} G^{\Lambda\omega}{}_{\sigma\rho} \nabla_{\omega} \hat{g}^{\sigma\rho} \\
- 2 \mathcal{F}(1) g^{\lambda\omega} \partial_{\omega} \Phi + \mathcal{A}^{\prime} \mathfrak{G}^{\lambda\omega}{}_{\mu\nu} \nabla_{\omega}\hat{g}^{\mu\nu}
\end{pmatrix} \,,
\end{equation}
\end{subequations}
where in addition to the quantities (\ref{definition_of_GGR}), (\hyperref[def_F_place]{\ref*{definition_of_F}}), (\ref{off_diagonal_metric}), we also defined
\begin{subequations}
\label{Einstein_frame_invariant_metric}
\begin{equation}
\tag{\theequation}
\hat{g}_{\mu\nu} \equiv \mathcal{A} g_{\mu\nu} \,, \quad
\hat{g}^{\sigma\rho} = \mathcal{A}^{-1} g^{\sigma\rho} \,
\end{equation}
\end{subequations}
which is the Einstein frame (invariant) metric [see Eq.\ ({\hypersetup{urlcolor=red}\href{https://arxiv.org/pdf/1411.1947.pdf#page=5}{18}}) in Ref.\ \cite{Jarv:2014hma}, and Eq.\ ({\hypersetup{urlcolor=red}\href{https://arxiv.org/pdf/1605.07033.pdf#page=5}{8}}) in Ref.\ \cite{Kuusk:2016rso}]. Moreover
\begin{equation}
\mathcal{I}_{3} \equiv \pm \int \sqrt{\mathcal{F}(1)} \mathrm{d} \Phi
\end{equation} 
is the Einstein frame (invariant) scalar field [see Eq.\ ({\hypersetup{urlcolor=red}\href{https://arxiv.org/pdf/1411.1947.pdf#page=4}{15}}) in Ref.\ \cite{Jarv:2014hma} and Eq.\ ({\hypersetup{urlcolor=red}\href{https://arxiv.org/pdf/1605.07033.pdf#page=4}{5b}}) in Ref.\ \cite{Kuusk:2016rso}, also Eqs.\ ({\hypersetup{urlcolor=red}\href{https://arxiv.org/pdf/1411.1947.pdf#page=10}{55}}), ({\hypersetup{urlcolor=red}\href{https://arxiv.org/pdf/1411.1947.pdf#page=11}{60}}) in Ref.\ \cite{Jarv:2014hma}]. Note that in that case we can transform to the Einstein frame, where $\mathcal{A} = 1$, and debraid the variables.


\subsection{Hamilton-like equations}

The \textquotedblleft{}Hamiltonian\textquotedblright{} is
\begin{subequations}
\label{Hamiltonian}
\begin{equation}
\tag{\theequation}
\mathscr{H} \equiv \frac{\kappa^2}{2 \sqrt{-g}} \Pi^{\lambda} \left( \mathscr{G}^{-1}_{\lambda\omega} \right) \Pi^{\omega} + \kappa^{-2} \ell^{-2} \sqrt{-g} \mathcal{V} - \sqrt{-g} \mathcal{L}_{\mathrm{m}} \,,
\end{equation}
\end{subequations}
where
\begin{equation}
\tag{\theequation a}
\Pi^{\lambda} \equiv \begin{pmatrix}
\Pi^{\Lambda}_{(g)} \\ \Pi^{\lambda}_{(\Phi)}
\end{pmatrix}
\end{equation}
gathers the \textquotedblleft{}generalized momenta\textquotedblright{}, and is transposed if necessary. A straightforward calculation verifies
\begin{equation}
\nabla_{\lambda} \Psi = \frac{\partial \mathscr{H} }{\partial \Pi^{\lambda}} \,.
\end{equation}
Calculating the equations for $\nabla_{\lambda}\Pi^{\lambda}$, and checking the consistency with Eqs.\ (\ref{eom_for_metric}) and (\ref{scalar_field_eom}), namely showing that up to choice of variables
\begin{subequations}
\label{Hamilton_like_equations}
\begin{align}
\nabla_{\lambda} \left( \Pi_{(g)}\right){}^{\lambda}{}_{\mu\nu}  + \frac{\partial \mathscr{H}}{\partial g^{\mu\nu}} \overset{(\ref{eom_for_metric})}{=}{} - \frac{\sqrt{-g}}{2\kappa^2} E^{(g)}_{\mu\nu} \,, \\
\nabla_{\lambda} \Pi^{\lambda}_{(\Phi)} + \frac{\partial \mathscr{H}}{\partial \Phi} \overset{(\ref{scalar_field_eom})}{=}{} - \frac{\sqrt{-g}}{2\kappa^2} E^{(\Phi)} \,,
\end{align}
\end{subequations}
is rather easy if one makes use of the result
\begin{equation}
\delta \left( \mathscr{G}^{-1}_{\lambda\omega} \right) = - \left( \mathscr{G}^{-1}_{\lambda\sigma} \right) \left( \delta \mathscr{G}^{\sigma\rho} \right) \left( \mathscr{G}^{-1}_{\rho\omega} \right) \,.
\end{equation}
Note that we do not need to calculate the expression explicitly, because the inverses $\left( \mathscr{G}^{-1}_{\lambda\omega} \right)$ contract with \textquotedblleft{}generalized momenta\textquotedblright{}, thus yielding up to a multiplier the \textquotedblleft{}generalized velocities\textquotedblright{}, analogously to the Lagrangian case. In principle, however, one can also calculate the variation of the inverse explicitly, by making use of
\begin{align}
\nonumber
&\delta \left( \mathcal{G}^{-1} \right){}_{\tau}{}^{\xi\zeta}{}_{\omega}{}^{\sigma\rho} = - \frac{1}{2} \Big\lbrace g_{\tau \alpha} \left( \mathcal{G}^{-1} \right)_{\beta}{}^{\xi\zeta}{}_{\omega}{}^{\sigma\rho} \\
\nonumber
&+ g_{\omega\alpha} \left(\mathcal{G}^{-1} \right){}_{\beta}{}^{\sigma\rho}{}_{\tau}{}^{\xi\zeta} - 2 \left( \mathcal{G}^{-1} \right){}_{\tau}{}^{\xi\zeta}{}_{ \omega }{}^{\mu(\sigma} \delta^{\rho)}_{\beta} g_{\alpha\mu} \\
&- 2 \left( \mathcal{G}^{-1} \right){}_{\omega}{}^{\sigma\rho}{}_{\tau}{}^{\mu(\xi} \delta^{\zeta)}_{\beta} g_{\alpha\mu} 
\Big\rbrace \delta g^{\alpha\beta} \,,
\end{align}
which can be shown via (\ref{inversion_of_G}) and (\hyperref[varying_G_place]{\ref*{varying_G}}). Note that for simplicity we assumed that the matter Lagrangian does not depend on the derivatives of the metric tensor, therefore
\begin{equation}
\mathcal{T}_{\mu\nu} = -\frac{2}{\sqrt{-g}} \frac{\partial \left( \sqrt{-g} \mathcal{L}_{\mathrm{m}} \right)}{\partial g^{\mu\nu}} \,.
\end{equation}

Unfortunately one cannot use a Poisson brackets like structure because the chain rule cannot be invoked. The field equations already contain contractions and by making use of these one cannot calculate neither
\begin{equation}
\frac{\partial \left( \hphantom{\Pi^{\Lambda}_{(g)}}\right) }{\partial \Pi^{\Lambda}_{(g)}} \nabla_{\sigma}\Pi^{\Lambda}_{(g)} \qquad \text{nor} \qquad \frac{\partial \left( \hphantom{\Pi_{(\Phi)}^{\lambda} } \right)}{\partial \Pi_{(\Phi)}^{\lambda}} \nabla_{\sigma} \Pi_{(\Phi)}^{\lambda} \,,
\end{equation}
unless perhaps in the case when there is a dependence only on one coordinate, in which case the necessity for contractions would drop somehow appropriately.

Let us point out that in such a Hamilton-like scheme we only obtain the equations (\ref{Hamilton_like_equations}), and hence there is no equivalent to the connection equation (\ref{connection_equation_with_derivatives}). We can, however, reproduce this equation by taking into account the diffeomorphism invariance of the action, see Subsec.\ \ref{subs:continuity_equation}. In Eqs.\ (\ref{Hamilton_like_equations}) the connection is present in the symmetric teleparallel covariant derivative which by a suitable choice of coordinates can be transformed to ordinary partial derivative. In the generic case such a transformation is permitted, and consistency must be checked only after one has chosen particular \textit{ans\"atze} for the metric and the scalar field. Let us recall that varying with respect to $\xi^{\sigma}$ is due to (\ref{varying_wrt_xi}) varying with respect to the coordinates $x^{\lambda}$.


\section{Summary\label{sec:Summary}}

In recent years teleparallel theories have gained more attention as alternative theories of gravity. While one mostly works in the torsion-based setting, there has been interest in the direction of symmetric teleparallelism, where instead of curvature or torsion gravity is effectively described by nonmetricity. In the current paper we extended the class of scalar-nonmetricity theories by coupling the quadratic five-parameter nonmetricity scalar to a  scalar field. This coupling resembles scalar-tensor theories where the scalar field is coupled to the metric tensor degree of freedom. As our previous work \cite{Jarv:2018bgs} indicates, when one considers as the quadratic nonmetricity scalar the equivalent for general relativity, one obtains a different theory than a simple scalar-curvature extension of general relativity. The current work on the one hand broadens this extension by five parameter generalization of the general relativity motivated quadratic nonmetricity scalar (the newer general relativity \cite{BeltranJimenez:2017tkd}), and on the other hand the inclusion of the boundary-term-motivated mixed kinetic term for $g_{\mu\nu}$ and $\Phi$ allows us to obtain an equivalent to the ordinary scalar-curvature theory as a particular subcase.

Much of the literature on symmetric teleparallelism is phrased in terms of differential forms (see, e.g., \cite{Nester:1998mp,Adak:2004uh,Adak:2005cd,Adak:2008gd,Mol2017}), and only recently coordinate basis and explicit formulation in terms of tensor components have gained more attention (see \cite{BeltranJimenez:2017tkd,Conroy:2017yln,Koivisto:2018aip,Jarv:2018bgs,BeltranJimenez:2018vdo,Hohmann:2018xnb,Harko:2018gxr,Hohmann:2018wxu,Soudi:2018dhv}). Thus, for the benefit of the reader, we included some foreknowledge in the Section \ref{sec:Foreknowledge}. As most remarkable results from this section, it is, firstly, interesting to observe that the variation of the metric-like object $\mathcal{G}^{\lambda}{}_{\mu\nu}{}^{\omega}{}_{\sigma\rho}$ in the contraction (\ref{definition_of_Q}) is given in terms of itself as expressed in (\hyperref[varying_G_place]{\ref*{varying_G}}). The hunch behind the result is the following. In the general relativity the Einstein tensor contracts to minus the Ricci scalar, i.e., minus the Einstein-Hilbert Lagrangian. We expect that in the nonmetricity based theory also at least part of the variation with respect to the metric contracts to minus $\mathcal{Q}$. Hence, in a sense we have to \textquotedblleft{}detach\textquotedblright{} the contraction $\mathcal{Q} = Q_{\lambda}{}^{\mu\nu} \mathcal{P}^{\lambda}{}_{\mu\nu}$ to yield (\hyperref[result_of_variation_of_G_place]{\ref*{result_of_variation_of_G}}). The result is also useful in the curvature-based general relativity, covered by (\ref{definition_of_GGR}), as we can first make the noncovariant split (\ref{split}) and then vary the Einstein Lagrangian (\hyperlink{Einstein_L_prime}{\ref*{Einstein_Lagrangian_prime}}). Secondly, let us point out that in many expressions the inclusion of the symmetric teleparallel connection is just a disguise, as there exists a purely Levi-Civita connection based version, see, e.g., (\ref{Riemann_and_Ricci}) for the Riemann tensor, and (\ref{BianchiId}) for the Bianchi identity. In the coincident gauge (\ref{zero_Gamma}) symmetric teleparallel covariant derivatives reduce to partial ones, and a rule of thumb is the following. Let us choose the coincident gauge (\ref{zero_Gamma}), and interpret the thereby obtained partial derivatives as regular partial derivatives, i.e., that do not transform covariantly by themselves. If the whole expression transforms as a tensor nevertheless, then this expression does not depend on the symmetric teleparallel connection in any coordinate system [see, e.g., Eq.\ (\ref{transformation})].

The action (\ref{action}) in Section \ref{sec:Action_and_field_equations} is motivated as follows. Firstly, the inclusion of the scalar field potential $\mathcal{V}$ in (\hyperref[L_Phi_place]{\ref*{L_Phi}}) in principle allows to describe both early and late time accelerated expansion of the Universe, as the potential behaves similarly to the cosmological constant. Secondly, the inclusion of the generic five-parameter dependent nonmetricity scalar $\mathcal{Q}$ in (\hyperref[Lagrangians]{\ref*{L_g}}) stems from the observation that the basic field equations (\ref{eom_for_metric}), (\ref{scalar_field_eom}) and (\ref{ConnectionEq}) have the same form regardless of the particular values of the coefficients $c_1$,$\ldots$,$c_5$. Thirdly, it is remarkable and at the same time expected, that the general-relativity-boundary-term-motivated Lagrangian (\hyperref[L_b_place]{\ref*{L_b}}) leads to general-relativity-motivated $P^{\lambda}{}_{\mu\nu}$ [definition (\ref{definition_of_GRP})] when varied with respect to the metric as on the third line of Eq.\ (\ref{eom_for_metric}), as well as when varied with respect to the connection which after some manipulation leads to Eq.\ (\ref{connection_equation_with_derivatives}).

The Hamilton-like formulation in Section \ref{sec:Hamilton_like_approach} first of all draws attention to the fact that nonvanishing nonmetricity immediately allows to introduce a manifestly covariant \textquotedblleft{}generalized velocity\textquotedblright{} for the metric. Note, that on the level discussed in the current paper, the variables are the \textquotedblleft{}generalized coordinates\textquotedblright{} $g^{\mu\nu}$, $\Phi$, the corresponding \textquotedblleft{}generalized momenta\textquotedblright{} $\Pi^{\Lambda}_{(g)}$, $\Pi^{\lambda}_{(\Phi)}$, and in addition the matter fields. The symmetric teleparallel connection is not explicitly present and this might ease solving the equations. A particularly interesting subcase is the equivalent to the scalar-(curvature)tensor theories (see, e.g., Ref.\ \cite{Jarv:2015kga}) given by $\epsilon = 1$ in the Lagrangian (\hyperref[L_b_place]{\ref*{L_b}}) while $c_i$-s are given by (\ref{GR_coefficients}). In fact, as the symmetric teleparallel connection drops out in this case, we have a curvature-based theory in the symmetric teleparallel disguise. Such a formulation in a sense allows an interpolation between curvature-based and nonmetricity-based scalar-tensor theories. The \textquotedblleft{}generalized momenta\textquotedblright{} for this particular theory, i.e., Eqs.\ (\ref{generalized_momenta_for_GR_coefficients}) are consistent with our previous knowledge as they turn out to be the momenta for the Einstein frame metric and scalar field, which describe the two types of propagating degrees of freedom \cite{Damour}. Last but not least, in order to construct a \textquotedblleft{}Hamiltonian\textquotedblright{} (\ref{Hamiltonian}), we must in principle invert the field space metric $\left( \mathscr{G}^{\lambda\omega} \right)$, defined by (\ref{Field_space_metric}). For the subcase under consideration the necessary and sufficient condition for the field space metric to be invertible is (\ref{field_space_metric_inverse_GR_condition}), which in this case ($\epsilon=1$), is the multiplier of the d\textquoteright{}Alembert operator in Eq.\ (\ref{scalar_field_eom_with_trace}), and generalizes the condition $\omega \neq - \frac{3}{2}$ for the Brans-Dicke parameter \cite{Damour,PhysRev.124.925}.

There are different directions for future work. One could study some specific applications, e.g., in order to distinguish the simplest scalar-nonmetricity and scalar-torsion theories \cite{Hohmann:2018rwf,Hohmann:2018ijr,Jarv:2015odu,Hohmann:2018vle,Hohmann:2018dqh} one could study perturbations on a cosmological background (see Ref.\ \cite{Golovnev:2018wbh}) or carry out the conventional Hamiltonian analysis. Similar studies could be carried out in order to compare the new and the newer general relativity (see Refs.\ \cite{Hohmann:2018jso} and \cite{Hohmann:2018wxu,Soudi:2018dhv} for recent references concerning the theories, respectively). From the curvature-based scalar-tensor theories it is known that the spontaneous scalarization effect has a considerable influence in the strong field regime, e.g., in astrophysical objects such as neutron stars, even if in the weak field regime the theory is indistinguishable from general relativity (see, e.g., \cite{Damour:1996ke,AltahaMotahar:2018djk} and references therein). It would be most intriguing to study, especially nowadays, the possible spontaneous scalarization and its consequences, in particular on the gravitational waves, also in the context of the family of scalar-nonmetricity theories proposed in the current paper. Another direction would be to study more general actions in the symmetric teleparallel framework, e.g., include more coupling functions or couplings to matter (for the latter, see \cite{Harko:2018gxr}), include the parity violating term, consider higher derivatives.


\begin{acknowledgments}
	We would like to thank our colleagues, especially Christian Pfeifer and Manuel Hohmann, and the participants of the Teleparallel Gravity Workshop in Tartu (\href{http://hexagon.fi.tartu.ee/~telegrav2018/}{TeleGrav2018}) for useful discussions and comments. Our deepest gratitude goes also to the anonymous referee for thorough reading and feedback. The work was supported by the Estonian Research Council through the Institutional Research	Funding project IUT02-27 and the Personal Research Funding project PUT790 (start-up project), as well as by the European Regional Development Fund through the Center of Excellence TK133 \textquotedblleft{}The Dark Side of the Universe\textquotedblright{}.
\end{acknowledgments}


\appendix

\section*{Appendixes\label{sec:Appendixes}}

\section{\texorpdfstring{Contractions of $\mathcal{G}^{\lambda}{}_{\mu\nu}{}^{\omega}{}_{\sigma\rho}$}{Contactions of G}\label{sec:AppendixA}}

Let us calculate the contractions of $\mathcal{G}^{\lambda}{}_{\mu\nu}{}^{\omega}{}_{\sigma\rho}$, defined by (\ref{definition_of_G}). A straightforward calculation yields
\begin{subequations}
	\begin{align}
	g^{\mu\nu} \mathcal{G}^{\lambda}{}_{\mu\nu}{}^{\omega}{}_{\sigma\rho} &= C_1 g_{\sigma\rho} g^{\lambda\omega} + C_2 \delta^{\hphantom{(}\lambda}_{(\sigma} \delta^{\omega}_{\rho)} \,, \\
	\delta^{\nu}_{\lambda} \mathcal{G}^{\lambda}{}_{\mu\nu}{}^{\omega}{}_{\sigma\rho} &= C_3 g_{\mu(\sigma} \delta^{\omega}_{\rho)} + C_4 \delta^{\omega}_{\mu} g_{\sigma\rho} \,, \\
	g_{\lambda\omega} \mathcal{G}^{\lambda}{}_{\mu\nu}{}^{\omega}{}_{\sigma\rho} &= C_5 g_{\mu\nu} g_{\sigma\rho} + C_6 g_{\sigma(\mu}g_{\nu)\rho} \,, \\
	g^{\mu\sigma} \mathcal{G}^{\lambda}{}_{\mu\nu}{}^{\omega}{}_{\sigma\rho} &= C_7 g^{\lambda\omega} g_{\nu\rho} + C_8 \delta^{\lambda}_{\rho} \delta^{\omega}_{\nu} + C_9 \delta^{\lambda}_{\nu} \delta^{\omega}_{\rho} \,,
	\end{align}
\end{subequations}
where
\begingroup
\allowdisplaybreaks
\begin{subequations}
	\label{def_C_1_C_2_place}
	\begin{alignat}{3}
	\label{definition_C_1_C_2}
	C_1 &\equiv c_1 + 4 c_3 + \frac{1}{2} c_5 \,, &\quad 
	C_2 &\equiv c_2 + c_4 + 2 c_5 \,, 
	\phantomsection
	\hypertarget{def_C_3_C_4_place}{}\\
	\label{definition_C_3_C_4}
	C_3 &\equiv c_1 + \frac{1}{2} c_2 + \frac{5}{2} c_4 + \frac{1}{2} c_5 \,, &\quad 
	C_4 &\equiv \frac{1}{2} c_2 + c_3 + \frac{5}{4} c_5 \,, \\
	C_5 &\equiv 4 c_3 + c_5 \,, &\quad 
	C_6 &\equiv 4 c_1 + c_2 + c_4 \,, \\
	C_7 &\equiv \frac{5}{2} c_1 + \frac{1}{4} c_2 + c_3 + \frac{1}{4} c_4 \,, &\quad
	C_8 &\equiv \frac{3}{2} c_2 + \frac{1}{2} c_5 \,, \\
	C_9 &\equiv \frac{3}{2} c_4 + \frac{1}{2} c_5 \,.
	\end{alignat}
\end{subequations}
\endgroup
The coefficients $C_2$, $C_3$, $C_4$, $C_5$, $C_7$ are linearly independent and form a basis. One can show that
\begin{subequations}
	\begin{align}
	C_1 &= - \frac{5}{2} C_2 + C_3 + 4 C_4 \,, \\
	C_6 &= - 9 C_2 + 4 C_3 + 16 C_4 - 4 C_5 \,, 
	\end{align}
\end{subequations}
while $C_8$ and $C_9$ are more complicated combinations, also including $C_7$.

The first four of these coefficients enter the theory through [see definition (\hyperlink{def_P_ex_place}{\ref*{definition_of_P_explicit}})]
\begin{subequations}
	\begin{align}
	\mathcal{P}^{\lambda} &\equiv \mathcal{P}^{\lambda}_{\hphantom{\lambda} \mu\nu} g^{\mu\nu} = C_1 Q^{\lambda} + C_2 \tilde{Q}^{\lambda} \,, \\
	\tilde{\mathcal{P}}_{\nu} &\equiv \mathcal{P}^{\lambda}_{\hphantom{\lambda} \mu\nu} \delta^{\mu}_{\lambda} = C_4 Q_{\nu} + C_3 \tilde{Q}_{\nu} \,.
	\end{align}
\end{subequations}
Also, if one considers the local Weyl rescaling of the metric
\begin{equation}
\label{conformal_transformation}
\bar{g}_{\mu\nu} = \mathrm{e}^{\Omega(\Phi)} g_{\mu\nu} \,, \qquad 
\bar{g}^{\mu\nu} = \mathrm{e}^{-\Omega(\Phi)} g^{\mu\nu}
\end{equation}
the nonmetricity tensor $Q_{\lambda\mu\nu}$ and its two contractions transform as
\begin{subequations}
\begin{align}
\bar{Q}_{\lambda\mu\nu} &\equiv \nabla_{\lambda} \bar{g}_{\mu\nu} = \mathrm{e}^{\Omega} \left( Q_{\lambda\mu\nu} + g_{\mu\nu} \partial_{\lambda} \Omega \right) \,, \\
\bar{Q}_{\lambda} &\equiv \bar{Q}_{\lambda\mu\nu} \bar{g}^{\mu\nu} = Q_{\lambda} + 4 \partial_{\lambda} \Omega \,, \\
\bar{\tilde{Q}}_{\lambda} &\equiv \bar{Q}_{\mu\nu\lambda} \bar{g}^{\mu\nu} = \tilde{Q}_{\lambda} + \partial_{\lambda} \Omega \,.
\end{align}
\end{subequations}
Thus, based on the definition (\ref{definition_of_Q_explicit}), it follows that
\begin{subequations}
\label{transformation_for_generic_Q}
\begin{align}
\nonumber
\bar{\mathcal{Q}} ={}& \mathrm{e}^{-\Omega} \mathcal{Q} + 2 \mathrm{e}^{-\Omega} C_1 Q^{\mu} \partial_{\mu} \Omega + 2 \mathrm{e}^{-\Omega} C_2 \tilde{Q}^{\mu} \partial_{\mu} \Omega \\
\tag{\theequation}
&{} + \mathrm{e}^{-\Omega}\left( 4 C_1 + C_2 \right) g^{\mu\nu} \partial_{\mu} \Omega \partial_{\nu} \Omega \,.
\end{align}
\end{subequations}
For GR motivated values (\ref{GR_Cs}) Eq.\ (\ref{transformation_for_generic_Q}) yields Eq. ({\hypersetup{urlcolor=red}\href{https://arxiv.org/pdf/1802.00492.pdf#page=5}{33}}) in Ref.\ \cite{Jarv:2018bgs}.


\section{\texorpdfstring{Inverting $\mathcal{G}^{\lambda}{}_{\mu\nu}{}^{\omega}{}_{\sigma\rho}$}{Inverting G}\label{inverse_metric}}

In order to invert $\mathcal{G}^{\lambda}{}_{\mu\nu}{}^{\omega}{}_{\sigma\rho}$, defined via (\ref{definition_of_G}), with respect to the Einstein product [see Definition {\hypersetup{urlcolor=red}\href{https://arxiv.org/pdf/1109.3830.pdf#page=3}{2.2}}, Eq.\ ({\hypersetup{urlcolor=red}\href{https://arxiv.org/pdf/1109.3830.pdf#page=3}{2.1}}) in Ref.\ \cite{Brazell}], i.e., to calculate 
\begin{subequations}
\label{inversion_of_G}
\begin{equation}
\tag{\theequation}
\left( \mathcal{G}^{-1} \right){}_{\tau}{}^{\xi\zeta}{}_{\lambda}{}^{\mu\nu} \,: \quad 
\left( \mathcal{G}^{-1} \right){}_{\tau}{}^{\xi\zeta}{}_{\lambda}{}^{\mu\nu} \mathcal{G}^{\lambda}{}_{\mu\nu}{}^{\omega}{}_{\sigma\rho} \equiv \delta^{\omega}_{\tau} \delta^{(\xi}_{(\sigma} \delta^{\zeta)}_{\rho)}
\end{equation}
\end{subequations}
explicitly, we make an \textit{ansatz} as
\begin{align}
\nonumber
\left( \mathcal{G}^{-1} \right){}_{\tau}{}^{\xi\zeta}{}_{\lambda}{}^{\mu\nu} \equiv{}& k_1 g^{\zeta(\mu} g^{\nu)\xi}  g_{\tau\lambda} + k_2 \delta_{\hphantom{(}\lambda}^{(\xi} g^{\zeta)(\mu} \delta_{\tau}^{\nu)} \\
\nonumber
&{} + k_3 g^{\xi\zeta} g_{\tau\lambda} g^{\mu\nu} + k_4 \delta_{\hphantom{(}\tau}^{(\xi} g^{\zeta)(\mu} \delta_{\lambda}^{\nu)} \\
\label{inverse_G_ansatz}
&{} + \frac{k_5}{2} g^{\xi\zeta} \delta_{\hphantom{(}\tau}^{(\mu} \delta_{\lambda}^{\nu)} + \frac{k_6}{2} g^{\mu\nu} \delta_{\hphantom{(}\tau}^{(\xi} \delta_{\lambda}^{\zeta)} \,.
\end{align}
A straightforward calculation leads us to the following system of linear algebraic equations
\begin{subequations}
\label{system_of_equations_for_k}
\begin{equation}
\tag{\theequation}
\begin{pmatrix}
%
c_1 & \frac{c_2}{2} & 0 & 0 & 0 & 0 \\
%
c_2 & c_1 + \frac{c_2}{2} & 0 & 0 & 0 & 0 \\
%
c_3 & \frac{c_5}{4} & C_1 & 0 & C_4 & 0 \\
%
c_4 & \frac{c_4 + c_5}{2} & 0 &
C_3 & 0 & C_2 \\
%
\frac{c_5}{2} & \frac{c_4}{2} & C_2 & 0 & C_3 & 0 \\
%
\frac{c_5}{2} & c_3 + \frac{c_5}{4} & 0 & C_4 & 0 & C_1
\end{pmatrix}
\begin{pmatrix}
k_1 \\ k_2 \\ k_3 \\ k_4 \\ \frac{k_5}{2} \\ \frac{k_6}{2}
\end{pmatrix}
=
\begin{pmatrix}
1 \\ 0 \\ 0 \\ 0 \\ 0 \\ 0
\end{pmatrix} \,.
\end{equation}
\end{subequations}
The matrix of the coefficients is regular, if
\begin{subequations}
\label{determinant}
\begin{equation}
\tag{\theequation}
\det = \left( c_{1}^{2} + \frac{1}{2} c_{1} c_{2} - \frac{1}{2} c_{2}^{2} \right) \left( C_{1} C_{3} - C_{2} C_{4} \right)^2 \neq 0 \,.
\end{equation}
\end{subequations}
The system (\ref{system_of_equations_for_k}) is solved by
\begingroup
\allowdisplaybreaks
\begin{subequations}
	\begin{align}
	%
	%
	k_1 ={}& \left( c_{1}^{2} + \frac{1}{2} c_1 c_2 - \frac{1}{2} c_{2}^{2} \right)^{-1} \left( c_1 + \frac{1}{2} c_2 \right) \,, \\
	%
	%
	k_2 ={}& \left( c_{1}^{2} + \frac{1}{2} c_1 c_2 - \frac{1}{2} c_{2}^{2} \right)^{-1} \left( -c_2 \right) \,, \\
	%
	%
	\nonumber
	k_3 ={}& \left[ \left( c_{1}^{2} + \frac{1}{2} c_1 c_2 - \frac{1}{2} c_{2}^{2} \right) \left( C_1 C_3 - C_2 C_4 \right)\right]^{-1} \\*
	\nonumber
	&{} \times \Bigg( -c_1^2 c_3 - c_1 c_2 c_3 + \frac{1}{2} c_1 c_2 c_5 - \frac{5}{2} c_1 c_3 c_4 \\*
	\nonumber
	&\qquad {} + \frac{5}{8} c_1 c_5^2 - \frac{1}{4} c_2^2 c_3 - \frac{1}{4} c_2^2 c_4 + \frac{1}{4} c_2^2 c_5 \\*
	&\qquad {} - \frac{7}{4} c_2 c_3 c_4 + \frac{7}{16} c_2 c_5^2 \Bigg) \,, \\
	%
	%
	\nonumber
	k_4 ={}& \left[ \left( c_{1}^{2} + \frac{1}{2} c_1 c_2 - \frac{1}{2} c_{2}^{2} \right) \left( C_1 C_3 - C_2 C_4 \right)\right]^{-1} \\*
	\nonumber
	&{} \times \Bigg( - c_1^2 c_4 + c_1 c_2 c_5 - 4 c_1 c_3 c_4 + c_1 c_5^2 - c_2^2 c_3  \\*
	&\qquad{} - c_2 c_3 c_4 + \frac{1}{4} c_2 c_5^2 \Bigg) \,, \\
	%
	%
	\nonumber
	\frac{k_5}{2} ={}& \left[ \left( c_{1}^{2} + \frac{1}{2} c_1 c_2 - \frac{1}{2} c_{2}^{2} \right) \left( C_1 C_3 - C_2 C_4 \right)\right]^{-1} \\*
	\nonumber
	&{} \times \Bigg( - \frac{1}{2} c_1^2 c_5 + c_1 c_2 c_3 + \frac{1}{2} c_1 c_2 c_4 - \frac{1}{4} c_1 c_2 c_5  \\*
	\nonumber 
	&\qquad{} + c_1 c_3 c_4 - \frac{1}{4} c_1 c_5^2 + \frac{1}{2} c_2^2 c_3 - \frac{1}{4} c_2^2 c_5 \\*
	&\qquad {} + \frac{5}{2} c_2 c_3 c_4 - \frac{5}{8} c_2 c_5^2 \Bigg) \,, \\
	%
	%
	\label{k_6_k_5}
	k_6 ={}& k_5 \,.
	\end{align}
\end{subequations}
\endgroup 
If the determinant (\ref{determinant}) is nonvanishing then the result (\ref{k_6_k_5}) enforces the symmetry
\begin{equation}
\left( \mathcal{G}^{-1} \right){}_{\tau}{}^{\xi\zeta}{}_{\lambda}{}^{\mu\nu} = \left( \mathcal{G}^{-1} \right){}_{\lambda}{}^{\mu\nu}{}_{\tau}{}^{\xi\zeta} \,,
\end{equation} 
as in (\ref{symmetry}).

For later use, let us define
\begin{subequations}
	\label{K_s}
	\begin{align}
	\label{definition_K_1}
	K_1 &\equiv k_1 + 4 k_3 + \frac{1}{2} k_6 = C_3 \left( C_1 C_3 - C_2 C_4 \right)^{-1} \,, \\
	\label{definition_K_2}
	K_2 &\equiv k_2 + k_4 + 2 k_5 =  - C_2 \left( C_1 C_3 - C_2 C_4 \right)^{-1} \,, \\
	K_3 &\equiv k_1 + \frac{1}{2} k_2 + \frac{5}{2} k_4 + \frac{1}{2} k_5 = C_1 \left( C_1 C_3 - C_2 C_4 \right)^{-1} \,, \\
	K_4 &\equiv \frac{1}{2} k_2 + k_3 + \frac{5}{4} k_6 = - C_4 \left( C_1 C_3 - C_2 C_4 \right)^{-1} \,,
	\end{align}
\end{subequations}
analogously to (\hyperref[def_C_1_C_2_place]{\ref*{definition_C_1_C_2}})-(\hyperlink{def_C_3_C_4_place}{\ref*{definition_C_3_C_4}}). Conveniently
\begin{equation}
K_1 K_3 - K_2 K_4 = \left( C_1 C_3 - C_2 C_4 \right)^{-1} \,.
\end{equation}


\subsection{\texorpdfstring{Inverting GR motivated $G^{\lambda}{}_{\mu\nu}{}^{\omega}{}_{\sigma\rho}$}{Inverting GR motivated G}}

For the general relativity case (\ref{definition_of_GGR})
\begin{subequations}
	\begin{align}
	\nonumber
	&\left( G^{-1} \right){}_{\tau}{}^{\xi\zeta}{}_{\lambda}{}^{\mu\nu} = 4 \delta_{\hphantom{(}\lambda}^{(\xi} g^{\zeta)(\mu} \delta_{\tau}^{\nu)} + \frac{2}{3} g^{\xi\zeta} g_{\tau\lambda} g^{\mu\nu} \\
	&\quad  - \frac{4}{3} \delta_{\hphantom{(}\tau}^{(\xi} g^{\zeta)(\mu} \delta_{\lambda}^{\nu)} - \frac{4}{3} g^{\xi\zeta} \delta_{\hphantom{(}\tau}^{(\mu} \delta_{\lambda}^{\nu)} - \frac{4}{3} g^{\mu\nu} \delta_{\hphantom{(}\tau}^{(\xi} \delta_{\lambda}^{\zeta)} \,,
	\end{align}
\end{subequations}
i.e., 
\begin{subequations}
\label{GR_motivated_inverse}
\begin{alignat}{5}
k_1 &= 0 \,, \quad &
k_2 &= 4 \,, \quad &
k_3 &= \frac{2}{3} \,, \\
k_4 &= - \frac{4}{3} \,, \quad &
\frac{k_5}{2} &= - \frac{4}{3} \,, \quad &
\frac{k_6}{2} &= - \frac{4}{3} \,. 
\end{alignat}
\end{subequations}


\subsection{\texorpdfstring{Coefficients $C_i$ and $K_i$ in GR motivated $\mathrlap{\text{case}}$}{Coefficients C\_i and K\_i in GR motivated case}}

Based on definitions (\hyperref[def_C_1_C_2_place]{\ref*{definition_C_1_C_2}}), (\hyperlink{def_C_3_C_4_place}{\ref*{definition_C_3_C_4}}), (\ref{K_s}), and numerical values (\ref{GR_coefficients}), (\ref{GR_motivated_inverse}), let us calculate
\begin{subequations}
\begin{alignat}{3}
\label{GR_Cs}
\left. C_1 \right|_{GR} &= \hphantom{-}\frac{1}{2} \,, \qquad &
\left. C_2 \right|_{GR} &= - \frac{1}{2} \,, \\
\left. C_3 \right|_{GR} &= - \frac{1}{4} \,, \qquad &
\left. C_4 \right|_{GR} &= - \frac{1}{8} \,, \\
\left. K_1 \right|_{GR} &= \hphantom{-}\frac{4}{3} \,, \qquad &
\left. K_2 \right|_{GR} &= - \frac{8}{3} \,, \\
\left. K_3 \right|_{GR} &= - \frac{8}{3} \,, \qquad &
\left. K_4 \right|_{GR} &= - \frac{2}{3} \,.
\end{alignat}
\end{subequations}


\section{\texorpdfstring{Inverting the field space metric $\left(\mathscr{G}^{\lambda\omega}\right)$}{Inverting the field space metric}\label{app:Inverting_field_G}}

In order to invert (\ref{Field_space_metric}), i.e., the field space metric $\left(\mathscr{G}^{\lambda\omega}\right)$, let us recall, how block matrices are inverted. From Wikipedia \cite{Wiki_block}
\begin{subequations}
\label{inverting_block_matrix}
\begin{equation}
\tag{\theequation}
\begin{pmatrix}
A & B \\ C & D
\end{pmatrix}^{-1} = \begin{pmatrix}
A^{-1} + A^{-1} B F^{-1} C A^{-1} & - A^{-1} B F^{-1} \\
- F^{-1} CA^{-1} & F^{-1}
\end{pmatrix}
\end{equation}
\end{subequations}
where
\begin{equation}
F = D - C A^{-1} B \,.
\end{equation}
In our case
\begin{subequations}
\label{F}
\begin{align}
\nonumber
F^{\xi\zeta} &= - \mathcal{B} g^{\xi\zeta} - \epsilon^2  \frac{\left( \mathcal{A}^{\prime} \right)^2 }{\mathcal{A}} \mathfrak{G}^{\xi\Lambda} \left( \mathcal{G}^{-1} \right)_{\Lambda \Omega} \mathfrak{G}^{\Omega\zeta} \\
\tag{\theequation}
&= - 2\mathcal{A} \mathfrak{F} g^{\xi\zeta} \,,
\end{align}
\end{subequations}
which is invertible, if the multiplier
\begin{subequations}
\label{multiplier}
\begin{equation}
\tag{\theequation}
\mathfrak{F} \equiv \frac{2 \mathcal{A} \mathcal{B} + \epsilon^2 \left( \mathcal{A}^{\prime} \right)^2  \frac{1}{4}\left[ 6 \left( K_1 - K_4 \right) - 3 \left( K_2 - K_3 \right) \right]}{4\mathcal{A}^2} \,,
\end{equation}
\end{subequations}
where
\begin{align}
\nonumber
&\frac{1}{8} \left[ 6 \left( K_1 - K_4 \right) - 3 \left( K_2 - K_3 \right) \right]  \\
&= \frac{9}{8} \left( k_1 - \frac{1}{2} k_2 + 2 k_3 + \frac{1}{2} k_4 - \frac{1}{2} k_5 - \frac{1}{2} k_6 \right) \,,
\end{align}
in front of $g^{\xi\zeta}$ is nonvanishing. In terms of the coefficients $c_1$, $\ldots$, $c_5$
\begin{align}
\nonumber
&\frac{1}{8} \left[ 6 \left( K_1 - K_4 \right) - 3 \left( K_2 - K_3 \right) \right]  \\
&= \frac{9}{8} \frac{\left( c_1 + c_2 + 2 c_3 +2 c_4 + 2 c_5 \right) }{C_1 C_3 - C_2 C_4} \,.
\end{align}
Hence, we see that dividing by zero can only occur, when (\ref{determinant}) vanishes, but in that case the coefficients $k_i$ cannot be determined via (\ref{system_of_equations_for_k}).

In the GR motivated case (\ref{GR_motivated_inverse}), or analogously (\ref{GR_coefficients}) we obtain
\begin{subequations}
\label{field_space_metric_inverse_GR_condition}
\begin{equation}
\tag{\theequation}
\mathfrak{F} = \frac{2 \mathcal{A} \mathcal{B} + \epsilon^2 3 \left( \mathcal{A}^{\prime} \right)^2 }{4\mathcal{A}^2} \neq 0 \,.
\end{equation}
\end{subequations}
\begin{itemize}
	\item[  i)] If $\epsilon = 0$ then this result accommodates the multiplier of the d\textquoteright{}Alembert operator in the scalar field equation of motion (\ref{scalar_field_eom}).
	
	\item[ ii)] If $\epsilon = 1$, then the multiplier is the same as (\hyperref[def_F_place]{\ref*{definition_of_F}}), i.e., the multiplier of the d\textquoteright{}Alembert operator in (\ref{scalar_field_eom_with_trace}) [see also definition ({\hypersetup{urlcolor=red}\href{https://arxiv.org/pdf/1411.1947.pdf#page=4}{12}}) in Ref.\ \cite{Jarv:2014hma}]. Under the assumptions this particular equation does not contain second derivatives of the metric tensor, because the conditions (\ref{conditions_for_scalar_field_eq_with_trace}) are fulfilled. Note that this case corresponds to the scalar-curvature theory \cite{Flanagan:2004bz}, and hence one can transform to the Einstein frame and decouple the \textquotedblleft{}generalized momenta\textquotedblright{} (\ref{generalized_momenta_for_GR_coefficients}), which then also contain (\ref{field_space_metric_inverse_GR_condition}).
	
	\item[iii)] If $\epsilon \neq 0$ and $\epsilon \neq 1$, then (\ref{field_space_metric_inverse_GR_condition}) differs from (\hyperref[def_F_place]{\ref*{definition_of_F}}) by $\epsilon^2$ multiplier.
\end{itemize}

The inverse for the field space metric (\ref{Field_space_metric}) reads
\begin{subequations}
\label{inverse_of_field_space_metric}
\begin{equation}
\tag{\theequation}
\left( \mathscr{G}^{-1}_{\omega\xi}  \right) \equiv \begin{pmatrix}
\left( \mathscr{G}^{-1}_{\omega\xi}  \right)_{11} & \left( \mathscr{G}^{-1}_{\omega\xi}  \right)_{12} \\
\left( \mathscr{G}^{-1}_{\omega\xi}  \right)_{21} & \left( \mathscr{G}^{-1}_{\omega\xi} \right)_{22}
\end{pmatrix} \,,
\end{equation}
\end{subequations}
where
\begin{subequations}
	\begin{align}
	\nonumber
	&\left( \mathscr{G}^{-1}_{\omega\xi}  \right)_{11} \equiv \mathcal{A}^{-1} \left( \mathcal{G}^{-1} \right)_{\Omega\Xi} \\
	&+ \epsilon^2 \left( \frac{\mathcal{A}^{\prime}}{\mathcal{A}} \right)^2 \left( \mathcal{G}^{-1} \right)_{\Omega \Gamma} \mathfrak{G}^{\Gamma \mu } \left( F^{-1} \right)_{\mu\nu} \mathfrak{G}^{\nu \Upsilon} \left( \mathcal{G}^{-1} \right)_{\Upsilon \Xi} \,, \\
	&\left( \mathscr{G}^{-1}_{\omega\xi}  \right)_{12} \equiv - \epsilon \frac{\mathcal{A}^{\prime}}{\mathcal{A}} \left( \mathcal{G}^{-1} \right)_{\Omega\Upsilon} \mathfrak{G}^{\Upsilon \mu} \left( F^{-1} \right)_{\mu\xi} \,, \\
	&\left( \mathscr{G}^{-1}_{\omega\xi}  \right)_{21} \equiv - \epsilon \frac{\mathcal{A}^{\prime}}{\mathcal{A}} \left( F^{-1} \right)_{\omega\mu} \mathfrak{G}^{\mu \Upsilon} \left( \mathcal{G}^{-1} \right)_{\Upsilon\Xi}  \,, \\
	&\left( \mathscr{G}^{-1}_{\omega\xi}  \right)_{22} \equiv \left( F^{-1} \right)_{\omega\xi} \,.
	\end{align}
\end{subequations}
A straightforward calculation verifies that indeed
\begin{equation}
\left( \mathscr{G}^{\lambda\omega} \right) \left( \mathscr{G}^{-1}_{\omega\xi} \right)= \begin{pmatrix} \Delta^{\Lambda}_{\Xi} & 0 \\
0 & \delta^{\lambda}_{\xi}
\end{pmatrix} \,,
\end{equation}
where
\begin{equation}
\Delta^{\Lambda}_{\Xi} \equiv \delta^{\lambda}_{\xi} \delta^{(\sigma}_{(\mu} \delta^{\rho)}_{\nu)} \,.
\end{equation}
Note that the prescription (\ref{inverting_block_matrix}) could be used recursively, and hence, if the momenta (\ref{gereralized_momenta}) would also include contributions from the matter Lagrangian $\mathcal{L}_{\mathrm{m}}$, then the inverse (\ref{inverse_of_field_space_metric}) could be used in the later steps of the recursion.


\section{\texorpdfstring{Block diagonal partitioning of $\left( \mathscr{G}^{\lambda\omega} \right)$}{Block diagonal partitioning of the field space metric}}

From the definition (\ref{Field_space_metric}) of the field space metric $\left( \mathscr{G}^{\lambda\omega} \right)$ one can observe that
\begin{align}
\nonumber
\left( \mathscr{G}^{\lambda\omega} \right)^{T} &\equiv \begin{pmatrix}
\mathcal{A} \mathcal{G}^{\Lambda \Omega} & \epsilon \mathcal{A}^{\prime} \mathfrak{G}^{\Lambda\omega } \\
\epsilon \mathcal{A}^{\prime} \mathfrak{G}^{\lambda\Omega} & - \mathcal{B} g^{\lambda\omega} \end{pmatrix}^{T}  \\
&\equiv
\begin{pmatrix}
\mathcal{A} \mathcal{G}^{\Omega\Lambda} & \epsilon \mathcal{A}^{\prime} \mathfrak{G}^{\omega\Lambda } \\
\epsilon \mathcal{A}^{\prime} \mathfrak{G}^{\Omega\lambda} & - \mathcal{B} g^{\omega\lambda} \end{pmatrix}
\,,
\end{align}
and due to that symmetry it is natural to seek for some diagonal partitioning procedure for such an object.

A visit to Mathematics Stack Exchange site \cite{MSEs} reveals the following. Let
\begin{equation}
M \equiv \begin{pmatrix}
A & B \\ C & D
\end{pmatrix}
\end{equation}
be a block matrix, then
\begin{subequations}
\label{block_diagonalization}
\begin{align}
\nonumber
&\begin{pmatrix}
I_1 & 0 \\ -CA^{-1} & I_2
\end{pmatrix}
\begin{pmatrix}
A & B \\ C & D
\end{pmatrix}
\begin{pmatrix}
I_1 & - A^{-1} B \\ 0 & I_2
\end{pmatrix}
\\
\tag{\theequation}
&\quad =
\begin{pmatrix}
A & 0 \\ 0 & D - C A^{-1} B
\end{pmatrix} \,,
\end{align}
\end{subequations}
where $I_1$ and $I_2$ are some suitable unit matrices. In our case $C=B^{T}$ and $\left(A^{-1}\right)^{T} = A^{-1}$. Under these conditions (\ref{block_diagonalization}) turns out to be a congruence transformation $P^T M P$ where
\begin{align}
P &= \begin{pmatrix}
I_1 & - A^{-1} B \\ 0 & I_2
\end{pmatrix} \,. \\[-0.4cm]
\intertext{Due to}
\label{P_inverse}
P^{-1} &= \begin{pmatrix}
I_1 & A^{-1} B \\ 0 & I_2
\end{pmatrix} 
\end{align}
Eq.\ (\ref{block_diagonalization}) is not a similarity transformation and thus the term diagonalization would not be suitable. However, for tensor components with two indices at the same vertical position, it is exactly the congruence transformation that corresponds to a change of the basis.

In our case
\begin{equation}
\left( P^{\omega}_{\xi} \right) \equiv \begin{pmatrix}
\Delta^{\Omega}_{\Xi} & - \epsilon \frac{ \mathcal{A}^{\prime}}{\mathcal{A}} \left( \mathcal{G}^{-1} \right)_{\Xi\Lambda} \mathfrak{G}^{\Lambda \omega} \\
0 & \delta^{\omega}_{\xi} 
\end{pmatrix} \,,
\end{equation}
and thus
\begin{equation}
\left( P^{\lambda}_{\xi} \right)^T \left( \mathscr{G}^{\xi\zeta} \right) \left( P^{\omega}_{\zeta} \right) = \begin{pmatrix}
\mathcal{A} \mathcal{G}^{\Lambda\Omega} & 0 \\
0 & F^{\lambda\omega}
\end{pmatrix} \,.
\end{equation}
In this diagonal partitioning scheme $F \equiv D - CA^{-1}B$ is already familiar from (\ref{F}).

The \textquotedblleft{}generalized velocities\textquotedblright{} (\hyperref[generalized_velocities_place]{\ref*{generalized_velocities}}) transform as
\begin{align}
P^{-1} \nabla \Psi = \begin{pmatrix}
\Delta^{\Omega}_{\Xi} & \epsilon \frac{\mathcal{A}^{\prime}}{\mathcal{A}} \left( \mathcal{G}^{-1} \right)_{\Xi \Lambda} \mathfrak{G}^{\Lambda\omega} \\
0 & \delta^{\omega}_{\xi}
\end{pmatrix}
\begin{pmatrix}
\nabla_{\Omega} g \\ \partial_{\omega} \Phi
\end{pmatrix} \,,
\end{align}
where
\begin{align}
\nonumber
&\left( \mathcal{G}^{-1} \right)_{\Xi \Lambda} \mathfrak{G}^{\Lambda\omega} \\
&= -\frac{1}{2} \left[ g^{\sigma\rho} \delta^{\omega}_{\xi} \left( K_1 - K_4 \right) + g^{\omega(\sigma} \delta^{\rho)}_{\xi} \left( K_2 - K_3 \right) \right] \,.
\end{align}

If the coefficients $k_1$, $\ldots$, $k_5$ are given by (\ref{GR_motivated_inverse}), and $\epsilon=1$ then we obtain the familiar result
\begin{equation}
\nabla_{\omega} \Psi \mapsto \begin{pmatrix}
\mathcal{A} \nabla_{\omega} \hat{g}^{\sigma\rho} \\ \partial_{\omega} \Phi
\end{pmatrix} \,,
\end{equation}
where $\hat{g}^{\sigma\rho}$ is defined via (\ref{Einstein_frame_invariant_metric}).


\vspace{0.6cm}

\section{\texorpdfstring{Different forms for equations for metric $g^{\mu\nu}$}{Different forms for equations for metric}\label{app:metric_eom}}

Since for a metric incompatible connection the covariant derivative does not commute with raising an index, one obtains
\begin{subequations}
\label{eom_for_metric_index_raised}
\begin{align}
\nonumber
g^{\omega\mu} &E^{(g)}_{\mu\nu} = \frac{2}{\sqrt{-g}} \nabla_{\lambda} \left( \sqrt{-g}\mathcal{A} \mathcal{P}^{\lambda\omega}{}_{\nu} \right) + \mathcal{A} \mathcal{P}^{\omega}{}_{\sigma\rho} Q_{\nu}{}^{\sigma\rho} \\
\nonumber
&{}- \frac{1}{2} \delta^{\omega}_{\nu} \mathcal A \mathcal{Q}  + \frac{1}{2} \delta^{\omega}_{\nu} \mathcal{B} g^{\sigma\rho} \partial_{\sigma} \Phi \partial_{\rho} \Phi - \mathcal{B} g^{\omega\mu} \partial_{\mu} \Phi \partial_{\nu} \Phi \\
\nonumber
&{} + \epsilon \left( \delta^{\omega}_{\nu} \DLC^{\sigma} \DLC_{\sigma} \mathcal{A} - \DLC^{\omega} \DLC_{\nu} \mathcal{A} - 2 P^{\lambda\omega}{}_{\nu} \partial_{\lambda} \mathcal{A} \right)
\\
\tag{\theequation}
&{} + \ell^{-2} \delta^{\omega}_{\nu} {\mathcal V} - \kappa^2 \mathcal{T}^{\omega}{}_{\nu} = 0 \,.
\end{align}
\end{subequations}
Additionally in Eq.\ (\ref{eom_for_metric}) one can use the Levi-Civita covariant derivative instead of the STP one
\begin{align}
\nonumber
&E^{(g)}_{\mu\nu} = 2 \DLC_{\lambda} \left( \mathcal{A} \mathcal{P}^{\lambda}{}_{\mu\nu} \right) - 2 \mathcal{A} Q_{\omega\lambda(\mu} \mathcal{P}^{\lambda\omega}{}_{\nu)} \\
\nonumber
&\;{} + 2 \mathcal{A} Q_{(\mu}{}^{\lambda\omega} \mathcal{P}_{|\lambda\omega|\nu)} + \mathcal{A} Q_{(\mu}{}^{\sigma\rho} \mathcal{P}_{\nu) \sigma\rho} - \frac{1}{2} g_{\mu\nu} \mathcal A \mathcal{Q} \\
\nonumber
&\;{} + \frac{1}{2} g_{\mu\nu} \mathcal{B} g^{\sigma\rho} \partial_{\sigma} \Phi \partial_{\rho} \Phi - \mathcal{B} \partial_{\mu} \Phi \partial_{\nu} \Phi + \ell^{-2} g_{\mu\nu} \mathcal{V} - \kappa^2 \mathcal{T}_{\mu\nu} \\
&\;{} + \epsilon \left( g_{\mu\nu} \DLC^{\sigma} \DLC_{\sigma} \mathcal{A} - \DLC_{\mu} \DLC_{\nu} \mathcal{A} - 2 P^{\lambda}{}_{\mu\nu} \partial_{\lambda}\mathcal{A} \right) = 0 \,.
\end{align}
Note that in such a form we must include symmetrizing parenthesis explicitly.

Let us consider the case $\epsilon = 0$, and $\mathcal{G}^{\lambda}{}_{\mu\nu}{}^{\omega}{}_{\sigma\rho} = G^{\lambda}{}_{\mu\nu}{}^{\omega}{}_{\sigma\rho}$ (see definitions (\ref{definition_of_G}) and (\ref{definition_of_GGR}), and Lagrangian (\hyperref[L_b_place]{\ref*{L_b}})), then one can write a more transparent form [cf.\ Eq.\ ({\hypersetup{urlcolor=red}\href{https://arxiv.org/pdf/1710.03116v1.pdf#page=4}{26}}) in the first version of Ref.\ \cite{BeltranJimenez:2017tkd}]
\begin{align}
\nonumber 
&{}-\mathcal{A} \Biggl( \left( \DLC_{\sigma} - \frac{1}{2} Q_{\sigma} \right) L^{\sigma}{}_{\mu\nu} + \frac{1}{2} \DLC_{\mu} Q_{\nu}  -L^{ \sigma }{}_{ \rho\mu } L^{\rho}{}_{ \sigma\nu } \\
\nonumber 
& \qquad \qquad {} - \frac{1}{2} g_{\mu \nu} \DLC_{\sigma} ( Q^{\sigma} - \tilde{Q}^{\sigma} ) + \frac{1}{2} g_{\mu \nu} Q \Biggr)
\\
\nonumber
&{} +\frac{1}{2}g_{\mu \nu} \Bigl(\mathcal B g^{\sigma\rho}\partial_{\sigma}\Phi\partial_{\rho}\Phi + 2 \ell^{-2} \mathcal{V} \Bigr)-\mathcal B\partial_{\mu}\Phi\partial_{\nu}\Phi \\
&{} - \partial_{\sigma} \mathcal{A} \Bigl( L^{\sigma}{}_{\mu\nu} -\frac{1}{2} g_{\mu\nu}( Q^{\sigma}-\tilde{Q}^{\sigma} ) + \frac{1}{2} \delta^{ \hphantom{(} \sigma }_{(\mu} Q_{\nu)} \Bigr) = \kappa^2 \mathcal{T}_{\mu \nu}
\,.
\end{align}
\vskip-1.5em
Note that due to (\ref{DQ_symmetry}) and (\ref{vanishing_R}) $\DLC_{\mu}Q_{\nu} = \DLC_{(\mu} Q_{\nu)}$.


\onecolumngrid


\begin{thebibliography}{99}

\bibitem{Aldrovandi:2013wha}
	R.\ Aldrovandi and J.\ G.\ Pereira \textquotedblleft{}Teleparallel Gravity: An Introduction\textquotedblright{} \href{https://www.springer.com/series/6001}{Fundam.\ Theor.\ Phys.} \textbf{173}, 214 pages (Springer Netherlands, Dordrecht, 2013) DOI:\href{https://dx.doi.org/10.1007/978-94-007-5143-9}{10.1007/978-94-007-5143-9}%
	%
	{\hypersetup{urlcolor = cyan}
		[\href{http://inspirehep.net/record/1262972}{\textsc{in}SPIRE}]
	}


\bibitem{Maluf:2013gaa}
	J.\ W.\ Maluf \textquotedblleft{}The teleparallel equivalent of general relativity\textquotedblright{} \textit{Ann.\ Phys.} \href{https://dx.doi.org/10.1002/andp.201200272}{\textbf{525}, 339--57} (2013)%
	%
	{\hypersetup{urlcolor = magenta}
		arXiv:\href{https://arxiv.org/abs/1303.3897}{1303.3897}}%
	%
	{\hypersetup{urlcolor = cyan}
		[\href{https://inspirehep.net/record/1224242}{\textsc{in}SPIRE}]
	}


\bibitem{Nester:1998mp} 
	J.\ M.\ Nester and H.-J. Yo \textquotedblleft{}Symmetric teleparallel general relativity\textquotedblright{} \textit{Chin.\ J.\ Phys.} \href{http://psroc.org/cjp/issues.php?vol=37&num=2}{\textbf{37}, 113--7} (1999)%
	%
	{\hypersetup{urlcolor = magenta}
		arXiv:\href{https://arxiv.org/abs/gr-qc/9809049}{gr-qc/9809049}}%
	%
	{\hypersetup{urlcolor = cyan}
		[\href{https://inspirehep.net/record/476449}{\textsc{in}SPIRE}]
	}


\bibitem{BeltranJimenez:2017tkd}
	J.\ Beltr\'{a}n Jim\'{e}nez, L.\ Heisenberg and T.\ S.\ Koivisto \textquotedblleft{}Coincident general relativity\textquotedblright{} \textit{Phys.\ Rev.} D \href{https://dx.doi.org/10.1103/PhysRevD.98.044048}{\textbf{98}, 044048} (2018)%
	%
	{\hypersetup{urlcolor = magenta}
		arXiv:\href{https://arxiv.org/abs/1710.03116}{1710.03116}}%
	%
	{\hypersetup{urlcolor = cyan}
		[\href{https://inspirehep.net/record/1629132}{\textsc{in}SPIRE}]
	}


\bibitem{PhysRev.124.925}
	C.\ H.\ Brans and R.\ H.\ Dicke \textquotedblleft{}Mach\textquoteright{}s principle and a relativistic theory of gravitation\textquotedblright{} \textit{Phys.\ Rev.} \href{https://dx.doi.org/10.1103/PhysRev.124.925}{\textbf{124}, 925--35} (1961)%
	%
	{\hypersetup{urlcolor = cyan}
		[\href{https://inspirehep.net/record/2450}{\textsc{in}SPIRE}]
	}


\bibitem{Flanagan:2004bz}
	\'{E}.\ \'{E}.\ Flanagan \textquotedblleft{}The conformal frame freedom in theories of gravitation\textquotedblright{} \textit{Class.\ Quantum Grav.} \href{https://dx.doi.org/10.1088/0264-9381/21/15/N02}{\textbf{21}, 3817--25} (2004)%
	%
	{\hypersetup{urlcolor = magenta}
		arXiv:\href{https://arxiv.org/abs/gr-qc/0403063}{gr-qc/0403063}}%
	%
	{\hypersetup{urlcolor = cyan}
		[\href{https://inspirehep.net/record/646324}{\textsc{in}SPIRE}]
	}	


\bibitem{Damour:1992we}
	T.\ Damour and G.\ Esposito-Far\`ese \textquotedblleft{}Tensor-multi-scalar theories of gravitation\textquotedblright{} \textit{Class.\ Quantum Grav.} \href{https://dx.doi.org/10.1088/0264-9381/9/9/015}{\textbf{9}, 2093--176} (1992)%
	%
	{\hypersetup{urlcolor = cyan}
		[\href{http://inspirehep.net/record/332753}{\textsc{in}SPIRE}]
	}


\bibitem{Horndeski:1974wa}
	G.\ W.\ Horndeski \textquotedblleft{}Second-order scalar-tensor field equations in a
	four-dimensional space\textquotedblright{} \textit{Int.\ J.\ Theor.\ Phys.} \href{https://dx.doi.org/10.1007/BF01807638}{\textbf{10}, 363--84} (1974)%
	%
	{\hypersetup{urlcolor = cyan}
		[\href{http://inspirehep.net/record/1189313}{\textsc{in}SPIRE}]
	}


\bibitem{Gleyzes:2014dya} 
	J.\ Gleyzes, D.\ Langlois, F.\ Piazza and F.\ Vernizzi
	\textquotedblleft{}New class of consistent scalar-tensor theories\textquotedblright{} \textit{Phys.\ Rev.\ Lett.} \href{https://dx.doi.org/10.1103/PhysRevLett.114.211101}{\textbf{114}, 211101} (2015)%
	%
	{\hypersetup{urlcolor = magenta}
		arXiv:\href{https://arxiv.org/abs/1404.6495}{1404.6495}}%
	%
	{\hypersetup{urlcolor = cyan}
		[\href{https://inspirehep.net/record/1292463}{\textsc{in}SPIRE}]
	}	


\bibitem{Belinchon:2016lwr}
	J.\ A.\ Belinch\'on, T.\ Harko and M.\ K.\ Mak \textquotedblleft{}Exact scalar–tensor cosmological models\textquotedblright{} \textit{Int.\ J.\ Mod.\ Phys.} D \href{https://dx.doi.org/10.1142/S0218271817500730}{\textbf{26}, 1750073} (2017)%
	%
	{\hypersetup{urlcolor = magenta}
		arXiv:\href{https://arxiv.org/abs/1612.05446}{1612.05446}}%
	%
	{\hypersetup{urlcolor = cyan}
		[\href{https://inspirehep.net/record/1504187}{\textsc{in}SPIRE}]
	}


\bibitem{Ade:2015lrj} 
	P.\ A.\ R.\ Ade \textit{et al.} [Planck Collaboration] \textquotedblleft{}Planck 2015 results. XX. Constraints on inflation\textquotedblright{} \textit{Astron.\ Astrophys.} \href{https://dx.doi.org/10.1051/0004-6361/201525898}{\textbf{594}, A20} (2016)%
	%
	{\hypersetup{urlcolor = magenta}
		arXiv:\href{https://arxiv.org/abs/1502.02114}{1502.02114}}%
	%
	{\hypersetup{urlcolor = cyan}
		[\href{https://inspirehep.net/record/1343460}{\textsc{in}SPIRE}]
	}	


\bibitem{Jarv:2018bgs} 
	L.\ J\"arv, M.\ R\"unkla, M.\ Saal and O.\ Vilson \textquotedblleft{}Nonmetricity formulation of general relativity and its scalar-tensor extension\textquotedblright{} \textit{Phys.\ Rev.} D \href{https://dx.doi.org/10.1103/PhysRevD.97.124025}{\textbf{97}, 124025} (2018)%
	%
	{\hypersetup{urlcolor = magenta}
		arXiv:\href{https://arxiv.org/abs/1802.00492}{1802.00492}}%
	%
	{\hypersetup{urlcolor = cyan}
		[\href{https://inspirehep.net/record/1652842}{\textsc{in}SPIRE}]
	}	


\bibitem{Hehl:1994ue} 
	F.\ W.\ Hehl, J.\ D.\ McCrea, E.\ W.\ Mielke and Y.\ Ne\textquoteright{}eman \textquotedblleft{}Metric-affine gauge theory of gravity: field equations, Noether identities, world spinors, and breaking of dilation invariance\textquotedblright{} \textit{Phys.\ Rept.} \href{https://dx.doi.org/10.1016/0370-1573(94)00111-F}{\textbf{258}, 1--171} (1995)%
	%
	{\hypersetup{urlcolor = magenta}
		arXiv:\href{https://arxiv.org/abs/gr-qc/9402012}{gr-qc/9402012}}%
	%
	{\hypersetup{urlcolor = cyan}
		[\href{https://inspirehep.net/record/37362}{\textsc{in}SPIRE}]
	}	


\bibitem{Neeman:1996zcr} 
	Y.\ Ne\textquoteright{}eman and F.\ W.\ Hehl \textquotedblleft{}Test matter in a spacetime with nonmetricity\textquotedblright{} \textit{Class.\ Quantum Grav.} \href{https://dx.doi.org/10.1088/0264-9381/14/1A/020}{\textbf{14}, A251--60} (1997)%
	%
	{\hypersetup{urlcolor = magenta}
		arXiv:\href{https://arxiv.org/abs/gr-qc/9604047}{gr-qc/9604047}}%
	%
	{\hypersetup{urlcolor = cyan}
		[\href{https://inspirehep.net/record/417988}{\textsc{in}SPIRE}]
	}


\bibitem{Puetzfeld:2007hr} 
	D.\ Puetzfeld and Y.\ N.\ Obukhov \textquotedblleft{}Propagation equations for deformable test bodies with microstructure in extended theories of gravity\textquotedblright{} \textit{Phys.\ Rev.} D \href{https://dx.doi.org/10.1103/PhysRevD.76.084025}{\textbf{76}, 084025} (2007); [Erratum: \textit{ibid.} \href{https://dx.doi.org/10.1103/PhysRevD.79.069902}{\textbf{79}, 069902} (2009)%
	%
	{\hypersetup{urlcolor = magenta}
		arXiv:\href{https://arxiv.org/abs/0707.2819}{0707.2819}}%
	%
	{\hypersetup{urlcolor = cyan}
		[\href{https://inspirehep.net/record/756242}{\textsc{in}SPIRE}]]
	}


\bibitem{Vitagliano} 
	V.\ Vitagliano \textquotedblleft{}The role of nonmetricity in metric-affine theories of gravity\textquotedblright{} \textit{Class.\ Quantum Grav.} \href{https://dx.doi.org/10.1088/0264-9381/31/4/045006}{\textbf{31}, 045006} (2014)%
	%
	{\hypersetup{urlcolor = magenta}
		arXiv:\href{https://arxiv.org/abs/1308.1642}{1308.1642}}%
	%
	{\hypersetup{urlcolor = cyan}
		[\href{https://inspirehep.net/record/1246899}{\textsc{in}SPIRE}]
	}


\bibitem{Ariki:2017qov}
	T.\ Ariki \textquotedblleft{}Field theory of hyperfluid\textquotedblright{} \textit{Class.\ Quantum Grav.} \href{https://dx.doi.org/10.1088/1361-6382/aa972d}{\textbf{35}, 035003} (2018)%
	%
	{\hypersetup{urlcolor = magenta}
		arXiv:\href{https://arxiv.org/abs/1701.00607}{1701.00607}}%
	%
	{\hypersetup{urlcolor = cyan}
		[\href{https://inspirehep.net/record/1507665}{\textsc{in}SPIRE}]
	}


\bibitem{Latorre:2017uve} 
	A.\ Delhom-Latorre, G.\ J.\ Olmo and M.\ Ronco \textquotedblleft{}Observable traces of non-metricity: new constraints on metric-affine gravity\textquotedblright{} \textit{Phys.\ Lett.} B \href{https://dx.doi.org/10.1016/j.physletb.2018.03.002}{\textbf{780}, 294--9} (2018)%
	%
	{\hypersetup{urlcolor = magenta}
		arXiv:\href{https://arxiv.org/abs/1709.04249}{1709.04249}}%
	%
	{\hypersetup{urlcolor = cyan}
		[\href{https://inspirehep.net/record/1623199}{\textsc{in}SPIRE}]
	}


\bibitem{Carroll:2004st}
	S.\ M.\ Carroll \textquotedblleft{}Spacetime and Geometry: An Introduction to General Relativity\textquotedblright{} 513 pages (Addison-Wesley, San Francisco, 2004) [\href{https://www.preposterousuniverse.com/spacetimeandgeometry/}{link}]%
	%
	{\hypersetup{urlcolor = cyan}
		[\href{http://inspirehep.net/record/650093}{\textsc{in}SPIRE}]%
	}
	%
	Read also a review by E.\ Poisson \textit{Class.\ Quantum Grav.} \href{https://doi.org/10.1088/0264-9381/22/20/B01}{\textbf{22}, 4385--6} (2005)
	

\bibitem{Blagojevic:2012bc} 
	M.\ Blagojevi\'c and F.\ W.\ Hehl \textquotedblleft{}Gauge theories of gravitation\textquotedblright{} 1--169 (2012)%
	%
	{\hypersetup{urlcolor = magenta}
		arXiv:\href{https://arxiv.org/abs/1210.3775}{1210.3775}}%
	%
	{\hypersetup{urlcolor = cyan}
		[\href{https://inspirehep.net/record/1190649}{\textsc{in}SPIRE}]
	}
	

\bibitem{Adak:2004uh} 
	M.\ Adak and \"O.\ Sert \textquotedblleft{}A solution to symmetric teleparallel gravity\textquotedblright{} \textit{Turk.\ J.\ Phys.} \href{http://journals.tubitak.gov.tr/physics/issue.htm?id=567}{\textbf{29}, 1--7} (2005)%
	%
	{\hypersetup{urlcolor = magenta}
		arXiv:\href{https://arxiv.org/abs/gr-qc/0412007}{gr-qc/0412007}}%
	%
	{\hypersetup{urlcolor = cyan}
		[\href{https://inspirehep.net/record/665866}{\textsc{in}SPIRE}]
	}


\bibitem{Adak:2005cd} 
	M.\ Adak, M.\ Kalay and \"O.\ Sert \textquotedblleft{}Lagrange formulation of the symmetric teleparallel gravity\textquotedblright{} \textit{Int.\ J.\ Mod.\ Phys.} D \href{https://dx.doi.org/10.1142/S0218271806008474}{\textbf{15}, 619--34} (2006)%
	%
	{\hypersetup{urlcolor = magenta}
		arXiv:\href{https://arxiv.org/abs/gr-qc/0505025}{gr-qc/0505025}}%
	%
	{\hypersetup{urlcolor = cyan}
		[\href{https://inspirehep.net/record/682021}{\textsc{in}SPIRE}]
	}
	

\bibitem{Adak:2008gd} 
	M.\ Adak, \"O.\ Sert, M.\ Kalay and M.\ Sari \textquotedblleft{}Symmetric teleparallel gravity: some exact solutions and spinor couplings\textquotedblright{} \textit{Int.\ J.\ Mod.\ Phys.} A \href{https://dx.doi.org/10.1142/S0217751X13501674}{\textbf{28}, 1350167} (2013)%
	%
	{\hypersetup{urlcolor = magenta}
		arXiv:\href{https://arxiv.org/abs/0810.2388}{0810.2388}}%
	%
	{\hypersetup{urlcolor = cyan}
		[\href{https://inspirehep.net/record/799353}{\textsc{in}SPIRE}]
	}


\bibitem{Mol2017}
	I.\ Mol \textquotedblleft{}The nonmetricity formulation of general relativity\textquotedblright{} \textit{Adv.\ Appl.\ Clifford Algebras} \href{https://dx.doi.org/10.1007/s00006-016-0749-8}{\textbf{27}, 2607--38} (2017)%
	%
	{\hypersetup{urlcolor = magenta}
		arXiv:\href{https://arxiv.org/abs/1406.0737}{1406.0737}}%
	%
	{\hypersetup{urlcolor = cyan}
		[\href{https://inspirehep.net/record/1298974}{\textsc{in}SPIRE}]
	}
	

\bibitem{Conroy:2017yln} 
	A.\ Conroy and T.\ S.\ Koivisto \textquotedblleft{}The spectrum of symmetric teleparallel gravity\textquotedblright{} 1--5 (2017)%
	%
	{\hypersetup{urlcolor = magenta}
		arXiv:\href{https://arxiv.org/abs/1710.05708}{1710.05708}}%
	%
	{\hypersetup{urlcolor = cyan}
		[\href{https://inspirehep.net/record/1630881}{\textsc{in}SPIRE}]
	}


\bibitem{Koivisto:2018aip}
	T.\ S.\ Koivisto \textquotedblleft{}An integrable geometrical foundation of gravity\textquotedblright{} \textit{Int.\ J.\ Geom.\ Meth.\ Mod.\ Phys.} \href{https://doi.org/10.1142/S0219887818400066}{1840006} (2018), DOI:\href{https://doi.org/10.1142/S0219887818400066}{https://doi.org/10.1142/S0219887818400066},%
	%
	{\hypersetup{urlcolor = magenta}
		arXiv:\href{https://arxiv.org/abs/1802.00650}{1802.00650}}%
	%
	{\hypersetup{urlcolor = cyan}
		[\href{https://inspirehep.net/record/1652843}{\textsc{in}SPIRE}]
	}


\bibitem{BeltranJimenez:2018vdo} 
	J.\ Beltr\'{a}n Jim\'{e}nez, L.\ Heisenberg and T.\ S.\ Koivisto \textquotedblleft{}Teleparallel Palatini theories\textquotedblright{} \textit{J.\ Cosmol.\ Astropart.\ Phys.} \href{https://doi.org/10.1088/1475-7516/2018/08/039}{08(2018)039} (2018)%
	%
	{\hypersetup{urlcolor = magenta}
		arXiv:\href{https://arxiv.org/abs/1803.10185}{1803.10185}%
	}
	%
	{\hypersetup{urlcolor = cyan}
		[\href{https://inspirehep.net/record/1664555}{\textsc{in}SPIRE}]
	}


\bibitem{Hohmann:2018xnb}
	M.\ Hohmann \textquotedblleft{}Polarization of gravitational waves in general teleparallel theories of gravity\textquotedblright{} 1--9 (2018) \textit{3rd Zeldovich Meeting: An international conference in honor of Ya.\ B.\ Zeldovich in Minsk, Minsk, Belarus, 23-27 of April, 2018,}%
	%
	{\hypersetup{urlcolor = magenta}
		arXiv:\href{https://arxiv.org/abs/1806.10429}{1806.10429}%
	}
	%
	{\hypersetup{urlcolor = cyan}
		[\href{https://inspirehep.net/record/1679917}{\textsc{in}SPIRE}]
	}


\bibitem{Harko:2018gxr} 
	T.\ Harko, T.\ S.\ Koivisto, F.\ S.\ N.\ Lobo, G.\ J.\ Olmo and D.\ Rubiera-Garc\'{\i}a \textquotedblleft{}Coupling matter in modified $Q$-gravity\textquotedblright{} \textit{to be published in Phys.\ Rev.} D, 1--12%
	%
	{\hypersetup{urlcolor = magenta}
		arXiv:\href{https://arxiv.org/abs/1806.10437}{1806.10437}%
	}
	%
	{\hypersetup{urlcolor = cyan}
		[\href{https://inspirehep.net/record/1679919}{\textsc{in}SPIRE}]
	}


\bibitem{Heisenberg:2018vsk} 
	L.\ Heisenberg \textquotedblleft{}A systematic approach to generalisations of general relativity and their cosmological implications\textquotedblright{} 1--193%
	%
	{\hypersetup{urlcolor = magenta}
		arXiv:\href{https://arxiv.org/abs/1807.01725}{1807.01725}%
	}
	%
	{\hypersetup{urlcolor = cyan}
		[\href{https://inspirehep.net/record/1681049}{\textsc{in}SPIRE}]
	}


\bibitem{Hohmann:2018wxu} 
	M.\ Hohmann, C.\ Pfeifer, J.\ L.\ Said and U.\ Ualikhanova \textquotedblleft{}Propagation of gravitational waves in symmetric teleparallel gravity theories\textquotedblright{} 1--9 (2018)%
	%
	{\hypersetup{urlcolor = magenta}
		arXiv:\href{https://arxiv.org/abs/1808.02894}{1808.02894}}%
	%
	{\hypersetup{urlcolor = cyan}
		[\href{https://inspirehep.net/record/1685985}{\textsc{in}SPIRE}]
	}


\bibitem{Adak} 
	M.\ Adak \textquotedblleft{}Gauge approach to the symmetric teleparallel gravity\textquotedblright{} to appear in \textit{Int.\ J.\ Geom.\ Meth.\ Mod.\ Phys.} 1850198 (2018), DOI:\href{https://doi.org/10.1142/S0219887818501980}{https://doi.org/10.1142/S0219887818501980},%
	%
	{\hypersetup{urlcolor = magenta}
		arXiv:\href{https://arxiv.org/abs/1809.01385}{1809.01385}}%
	%
	{\hypersetup{urlcolor = cyan}
		[\href{https://inspirehep.net/record/1692856}{\textsc{in}SPIRE}]
	}


\vskip0.2em   
\bibitem[$32^\text{1}\!$/$\!{}_\text{3}$]{Soudi:2018dhv} 
	I.\ Soudi, G.\ Farrugia, V.\ Gakis, J.\ L.\ Said and E.\ N.\ Saridakis \textquotedblleft{}Polarization of gravitational waves in symmetric teleparallel theories of gravity and their modifications\textquotedblright{} 1--11 (2018)
		%
	{\hypersetup{urlcolor = magenta}
		arXiv:\href{https://arxiv.org/abs/1810.08220}{1810.08220}}%
	%
	{\hypersetup{urlcolor = cyan}
		[\href{https://inspirehep.net/record/1699376}{\textsc{in}SPIRE}]
	}

\setcounter{NAT@ctr}{32}

	
\bibitem{Golovnev:2017dox} 
	A.\ Golovnev, T.\ S.\ Koivisto and M.\ Sandstad \textquotedblleft{}On the covariance of teleparallel gravity theories\textquotedblright{} \textit{Class.\ Quantum Grav.} \href{https://dx.doi.org/10.1088/1361-6382/aa7830}{\textbf{34}, 145013} (2017)%
	%
	{\hypersetup{urlcolor = magenta}
		arXiv:\href{https://arxiv.org/abs/1701.06271}{1701.06271}%
	}
	%
	{\hypersetup{urlcolor = cyan}
		[\href{https://inspirehep.net/record/1510253}{\textsc{in}SPIRE}]
	}


\bibitem{Hohmann:2018rwf} 
	M.\ Hohmann, L.\ J\"arv and U.\ Ualikhanova \textquotedblleft{}Covariant formulation of scalar-torsion gravity\textquotedblright{} \textit{Phys.\ Rev.} D \href{https://dx.doi.org/10.1103/PhysRevD.97.104011}{\textbf{97}, 104011} (2018)%
	%
	{\hypersetup{urlcolor = magenta}
		arXiv:\href{https://arxiv.org/abs/1801.05786}{1801.05786}%
	}
	%
	{\hypersetup{urlcolor = cyan}
		[\href{https://inspirehep.net/record/1648600}{\textsc{in}SPIRE}]
	}


\bibitem{Poplawski:2009fb} 
	N.\ J.\ Pop{\l}awski \textquotedblleft{}Spacetime and fields\textquotedblright{} 1--114 (2009)%
	%
	{\hypersetup{urlcolor = magenta}
		arXiv:\href{https://arxiv.org/abs/0911.0334}{0911.0334}%
	}
	%
	{\hypersetup{urlcolor = cyan}
		[\href{https://inspirehep.net/record/835783}{\textsc{in}SPIRE}]
	}
	

\bibitem{Bao:2000}
	D.\ Bao, S.-S.\ Chern and Z.\ Shen \textquotedblleft{}An Introduction to Riemann-Finsler Geometry\textquotedblright{} Graduate Texts in Mathematics \textbf{200}, 431 pages, edited by S.\ Axler, F.\ W.\ Gehring and K.\ A.\ Ribet (Springer-Verlag, New York, 2000) DOI:\href{https://dx.doi.org/10.1007/978-1-4612-1268-3}{10.1007/978-1-4612-1268-3}%


\bibitem{Ortin:2015hya} 
	T.\ Ort{\'{\i}}n \textquotedblleft{}Gravity and Strings\textquotedblright{} Cambridge Monographs on Mathematical Physics, 684 pages (Cambridge University Press, Cambridge, 2015) DOI:\href{https://dx.doi.org/10.1017/CBO9781139019750}{10.1017/CBO9781139019750}%
	%
	{\hypersetup{urlcolor = cyan}
		[\href{http://inspirehep.net/record/1383727}{\textsc{in}SPIRE}]
	}


\bibitem{Brazell}
	M.\ Brazell, N.\ Li, C.\ Navasca and C.\ Tamon \textquotedblleft{}Tensor and matrix inversions with applications\textquotedblright{} 1--24 (2011)%
	%
	{\hypersetup{urlcolor = magenta}
		arXiv:\href{https://arxiv.org/abs/1109.3830}{1109.3830}
	}


\vskip0.2em   
\bibitem[$38^\text{3}\!$/$\!{}_\text{4}$]{Burton:1997sj}
	H.\ Burton and R.\ B.\ Mann \textquotedblleft{}Palatini variational principle for an extended Einstein-Hilbert action\textquotedblright{} \textit{Phys.\ Rev.} D \href{https://doi.org/10.1103/PhysRevD.57.4754}{\textbf{57}, 4754--9} (1998)
	%
	{\hypersetup{urlcolor = magenta}
		arXiv:\href{https://arxiv.org/abs/gr-qc/9711003}{gr-qc/9711003}%
	}
	%
	{\hypersetup{urlcolor = cyan}
		[\href{https://inspirehep.net/record/450608}{\textsc{in}SPIRE}]
	}

\setcounter{NAT@ctr}{38}


\bibitem{So:2006pm} 
	L.\ L.\ So and J.\ M.\ Nester \textquotedblleft{}On source coupling and the teleparallel equivalent to GR\textquotedblright{} \href{https://doi.org/10.1142/9789812704030_0140}{1498--1500} (2006) \textit{Recent developments in theoretical and experimental general relativity, gravitation, and relativistic field theories. Proceedings, 10th Marcel Grossmann Meeting, MG10, Rio de Janeiro, Brazil, 20-26 of July, 2003}%
	%
	{\hypersetup{urlcolor = magenta}
		arXiv:\href{https://arxiv.org/abs/gr-qc/0612062}{gr-qc/0612062}%
	}
	%
	{\hypersetup{urlcolor = cyan}
		[\href{https://inspirehep.net/record/733986}{\textsc{in}SPIRE}]
	}


\bibitem{Jarv:2015kga}
	L.\ J\"arv, P.\ Kuusk, M.\ Saal and O.\ Vilson \textquotedblleft{}Transformation properties and general relativity regime in scalar–tensor theories\textquotedblright{} \textit{Class.\ Quantum Grav.} \href{https://dx.doi.org/10.1088/0264-9381/32/23/235013}{\textbf{32}, 235013} (2015)%
	%
	{\hypersetup{urlcolor = magenta}
		arXiv:\href{https://arxiv.org/abs/1504.02686}{1504.02686}%
	}
	%
	{\hypersetup{urlcolor = cyan}
		[\href{https://inspirehep.net/record/1358926}{\textsc{in}SPIRE}]
	}

\bibitem{Schuller:2011nh} 
	F.\ P.\ Schuller \textquotedblleft{}All spacetimes beyond Einstein (Obergurgl Lectures)\textquotedblright{} 1--44 (2011)%
	%
	{\hypersetup{urlcolor = magenta}
		arXiv:\href{https://arxiv.org/abs/1111.4824}{1111.4824}%
	}
	%
	{\hypersetup{urlcolor = cyan}
		[\href{https://inspirehep.net/record/946780}{\textsc{in}SPIRE}]
	}


\bibitem{Bettoni:2015wta} 
	D.\ Bettoni and M.\ Zumalac{\'a}rregui \textquotedblleft{}Kinetic mixing in scalar-tensor theories of gravity\textquotedblright{} \textit{Phys.\ Rev.} D \href{https://dx.doi.org/10.1103/PhysRevD.91.104009}{\textbf{91}, 104009} (2015)%
	%
	{\hypersetup{urlcolor = magenta}
		arXiv:\href{https://arxiv.org/abs/1502.02666}{1502.02666}%
	}
	%
	{\hypersetup{urlcolor = cyan}
		[\href{https://inspirehep.net/record/1343728}{\textsc{in}SPIRE}]
	}


\bibitem{Hohmann:2018ijr} 
	M.\ Hohmann \textquotedblleft{}Scalar-torsion theories of gravity. III. Analogue of scalar-tensor gravity and conformal invariants\textquotedblright{} \textit{Phys.\ Rev.} D \href{https://dx.doi.org/10.1103/PhysRevD.98.064004}{\textbf{98}, 064004} (2018)%
	%
	{\hypersetup{urlcolor = magenta}
		arXiv:\href{https://arxiv.org/abs/1801.06531}{1801.06531}%
	}
	%
	{\hypersetup{urlcolor = cyan}
		[\href{https://inspirehep.net/record/1649115}{\textsc{in}SPIRE}]
	}
	

\bibitem{Hohmann:2015pva} 
	M.\ Hohmann \textquotedblleft{}Spacetime and observer space symmetries in the language of Cartan geometry\textquotedblright{} \textit{J.\ Math.\ Phys.} \href{https://dx.doi.org/10.1063/1.4961152}{\textbf{57}, 082502} (2016)%
	%
	{\hypersetup{urlcolor = magenta}
		arXiv:\href{https://arxiv.org/abs/1505.07809}{1505.07809}}%
	%
	{\hypersetup{urlcolor = cyan}
		[\href{https://inspirehep.net/record/1373517}{\textsc{in}SPIRE}]
	}


\bibitem{Jarv:2014hma} 
	L.\ J\"arv, P.\ Kuusk, M.\ Saal and O.\ Vilson \textquotedblleft{}Invariant quantities in the scalar-tensor theories of gravitation\textquotedblright{} \textit{Phys.\ Rev.} D \href{https://dx.doi.org/10.1103/PhysRevD.91.024041}{\textbf{91}, 024041} (2015)%
	%
	{\hypersetup{urlcolor = magenta}
		arXiv:\href{https://arxiv.org/abs/1411.1947}{1411.1947}}%
	%
	{\hypersetup{urlcolor = cyan}
		[\href{https://inspirehep.net/record/1326631}{\textsc{in}SPIRE}]
	}


\bibitem{Kuusk:2016rso} 
	P.\ Kuusk, M.\ R\"unkla, M.\ Saal and O.\ Vilson \textquotedblleft{}Invariant slow-roll parameters in scalar–tensor theories\textquotedblright{} \textit{Class.\ Quantum Grav.} \href{https://dx.doi.org/10.1088/0264-9381/33/19/195008}{\textbf{33}, 195008} (2016)%
	%
	{\hypersetup{urlcolor = magenta}
		arXiv:\href{https://arxiv.org/abs/1605.07033}{1605.07033}}%
	%
	{\hypersetup{urlcolor = cyan}
		[\href{https://inspirehep.net/record/1464827}{\textsc{in}SPIRE}]
	}


\bibitem{Damour} 
	T.\ Damour and K.\ Nordtvedt \textquotedblleft{}General relativity as a cosmological attractor of tensor-scalar theories\textquotedblright{} \textit{Phys.\ Rev.\ Lett.} \href{https://dx.doi.org/10.1103/PhysRevLett.70.2217}{\textbf{70}, 2217--9} (1993)%
	%
	{\hypersetup{urlcolor = cyan}
		[\href{http://inspirehep.net/record/342950}{\textsc{in}SPIRE}]
	}


\bibitem{Jarv:2015odu} 
	L.\ J\"arv and A.\ Toporensky \textquotedblleft{}General relativity as an attractor for scalar-torsion cosmology\textquotedblright{} \textit{Phys.\ Rev.} D \href{https://dx.doi.org/10.1103/PhysRevD.93.024051}{\textbf{93}, 024051} (2016)%
	%
	{\hypersetup{urlcolor = magenta}
		arXiv:\href{https://arxiv.org/abs/1511.03933}{1511.03933}}%
	%
	{\hypersetup{urlcolor = cyan}
		[\href{http://inspirehep.net/record/1404138}{\textsc{in}SPIRE}]
	}


\bibitem{Hohmann:2018vle} 
	M.\ Hohmann \textquotedblleft{}Scalar-torsion theories of gravity. I. General formalism and conformal transformations\textquotedblright{} \textit{Phys.\ Rev.} D \href{https://dx.doi.org/10.1103/PhysRevD.98.064002}{\textbf{98}, 064002} (2018)%
	%
	{\hypersetup{urlcolor = magenta}
		arXiv:\href{https://arxiv.org/abs/1801.06528}{1801.06528}}%
	%
	{\hypersetup{urlcolor = cyan}
		[\href{https://inspirehep.net/record/1649094}{\textsc{in}SPIRE}]
	}
	

\bibitem{Hohmann:2018dqh} 
	M.\ Hohmann and C.\ Pfeifer \textquotedblleft{}Scalar-torsion theories of gravity. II. $L\mathbf{(}T,\text{ }X,\text{ }Y,\text{ }\ensuremath{\phi}\mathbf{)}$ theory\textquotedblright{} \textit{Phys.\ Rev.} D \href{https://dx.doi.org/10.1103/PhysRevD.98.064003}{\textbf{98}, 064003} (2018)%
	%
	{\hypersetup{urlcolor = magenta}
		arXiv:\href{https://arxiv.org/abs/1801.06536}{1801.06536}}%
	%
	{\hypersetup{urlcolor = cyan}
		[\href{https://inspirehep.net/record/1649118}{\textsc{in}SPIRE}]
	}


\bibitem{Golovnev:2018wbh} 
	A.\ Golovnev and T.\ S.\ Koivisto \textquotedblleft{}Cosmological perturbations in modified teleparallel gravity models\textquotedblright{} 1--18 (2018)%
	%
	{\hypersetup{urlcolor = magenta}
		arXiv:\href{https://arxiv.org/abs/1808.05565}{1808.05565}}%
	%
	{\hypersetup{urlcolor = cyan}
		[\href{https://inspirehep.net/record/1687823}{\textsc{in}SPIRE}]
	}


\bibitem{Hohmann:2018jso} 
	M.\ Hohmann, M.\ Kr\v{s}\v{s}\'ak, C.\ Pfeifer and U.\ Ualikhanova \textquotedblleft{}Propagation of gravitational waves in teleparallel gravity theories\textquotedblright{} 1--10 (2018)%
	%
	{\hypersetup{urlcolor = magenta}
		arXiv:\href{https://arxiv.org/abs/1807.04580}{1807.04580}}%
	%
	{\hypersetup{urlcolor = cyan}
		[\href{https://inspirehep.net/record/1681973}{\textsc{in}SPIRE}]
	}


\bibitem{Damour:1996ke} 
	T.\ Damour and G.\ Esposito-Far\`ese \textquotedblleft{}Tensor--scalar gravity and binary--pulsar experiments\textquotedblright{} \textit{Phys.\ Rev.} D \href{https://dx.doi.org/10.1103/PhysRevD.54.1474}{{\bf 54}, 1474--91} (1996)%
	%
	{\hypersetup{urlcolor = magenta}
		arXiv:\href{https://arxiv.org/abs/gr-qc/9602056}{gr-qc/9602056}}%
	%
	{\hypersetup{urlcolor = cyan}
		[\href{https://inspirehep.net/record/416312}{\textsc{in}SPIRE}]
	}


\bibitem{AltahaMotahar:2018djk} 
	Z.\ Altaha Motahar, J.\ L.\ Bl\'{a}zquez-Salcedo, B.\ Kleihaus and J.\ Kunz \textquotedblleft{}Axial quasinormal modes of scalarized neutron stars with realistic equations of state\textquotedblright{} \textit{Phys.\ Rev.} D \href{https://dx.doi.org/10.1103/PhysRevD.98.044032}{{\bf 98}, 044032} (2018)%
	%
	{\hypersetup{urlcolor = magenta}
		arXiv:\href{https://arxiv.org/abs/1807.02598}{1807.02598}}%
	%
	{\hypersetup{urlcolor = cyan}
		[\href{https://inspirehep.net/record/1681550}{\textsc{in}SPIRE}]
	}


\bibitem{Wiki_block}
	Wikipedia contributors \textquotedblleft{}Block matrix: Block matrix inversion\textquotedblright{} \textit{Wikipedia, The Free Encyclopedia} online, accessed 30 May 2018 [\href{https://en.wikipedia.org/wiki/Block_matrix#Block_matrix_inversion}{link}]%
	
	The necessary result presented in the Wikipedia page can easily be verified, and the source is solely for inspiration.


\bibitem{MSEs}
	Question \textquotedblleft{}Diagonalizing a block matrix\textquotedblright{} answered by the user \textit{hardmath} in the Mathematics Stack Exchange forum [\href{https://math.stackexchange.com/questions/975534/diagonalizing-a-block-matrix}{link}]%
	
	The necessary result from the Mathematics Stack Exchange forum can easily be verified, and the source is solely for inspiration. 

\end{thebibliography}
\end{document}